\RequirePackage{lineno}
\documentclass[10pt,
aps,
pra,
twocolumn,
superscriptaddress,
showpacs,
altaffilletter,
nofootinbib,
amsmath,
 amssymb,
]{revtex4-2}

\usepackage{amsmath,amssymb}   \usepackage{graphicx}  \usepackage{bm}        \usepackage{dcolumn}   \usepackage{xspace}    \usepackage{color}     \usepackage{url}       \usepackage[alsoload=hep]{siunitx}   \usepackage{appendix}
\usepackage{caption}    
\usepackage{subcaption} \usepackage{multirow}
\usepackage[hidelinks]{hyperref}  \usepackage[nolist]{acronym} \usepackage{tikz}
\usetikzlibrary{calc}
\usepackage{gensymb}
\usepackage[export]{adjustbox}
\usetikzlibrary{shapes,arrows,positioning}

\newcommand{\amu}{\ensuremath{a^{}_{\mu}}\xspace}
\newcommand{\gm}{\ensuremath{g\!-\!2}\xspace}
\newcommand{\gmtwo}{\gm}
\newcommand{\oa}{\ensuremath{\omega^{}_a}\xspace}

\newcommand{\opmeas}{\omega_{p}^{\pp}\xspace}
\newcommand{\opprime}{\ensuremath{\omega'^{}_p}\xspace}
\newcommand{\opprimeatTexp}{\ensuremath{\omega'^{}_p(\Tr)}\xspace}
\newcommand{\opprimetilde}{\ensuremath{\tilde{\omega}'^{}_p}\xspace}

\newcommand{\opprimetildeatTexp}{\ensuremath{\tilde{\omega}'^{}_p(\Tr)}\xspace}

\newcommand{\BNLexperiment}{\ac{BNL} E821 experiment\xspace}
\newcommand{\FNALexperiment}{\ac{FNAL} Muon \gm Experiment\xspace}
\newcommand{\RunOne}{Run-1\xspace}
\newcommand{\RunTwo}{Run-2\xspace}
\newcommand{\RunThree}{Run-3\xspace}

\newcommand{\RunOneA}{Run-1a\xspace}
\newcommand{\RunOneB}{Run-1b\xspace}
\newcommand{\RunOneC}{Run-1c\xspace}
\newcommand{\RunOneD}{Run-1d\xspace}
\newcommand{\hethree}{\ensuremath{^3\text{He}}}

\newcommand{\pp}{\mathrm{cp}}

\newcommand{\opprimeq}{\ensuremath{\omega'^{}_{p,q}}\xspace}
\newcommand{\opprimej}{\ensuremath{\omega'^{}_{p,j}}\xspace}
\newcommand{\opprimeqj}{\ensuremath{\omega'^{}_{p,q,j}}\xspace}

\newcommand{\opprimetildeq}{\ensuremath{\tilde\omega'^{}_{p,q}}\xspace}
\newcommand{\opprimetildeqj}{\ensuremath{\tilde\omega'^{}_{p,q,j}}\xspace}
\newcommand{\Tr}{\ensuremath{T^{}_{r}\xspace}}

\newcommand{\sigmamuq}{\sigma^\mu_{q}}
\newcommand{\sigmamuqj}{\sigma^\mu_{q,j}}

\newcommand{\mi}{\ensuremath{m^{}_{i}}\xspace}
\newcommand{\mone}{\ensuremath{m^{}_{1}}\xspace}
\newcommand{\mtwo}{\ensuremath{m^{}_{2}}\xspace}
\newcommand{\mthree}{\ensuremath{m^{}_{3}}\xspace}
\newcommand{\mfour}{\ensuremath{m^{}_{4}}\xspace}
\newcommand{\mfive}{\ensuremath{m^{}_{5}}\xspace}
\newcommand{\msix}{\ensuremath{m^{}_{6}}\xspace}
\newcommand{\mseven}{\ensuremath{m^{}_{7}}\xspace}

\newcommand{\authornote}[1]{{\let\thempfn\relax \footnotetext[0]{$\diamond${ }#1}}}
 \newcolumntype{L}[1]{>{\raggedright\arraybackslash}p{#1}}
\newcolumntype{C}[1]{>{\centering\arraybackslash}p{#1}}
\newcolumntype{R}[1]{>{\raggedleft\arraybackslash}p{#1}}

\DeclareSIUnit\gauss{G}

\usepackage{etoolbox}

\ExplSyntaxOn
\cs_set_eq:NN \__siunitx_unit_output_number_sep:
  \__siunitx_unit_output_number_sep_aux:
\ExplSyntaxOff

\newcommand{\dd}{{\mathrm{d}}}

\newcommand{\grad}{\nabla}
\newcommand{\Ba}{{\bf a}}

\newcommand{\Bc}{{\bf c}}

\newcommand{\Bm}{{\bf m}}

\newcommand{\Br}{{\bf r}}
\newcommand{\Bs}{{\bf s}}

\newcommand{\Bx}{{\bf x}}

\newcommand{\BB}{{\bf B}}

\newcommand{\BJ}{{\bf J}}

\newcommand{\Bomega}{{\boldsymbol \omega}}
\newcommand{\BSigma}{\boldsymbol \Sigma}
\newcommand{\pdiff}[2]{\frac{\partial #1}{\partial #2}}
\newcommand{\diff}[2]{\frac{\mathrm{d} #1}{\mathrm{d} #2}}
\newcommand{\integral}[4]{\int_{#3}^{#4}\!\mathrm{d} #2 \; #1}
\newcommand{\avg}[1]{\left\langle #1 \right\rangle}
\newcommand{\norm}[1]{\left| #1 \right|}

\newcommand\x{{\bf \hat x}}
\newcommand\y{{\bf \hat y}}

\newcommand{\cm}{~\mathrm{cm}}
\newcommand{\ppb}{~\mathrm{ppb}}

\newcommand{\tr}{\mathrm{tr}}

\newcommand{\fp}{\mathrm{fp}}
\renewcommand{\cp}{\mathrm{cp}}

\newcommand{\ho}{\mathrm{ho}}

\newcommand{\mtr}{m^{\tr}}
\newcommand{\Bmtr}{\Bm^{\tr}}
\newcommand{\mfp}{m^{\mathrm{\fp}}}
\newcommand{\Bmfp}{\Bm^{\mathrm{\fp}}}

\newcommand{\st}{\ensuremath{s}}

\newcommand{\Bmfpst}{\Bmfp_\st}
\newcommand{\Bmtrst}{\Bmtr_\st}
\newcommand{\BJst}{\BJ^{}_\st}
\newcommand{\Jstij}{J^{}_{\st,\,ij}}
 
\setkeys{Gin}{clip=true,keepaspectratio=true} 

\begin{document}

   \title{\texorpdfstring{Magnetic Field Measurement and Analysis for the Muon \gm Experiment at Fermilab}{Magnetic Field Measurement and  Analysis for the Muon g-2 Experiment at Fermilab}}

    \affiliation{Argonne National Laboratory, Lemont, Illinois, USA}
\affiliation{Boston University, Boston, Massachusetts, USA}
\affiliation{Brookhaven National Laboratory, Upton, New York, USA}
\affiliation{Budker Institute of Nuclear Physics, Novosibirsk, Russia}
\affiliation{Center for Axion and Precision Physics (CAPP) / Institute for Basic Science (IBS), Daejeon, Republic of Korea}
\affiliation{Cornell University, Ithaca, New York, USA}
\affiliation{Fermi National Accelerator Laboratory, Batavia, Illinois, USA}
\affiliation{INFN Gruppo Collegato di Udine, Sezione di Trieste, Udine, Italy}
\affiliation{INFN, Laboratori Nazionali di Frascati, Frascati, Italy}
\affiliation{INFN, Sezione di Napoli, Napoli, Italy}
\affiliation{INFN, Sezione di Pisa, Pisa, Italy}
\affiliation{INFN, Sezione di Roma Tor Vergata, Roma, Italy}
\affiliation{INFN, Sezione di Trieste, Trieste, Italy}
\affiliation{Istituto Nazionale di Ottica - Consiglio Nazionale delle Ricerche, Pisa, Italy}
\affiliation{Department of Physics and Astronomy, James Madison University, Harrisonburg, Virginia, USA}
\affiliation{Institute of Physics and Cluster of Excellence PRISMA+, Johannes Gutenberg University Mainz, Mainz, Germany}
\affiliation{Joint Institute for Nuclear Research, Dubna, Russia}
\affiliation{Department of Physics, Korea Advanced Institute of Science and Technology (KAIST), Daejeon, Republic of Korea}
\affiliation{Lancaster University, Lancaster, United Kingdom}
\affiliation{Michigan State University, East Lansing, Michigan, USA}
\affiliation{North Central College, Naperville, Illinois, USA}
\affiliation{Northern Illinois University, DeKalb, Illinois, USA}
\affiliation{Northwestern University, Evanston, Illinois, USA}
\affiliation{Regis University, Denver, Colorado, USA}
\affiliation{Scuola Normale Superiore, Pisa, Italy}
\affiliation{School of Physics and Astronomy, Shanghai Jiao Tong University, Shanghai, China}
\affiliation{Tsung-Dao Lee Institute, Shanghai Jiao Tong University, Shanghai, China}
\affiliation{Institut für Kern- und Teilchenphysik, Technische Universit\"at Dresden, Dresden, Germany}
\affiliation{Universit\`a del Molise, Campobasso, Italy}
\affiliation{Universit\`a di Cassino e del Lazio Meridionale, Cassino, Italy}
\affiliation{Universit\`a di Napoli, Napoli, Italy}
\affiliation{Universit\`a di Pisa, Pisa, Italy}
\affiliation{Universit\`a di Roma Tor Vergata, Roma, Italy}
\affiliation{Universit\`a di Trieste, Trieste, Italy}
\affiliation{Universit\`a di Udine, Udine, Italy}
\affiliation{Department of Physics and Astronomy, University College London, London, United Kingdom}
\affiliation{University of Illinois at Urbana-Champaign, Urbana, Illinois, USA}
\affiliation{University of Kentucky, Lexington, Kentucky, USA}
\affiliation{University of Liverpool, Liverpool, United Kingdom}
\affiliation{Department of Physics and Astronomy, University of Manchester, Manchester, United Kingdom}
\affiliation{Department of Physics, University of Massachusetts, Amherst, Massachusetts, USA}
\affiliation{University of Michigan, Ann Arbor, Michigan, USA}
\affiliation{University of Mississippi, University, Mississippi, USA}
\affiliation{University of Oxford, Oxford, United Kingdom}
\affiliation{University of Rijeka, Rijeka, Croatia}
\affiliation{Department of Physics, University of Texas at Austin, Austin, Texas, USA}
\affiliation{University of Virginia, Charlottesville, Virginia, USA}
\affiliation{University of Washington, Seattle, Washington, USA}
\author{T.~Albahri}  \affiliation{University of Liverpool, Liverpool, United Kingdom}
\author{A.~Anastasi} \thanks{Deceased} \affiliation{INFN, Sezione di Pisa, Pisa, Italy}
\author{K.~Badgley}  \affiliation{Fermi National Accelerator Laboratory, Batavia, Illinois, USA}
\author{S.~Bae{\ss}ler} \altaffiliation[Also at ]{Oak Ridge National Laboratory}  \affiliation{University of Virginia, Charlottesville, Virginia, USA}
\author{I.~Bailey} \altaffiliation[Also at ]{The Cockcroft Institute of Accelerator Science and Technology}  \affiliation{Lancaster University, Lancaster, United Kingdom}
\author{V.~A.~Baranov}  \affiliation{Joint Institute for Nuclear Research, Dubna, Russia}
\author{E.~Barlas-Yucel}  \affiliation{University of Illinois at Urbana-Champaign, Urbana, Illinois, USA}
\author{T.~Barrett}  \affiliation{Cornell University, Ithaca, New York, USA}
\author{F.~Bedeschi}  \affiliation{INFN, Sezione di Pisa, Pisa, Italy}
\author{M.~Berz}  \affiliation{Michigan State University, East Lansing, Michigan, USA}
% ...
\collaboration{The Muon \gmtwo Collaboration} \noaffiliation
\vskip 0.25cm

\begin{abstract}
    The \FNALexperiment has measured the anomalous precession frequency $\amu \equiv (g^{}_{\mu}\!-\!2)/2$ of the muon to a combined precision of 0.46 parts per million with data collected during its first physics run in 2018. This paper documents the measurement of the magnetic field in the muon storage ring. The magnetic field is monitored by \acl{NMR} systems and calibrated in terms of the equivalent proton spin precession frequency in a spherical water sample at \SI{34.7}{\celsius}. The measured field is weighted by the muon distribution resulting in \opprimetilde, the denominator in the ratio \oa/\opprimetilde that together with known fundamental constants yields \amu. The reported uncertainty on \opprimetilde for the \RunOne data set is \SI{114}{ppb} consisting of uncertainty contributions from frequency extraction, calibration, mapping, tracking, and averaging of \SI{56}{ppb}, and contributions from fast transient fields of \SI{99}{ppb}.
\end{abstract}

   \maketitle
   \tableofcontents
   
   \begin{acronym}
  \acro{2D}{two-dimensional}
  \acro{3D}{three-dimensional}

\acro{ADC}{analog-to-digital converter}  
  \acro{ANL}{Argonne National Laboratory}

\acro{BNL}{Brookhaven National Laboratory}
  \acro{BSM}{Beyond the Standard Model}
  
\acro{COD}{closed orbit distortion}
  \acro{CPU}{central processing unit}
  \acro{ctag}{calorimeter tag}

\acro{DAC}{digital-to-analog converter}  
  \acro{DAQ}{data acquisition}  
  \acro{DQC}{data quality cuts}  

\acro{ESQ}{electrostatic quadrupole}

\acro{FFT}{fast fourier transform}
  \acro{FID}{free induction decay}
  \acro{FNAL}{Fermi National Accelerator Laboratory}
  \acro{FPGA}{field programmable gate arrays}

\acro{GPS}{global positioning system}
  \acro{GPU}{graphic processing unit}
  \acro{GUI}{graphic user interface}

\acro{HV}{high voltage}
\acro{IRIG-B}{inter-range instrumentation group code B}
  \acro{IC}{integrated circuit}

\acro{LVDS}{low-voltage differential signaling}
  \acro{LED}{light emmitting diode}

\acro{MIDAS}{maximum integrated data acquisition system}
 \acro{MRI}{magnetic resonance imaging}

\acro{NMR}{nuclear magnetic resonance}

\acro{ODB}{online database}

\acro{ppb}{parts per billion}
  \acro{ppm}{parts per million}
  \acro{ppt}{parts per trillion}
  \acro{PEEK}{polyether ether ketone}
  \acro{PLL}{phase-locked loop}
  \acro{POT}{potentiometer}
  \acro{QCD}{quantum chromodynamics}
  \acro{QED}{quantum electrodynamics}
  \acro{PID}{proportional–integral–derivative}
  \acro{PBSC}{polarizing beam splitter cube}

\acro{RF}{radio frequency}
  \acro{RMS}{root mean square}

\acro{SCC}{surface correction coils}
  \acro{SPI}{serial peripheral interface}
  \acro{SM}{Standard Model}

\acro{TDR}{technical design report}
  \acro{TI}{Texas Instrument}
  \acro{TTL}{transistor–transistor logic}
  \acro{TGG}{terbium gallium garnet}

\acro{UTC}{universal time coordinated}

\acro{VTM}{virtual trolley measurement}

\end{acronym}

    \graphicspath{{./introduction/figures/}}

\section{Introduction}
The Muon \gm collaboration reports a new measurement of the positive muon magnetic anomaly \amu$=(g^{}_{\mu}\!-\!2)/2$ \cite{prl2021}. The result is based on the \RunOne data set analysis, collected from March through July of 2018. The data are divided into four subsets grouped by different operating parameters of the experiment. These data subsets are analyzed separately and give consistent results for \amu. The combined \RunOne\ result is
\begin{linenomath}
        \begin{align}
                \amu(\textrm{FNAL}) = 116\,592\,040\,(54) \times 10^{-11} (0.46~\textrm{ppm}).
        \end{align}
\end{linenomath}
 {
Three companion papers to Ref. \cite{prl2021} provide the details for the key inputs to this result. Reference \cite{E989wapaper} details the analysis of the precision determination of the anomalous spin-precession frequency, \oa. Reference \cite{E989SRBDpaper} provides corrections to the \amu measurement that arise from effects of the muon beam dynamics. This paper provides data reconstruction, analysis, and systematic uncertainties of the measurement of the magnetic field in the muon storage ring.

The goal of the \acf{FNAL} Muon \gm Experiment is the determination of the muon magnetic anomaly with high precision~\cite{Grange:2015fou}.
There is great interest in this quantity because the standard model of particle physics is incomplete; this quantity is sensitive to potential new physics contributions not present in the current calculations.
The previous experiment at \ac{BNL}~\cite{Bennett:2006fi} shows a tension between the theoretical expectation and the experimental result of about 3.7\,$\sigma$~\cite{Aoyama:2020ynm}.
Since \amu is sensitive to a wide array of potential new physics contributions, both experimentalists \cite{prl2021} and theorists \cite{Aoyama:2020ynm,Aoyama:2012wk,Aoyama:2019ryr,Czarnecki:2002nt,Gnendiger:2013pva,Davier:2017zfy,Keshavarzi:2018mgv,Colangelo:2018mtw,Hoferichter:2019gzf,Davier:2019can,Keshavarzi:2019abf,Kurz:2014wya,Melnikov:2003xd,Masjuan:2017tvw,Colangelo:2017fiz,Hoferichter:2018kwz,Gerardin:2019vio,Bijnens:2019ghy,Colangelo:2019uex,Blum:2019ugy,Colangelo:2014qya} have worked to reduce their uncertainties. 
Contributions to \amu from \ac{QED}, electroweak theory, and \ac{QCD} loops have also been calculated to higher precision~\cite{Aoyama:2020ynm}.
This new result, from the \RunOne data set, differs by $3.3\,\sigma$ from the standard model prediction and agrees with the BNL E821 measurement. The combined experimental average results in a $4.2\,\sigma$ discrepancy with the theoretical calculation.

\subsection{The Muon \texorpdfstring{$g-2$}{g-2} Experiment}
\label{subsec:Introduction_Experiment}

In this experiment, pulses of polarized muons are injected with momentum $p=\SI{3.094}{\giga\electronvolt/c}$ into the magnetic storage ring shown in Fig.~\ref{fig:ring}. In the highly uniform vertical magnetic field of magnitude $\norm{\BB}\approx\SI{1.45}{\tesla}$, the muons circulate with a mean radius of \SI{7.112}{\meter} at the cyclotron frequency $\omega^{}_c / (2\pi) = \SI{6.7}{\mega\hertz}$. Their spin-precession frequency $\omega^{}_s / (2\pi)$ is the combination of their Larmor and Thomas precession, and differs slightly from the cyclotron frequency. The difference between these two frequencies is the rate at which the muons' helicity precesses, and is called the anomalous spin-precession frequency. For a muon in a uniform vertical magnetic field and an ideal horizontal orbit, the experimentally observed anomalous spin-precession frequency is
\begin{linenomath}\begin{equation}
		\Bomega^{}_a = \Bomega^{}_s - \Bomega^{}_c = - a^{}_{\mu}\frac{q}{m_{\mu}}\BB.\label{eq:amu_ideal}
\end{equation}\end{linenomath}
The measurement of both the magnitude of the anomalous spin-precession frequency $\oa=\norm{\Bomega^{}_{a}}$ and the storage ring magnetic field $\BB$ can be used to calculate \amu. Additional terms modifying Eq.~\eqref{eq:amu_ideal} originate in the experiment due to the electric focusing fields that are needed for vertical muon confinement and from muon motion that is not entirely perpendicular to $\norm{\BB}$. While the choice of the momentum strongly suppresses these additional terms, small corrections are applied when calculating \amu \cite{E989SRBDpaper}. Furthermore, the presence of an electric dipole moment of the muon would give rise to additional terms in Eq.~\eqref{eq:amu_ideal} \cite{PhysRevD.80.052008}.

\begin{figure}[ht]\centering
        \includegraphics[width=3.375in]{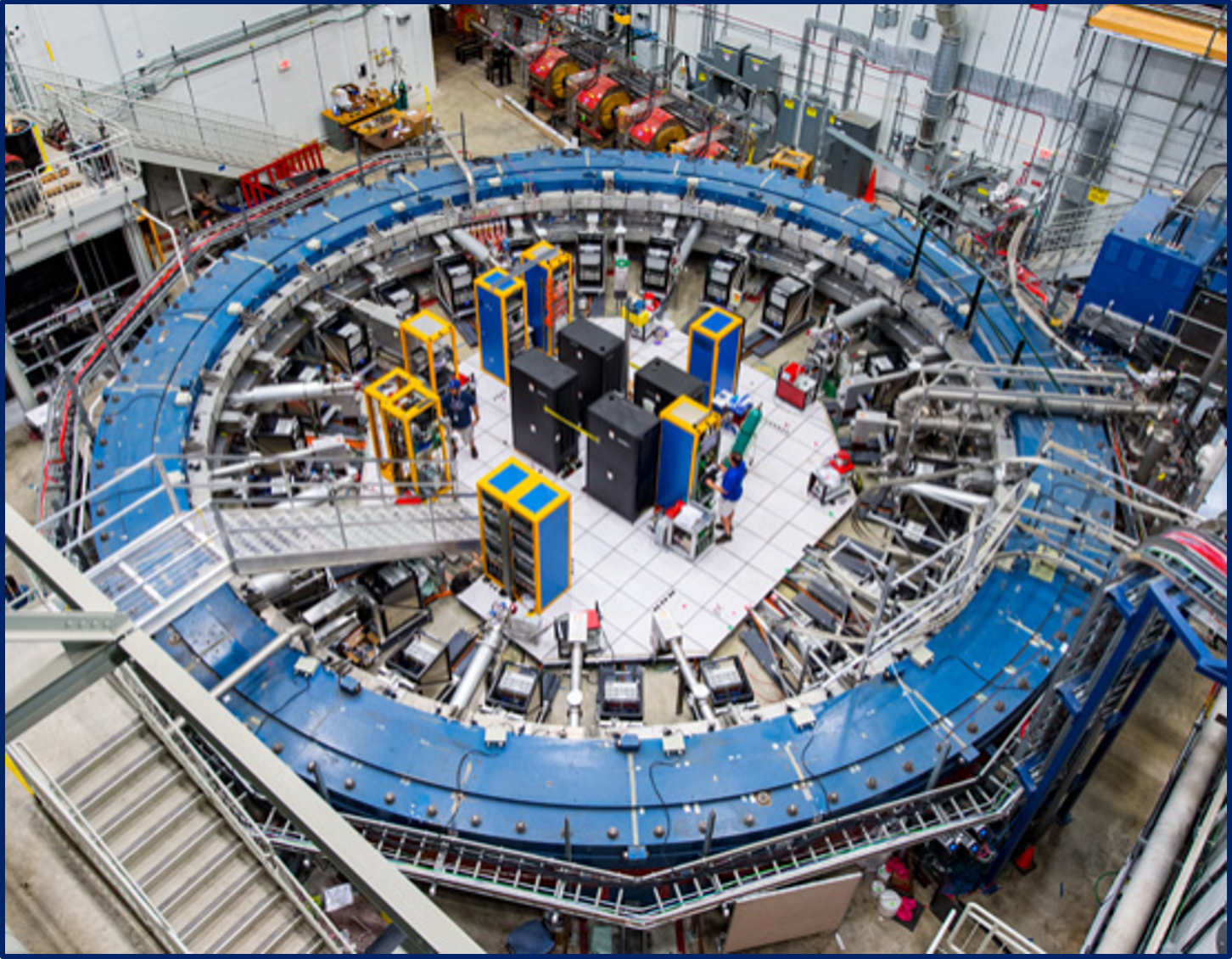}\caption{An image of the storage ring as prepared for \RunOne. Credit: Fermilab.}
        \label{fig:ring}\end{figure}

\begin{figure}[ht]
        \centering
        \includegraphics[width=3.375in]{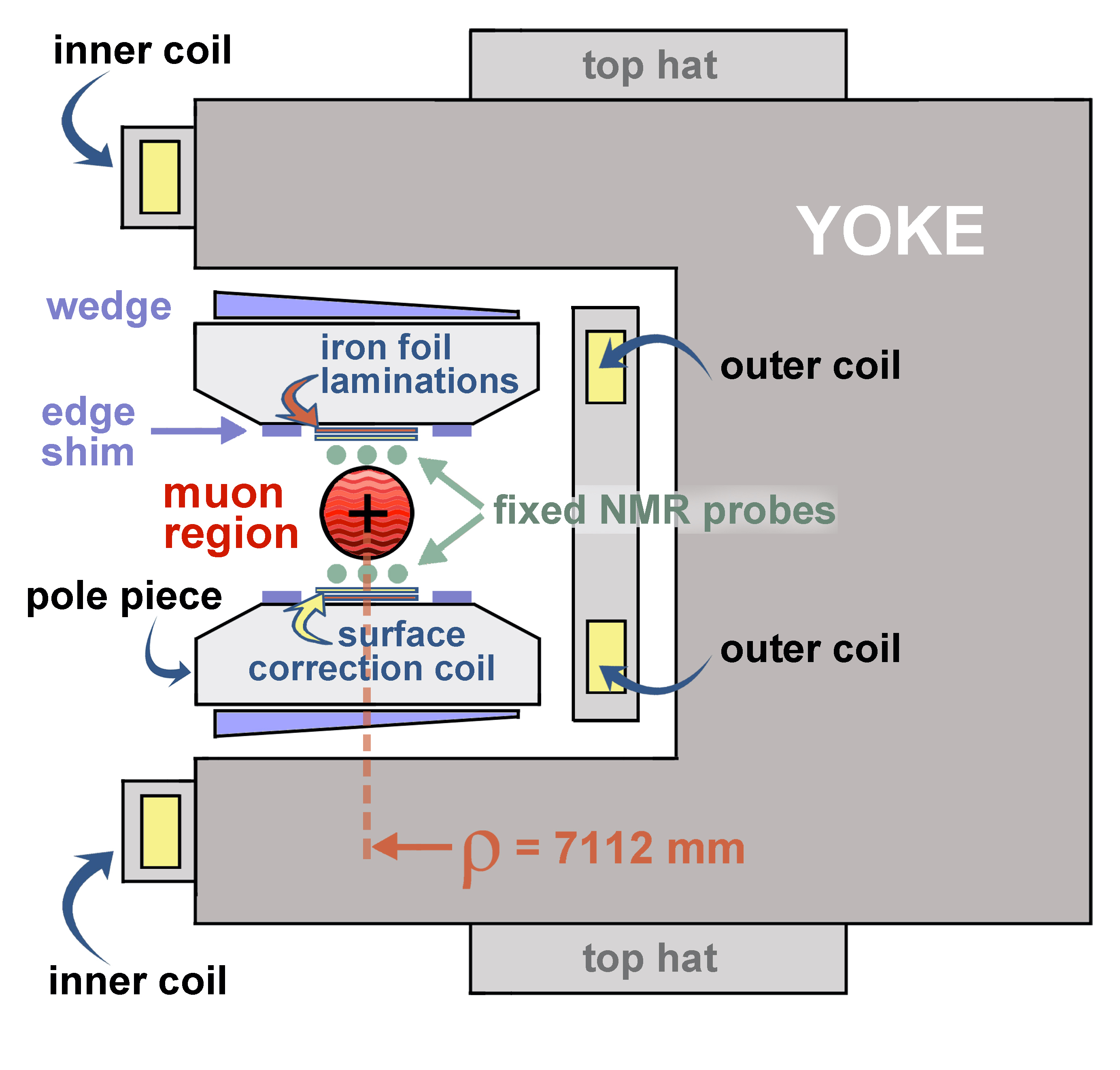}
        \caption{A cross section of the storage ring magnet featuring the components used to generate the highly uniform \SI{1.45}{\tesla} magnetic field in the \RunOne configuration.}
        \label{fig:ring_cross_section}
\end{figure}

The experiment was designed to balance the statistical and systematic uncertainties to reach its precision goal. The measurement of \oa \cite{E989wapaper} is based on the time dependence of the decay positrons above an energy threshold measured in 24 electromagnetic calorimeters \cite{fienberg:14, kaspar:17, khaw:19} with gain stabilized by a laser system \cite{anastasi:19}. Two in-vacuum straw trackers \cite{tracker_reference} provide the detailed information about the distribution of the muons in the storage ring that determines how the magnetic field is weighted and inform the beam-dynamics corrections to \amu \cite{E989SRBDpaper}.

A central component of the experiment is the precision superconducting magnetic storage ring that generates the magnetic field. Its main elements were designed for the \BNLexperiment and detailed in \cite{Danby:2001eh}.
The temporal stability and spatial homogeneity of the magnetic field are essential to the experiment.
Because the muon precession frequency is proportional to the strength of the magnetic field, we require that the average magnetic field experienced by the muons remain stable on the scale of \ac{ppm} throughout the experiment.
A very homogeneous field is required to minimize the uncertainty of the magnetic field maps caused by any nonuniformities in the muon distribution.

The magnet, operated in non-persistent mode, had a current of $\sim$\SI{5170}{\ampere}. Over long timescales, the magnetic field's stability is driven by thermal expansion and contraction of the magnet steel in response to temperature changes in the experimental hall. The magnetic field is stabilized by feedback to the magnet current supply from a set of \ac{NMR} magnetometers, described in Sec.~\ref{subsubsec:Introduction_Hardware_Systems}, distributed around the ring.

The homogeneity of the magnetic field required shimming with a suite of movable elements labeled in Fig.~\ref{fig:ring_cross_section} that can fine tune the magnetic field in localized regions during data collection periods.
Precision positioning of the 72 pole pieces (36 each upper and lower) drives the overall field strength, while their pitch with respect to horizontal drives the linear gradients. Additional pieces of iron were positioned along the surfaces of the pole pieces (edge shims and iron foil laminations), in the air gap between the pole pieces and yoke (wedges), and the top and bottom of the 24 yoke pieces (top hats). These were used to fine tune the average field as a function of azimuth and control gradients in the direction transverse to the beam propagation.
A set of coils, called \acf{SCC}, are installed on the surface of the pole pieces.
The \ac{SCC} consists of 100 individually powered, concentric coils on each of the upper and lower pole surfaces.
Specific current distributions were used to minimize the field variations across the beam aperture to better than \SI{1}{ppm} when averaged over the storage ring azimuth, and updated periodically in response to magnetic field drifts.
Shimming resulted in a field homogeneity over the storage volume of roughly 14\,ppm RMS, a threefold improvement~\cite{Smith:phd} in the azimuthal variation of the average field compared to the \BNLexperiment \cite{Bennett:2006fi}.

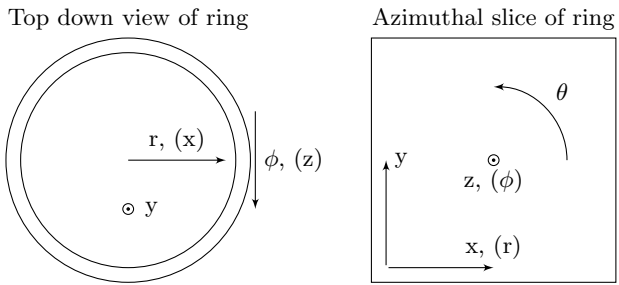
\begin{figure}
	\centering
	\begin{tikzpicture}[scale=0.065]
		\node[anchor=south] at (0,25) {Top down view of ring};
		\draw (0,0) circle (25);
		\draw (0,0) circle (22);
		\draw[-latex'] (0,0) -- (20,0);
		\node[anchor=south] at (10,0) {r, (x)};
		\draw[-latex'] (26,10) -- (26,-10);
		\node[anchor=west] at (26,0) {$\phi$, (z)};
		\draw[fill=black] (0,-10) circle (0.25) node[anchor=west] {~y};
		\draw (0,-10) circle (1.1);
	\end{tikzpicture}
	\quad
	\begin{tikzpicture}[scale=0.065]
		\node[anchor=south] at (0,25) {Azimuthal slice of ring};
		\draw (-25,25) rectangle (25,-25);
		\draw[-latex'] (-22,-21.5) -- (-22, 0) node[anchor=west] {y};
		\draw[-latex'] (-21.5,-22) -- (0, -22) node[anchor=south] {x, (r)};
		\draw[fill=black] (0,0) circle (0.25) node[anchor=north] {z, ($\phi$)};
		\draw[-latex'] (0:15) arc (0:90:15);
		\node[anchor=center] at (14,14) {$\theta$};
		\draw (0,0) circle (1.1);
	\end{tikzpicture}
	\caption{The coordinate systems used in this paper. The muon beam nominal orbital radius is at $r = \SI{7.112}{\meter}$ in the $ry\phi$ basis, equivalent to $x=\SI{0}{\centi\meter}$ in the $xyz$ basis.}
	\label{fig:coordinates}	
\end{figure}

Figure~\ref{fig:coordinates} shows the coordinate systems we use in this paper. There are two primary reference frames: a top-down view of the entire storage ring used mostly for considering azimuthally dependent effects, and a cross section through the ring used for considering the radially and vertically dependent effects. The coordinate $y$ always refers to the direction of the axis of the storage ring in both systems. The coordinate $r$ in the top-down system is replaced by the coordinate $x$ in the cross-section system. They are related by $x = r - \SI{7.112}{\meter}$. The azimuthal angle in the top-down system is represented by $\phi$. In the cross-section system, it is replaced by $z$.

\subsection{Measuring the Magnetic Field}
\label{subsec:Introduction_Measuring_Magnetic_Field}

Equation~\eqref{eq:amu_ideal} shows that determining $a_{\mu}$ from $\oa$ requires precise knowledge of the magnetic field magnitude experienced by the muons, which we measured with pulsed proton \ac{NMR}. This technique, pioneered by Bloch \cite{PhysRev.69.127} and Purcell \cite{PhysRev.69.37}, has been employed since the 1950s~\cite{Becker:1993} across a wide range of chemical and physical applications, routinely demonstrating accuracy and precision at the \ac{ppm} and even \ac{ppb} scales. The \ac{NMR} devices (or magnetometers) are called probes. A careful sequence of calibrations and synchronizations is performed to relate the magnetic field to the Larmor precession frequency of protons shielded in a spherical water sample at a reference temperature $T$. The average field over the muon distribution weighted by the detected decays over time is $\tilde B$. The frequency measurements determine $\tilde B$ when combined with the shielded proton magnetic moment $\mu'^{}_p(T)$ via
\begin{linenomath}\begin{equation}
	\tilde{B}=\frac{\hbar \opprimetilde(T)}{2\mu'^{}_p(T)} = \frac{\hbar \tilde{\omega}'^{}_p(T)}{2} \frac{\mu^{}_e(H)}{\mu'^{}_p(T)}\frac{\mu^{}_e}{\mu^{}_e(H)}\frac{1}{\mu^{}_e}.
	\label{eq:Btoomega}
\end{equation}\end{linenomath}
Here, $\mu^{}_e(H)/\mu'^{}_p(T)$ is the ratio of the magnetic moments of an electron bound in hydrogen to that of a proton shielded in a spherical water sample, measured to \SI{10.5}{ppb} at a water temperature $\Tr=\SI{34.7}{\celsius}$ ~\cite{Phillips:1977}. The bound-state QED corrections that determine the magnetic moment ratio of the electron bound in hydrogen versus a free electron $\mu^{}_e(H)/\mu^{}_e$ are considered essentially exact \cite{Mohr:2015ccw}, and the electron magnetic moment $\mu^{}_e$ is known to \SI{0.3}{ppb} \cite{Mohr:2015ccw}. Combining Eqs.~\eqref{eq:amu_ideal}, \eqref{eq:Btoomega}, and $\mu^{}_e = \frac{g^{}_e}{2} \frac{e}{m^{}_e} \frac{\hbar}{2}$ yields
\begin{linenomath}\begin{align}
	a^{}_{\mu} = \frac{\omega^{}_a}{\opprimetilde(\Tr)} \frac{\mu'^{}_p(\Tr)}{\mu^{}_e(H)} \frac{{\mu^{}_e(H)}}{\mu^{}_e} \frac{m^{}_{\mu}}{m^{}_e} \frac{g^{}_e}{2}.\label{eq:amu}
\end{align}\end{linenomath}
 The ratio of the mass of the muon and the mass of the electron $m^{}_\mu/m^{}_e$ is known to \SI{22}{ppb} from the measurement of the hyperfine splitting of muonium~\cite{Liu:1999iz} and bound-state QED \cite{Mohr:2015ccw}. Finally, the $g$ factor of the electron $g^{}_e$ is known to \SI{0.28}{ppt} \cite{Hanneke:2011}.

To determine $\opprimetildeatTexp$, we perform a sequence of measurements  with proton-rich magnetometers:
\begin{enumerate}
	\item The 17 \ac{NMR} probes of the in-vacuum trolley are calibrated in terms of the equivalent \opprimeatTexp with a precision calibration probe containing a pure water sample. The calibration probe's precise measurements are corrected for material effects, temperature, and field variations during the calibration to achieve high accuracy and precision.
	\item The magnetic field in the muon storage volume is mapped using the trolley approximately every three days. The result is called a trolley map or field map.
	\item The 378 fixed \ac{NMR} probes, located in 72 azimuthal stations, are synchronized to the trolley measurements. These fixed probes are located above and below the storage volume and regularly spaced around the ring to track the field's evolution between trolley maps.
	
	\item The magnetic-field maps are weighted by the temporal and spatial distributions of those muons included in the $\omega_{a}$ measurement.
        \item Corrections are applied for the presence of fast transient fields generated by pulsed muon injection systems that are not resolved by the asynchronous magnetic-field tracking and not present during the trolley measurements.
\end{enumerate}

\subsection{Hardware Systems}\label{subsubsec:Introduction_Hardware_Systems}

The precision calibration probe employed in the first step of the measurement sequence is shown in Fig.~\ref{fig:pp-schematic}. This probe is highly symmetric and uses an ultrapure, cylindrical water sample. It is constructed from a combination of paramagnetic and diamagnetic materials so that the total correction due to its intrinsic magnetic influence is less than \SI{10}{ppb} \cite{pp_paper}. The calibration probe's total uncertainty on the corrections is less than \SI{20}{ppb}, corroborated through cross calibrations with both a spherical water sample \cite{Fei:1997sd} and \hethree~\cite{Farooq2020}. The calibration probe is used to generate calibration constants for each of the trolley probes. It is operated inside the vacuum chambers and mounted on a \ac{3D} translation stage that allows it to match each trolley probe's position using applied magnetic-field gradients. Details of the calibration procedure are given in Sec.~\ref{sec:pp-trly-calib}.

\begin{figure}[ht]
        \includegraphics[width=\linewidth]{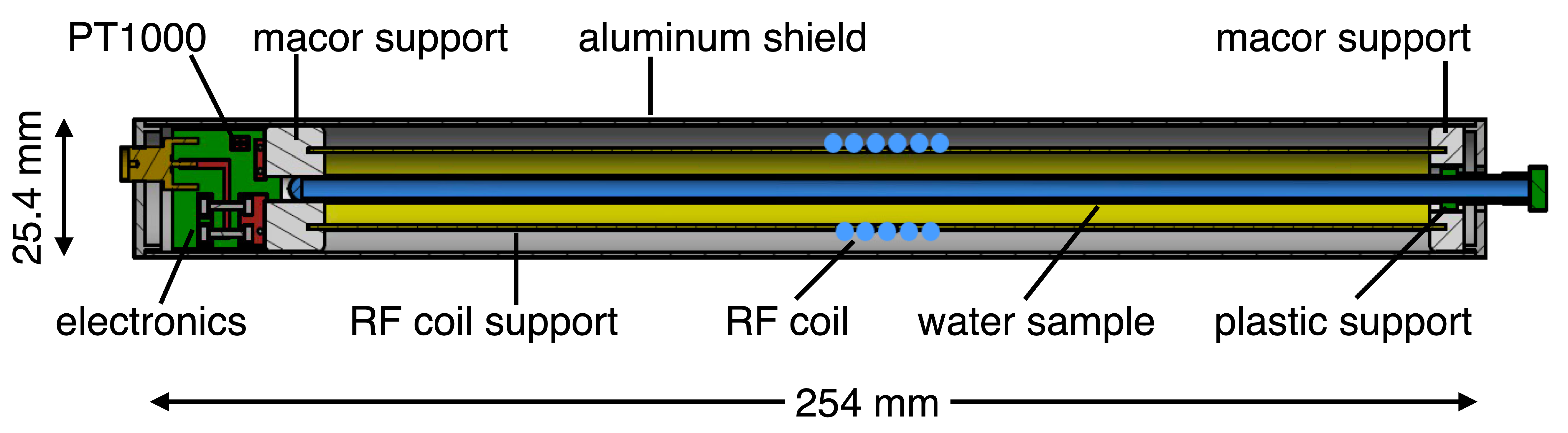}
	\caption{Schematic drawing of the calibration probe used to calibrate the trolley probe measurements.}
	\label{fig:pp-schematic}
\end{figure}

Figure~\ref{fig:fxpr-schematic} shows the design of the trolley and fixed probes, which are based on a similar design from the \BNLexperiment \cite{PRIGL1996118}. The cylindrical sample volume in each probe is filled with petroleum jelly, chosen for its low volatility. The trolley shell and its mechanical hardware for the motion (rails and drums) were from the \BNLexperiment and the trolley electronics, position encoders, and controllers were upgraded for this experiment as detailed in \cite{Corrodi:2020}.

\begin{figure}[ht]
	\includegraphics[width=\linewidth]{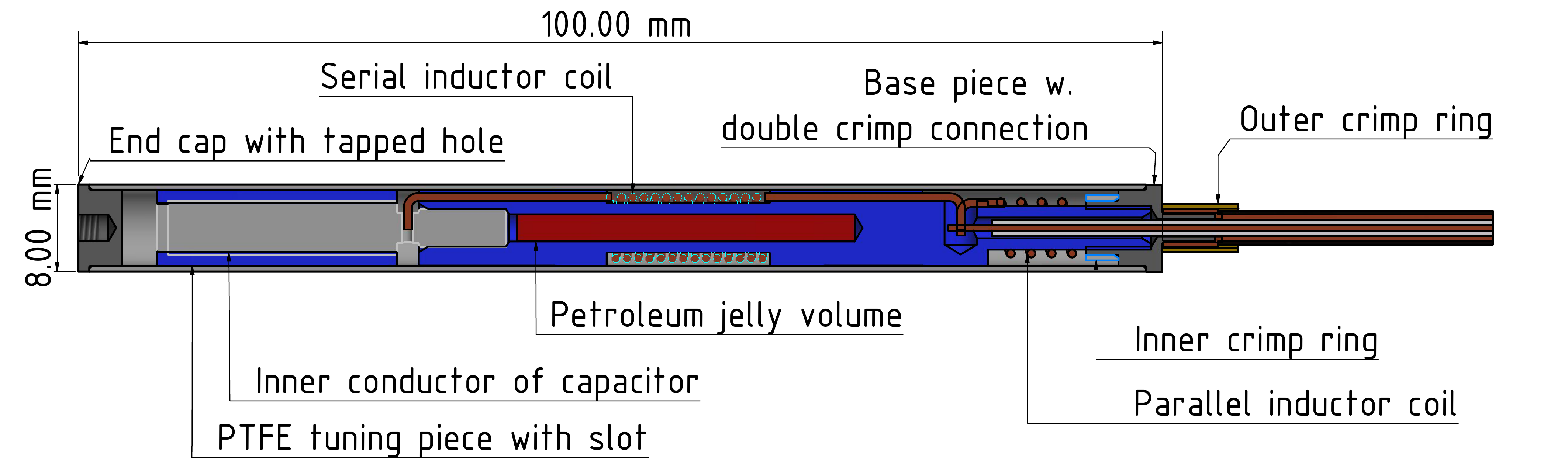} 
	\caption{Schematic drawing of the \ac{NMR} probe for field mapping and monitoring.}
	\label{fig:fxpr-schematic}
\end{figure}

The calibrated trolley is used to produce detailed field maps over the entire azimuth of the storage ring.
The muon storage region extends in the $x$ and $y$ directions to $\pm$\SI{4.5}{\centi\meter}, defined by a set of five circular collimators placed at various azimuthal positions around the storage ring.
In order to determine the magnetic field in the muon storage region, the trolley's 17 \ac{NMR} probes are arranged in the configuration shown in Fig.~\ref{fig:trolley_layout_basic}.
The trolley is pulled by two cables along rails in the storage ring vacuum chamber, and the field is sampled in $\sim$9000 azimuthal locations.
The analysis of the trolley maps is detailed in Sec.~\ref{sec:Trolley_Data_Analysis}.

\begin{figure}
    \centering
	\begin{tikzpicture}[scale=0.065]
	\draw [dotted] (-50,0) -- (50,0);
	\draw [dotted] (-55,77) -- (55,77);
	\draw [dotted] (-50,-77) -- (50,-77);
	\draw [dotted] (0,-80) -- (0,80);
	\draw [dotted] (-17.5, 0) -- (-17.5, -50);
	\draw [dotted] (-35, 0) -- (-35, -50);
	\draw [dotted] (45, 0) -- (45, -50);
	\draw [dotted] (30, -50) -- (30, -77);
	\draw [dotted] (-50, 45) -- (0, 45);
	\draw[black, fill=white] (0,0)           circle (4.3) node[anchor=center] {1};
	\draw[black, fill=white] (0,-17.5)       circle (4.3) node[anchor=center] {2};
	\draw[black, fill=white] (17.5 ,0)       circle (4.3) node[anchor=center] {3};
	\draw[black, fill=white] (0, 17.5)       circle (4.3) node[anchor=center] {4};
	\draw[black, fill=white] (-17.5 ,0)      circle (4.3) node[anchor=center] {5};
	\draw[black, fill=white] (0, -35.)       circle (4.3) node[anchor=center] {6};
	\draw[black, fill=white] (17.5,-30.31)   circle (4.3) node[anchor=center] {7};
	\draw[black, fill=white] (30.31,-17.5)   circle (4.3) node[anchor=center] {8};
	\draw[black, fill=white] (35.,0)         circle (4.3) node[anchor=center] {9};
	\draw[black, fill=white] (30.31, 17.5)   circle (4.3) node[anchor=center] {10};
	\draw[black, fill=white] (17.5,30.31)    circle (4.3) node[anchor=center] {11};
	\draw[black, fill=white] (0, 35. )       circle (4.3) node[anchor=center] {12};
	\draw[black, fill=white] (-17.5,30.3)    circle (4.3) node[anchor=center] {13};
	\draw[black, fill=white] (-30.31,17.5)   circle (4.3) node[anchor=center] {14};
	\draw[black, fill=white] (-35,0)         circle (4.3) node[anchor=center] {15};
	\draw[black, fill=white] (-30.31, -17.5) circle (4.3) node[anchor=center] {16};
	\draw[black, fill=white] (-17.5,-30.31)  circle (4.3) node[anchor=center] {17};
	\draw[black] (0,0)           circle (45) node[anchor=center] { };
	\draw[black, fill=lightgray] (-30, 77)   circle (4.3) node[anchor=center] {TI};
	\draw[black, fill=white] (0,77)      circle (4.3) node[anchor=center] {TM};
	\draw[black, fill=white] (30,77)     circle (4.3) node[anchor=center] {TO};
	\draw[black, fill=lightgray] (-30,-77)   circle (4.3) node[anchor=center] {BI};
	\draw[black, fill=white] (0,-77)     circle (4.3) node[anchor=center] {BM};
	\draw[black, fill=white] (30,-77)    circle (4.3) node[anchor=center] {BO};
\draw[<->, thick] (-30,65) -- ( 0,65) node[midway, fill=white] {\SI{30}{\milli\meter}};
	\draw[<->, thick] (. 0,65) -- (30,65) node[midway, fill=white] {\SI{30}{\milli\meter}};
	\draw[->, thick] (-54,-50) -- (50,-50) node[yshift=-25, xshift=-15] {X [mm]};
	\draw[->, thick] (-50,-80) -- (-50,90) node[xshift=-40, yshift=-15, rotate=90] {Y [mm]};
	\draw (0, -48) -- (0, -52)         node[below] {0.0};
	\draw (-17.5, -48) -- (-17.5, -52) node[below] {-17.5};
	\draw (-35, -48) -- (-35, -52)     node[below] {-35.0};
	\draw (45, -48) -- (45, -52)     node[below] {45.0};
	\draw (30, -48) -- (30, -52)     node[below] {30.0};
	\draw (-48, -77) -- (-52, -77) node[left] {-77.0};
	\draw (-48,  77) -- (-52,  77) node[left] { 77.0};
	\draw (-48,  0) -- (-52,  0) node[left] { 0.0};
	\draw (-48,  45) -- (-52,  45) node[left] { 45.0};
\end{tikzpicture}
	\caption{The layout of the 17 probes in the trolley. Positive $x$ is towards higher radius. The fixed probe locations on the top (T) and bottom (B) of the storage region are shown as well. For the fixed probes, the six-probe stations have probes in the inner (I), middle (M), and outer (O) positions. In the four-probe stations, only the middle and outer probes are present.}
	\label{fig:trolley_layout_basic}
\end{figure}

The trolley system includes electronics to control the \ac{NMR} sequence and to read out the digitized \ac{FID} signals. The initial $\sim$\SI{61.79}{\mega\hertz} signal, corresponding to $\norm{\BB} \approx \SI{1.45}{\tesla}$, is mixed down to approximately \SI{50}{\kilo\hertz} prior to digitization and transferred through an electronic interface to a \ac{DAQ} computer. A bar code scanner on the trolley reads marks etched into the bottom of the storage ring vacuum chambers that are analyzed to determine the trolley's azimuthal position. 

In order to measure the field experienced by the muons, ideally the trolley maps would be taken under the identical conditions that exist during muon injections.
In reality, three main configuration changes are needed for field mapping: i) the pulsed beam injection systems [kicker and \ac{ESQ}] are switched off, ii) the beam collimators are moved from their regular positions because they would physically interfere with the trolley, and iii) the garage rail is moved into the storage region to insert the trolley. Dedicated measurements and calculations were made to correct for these modified conditions and are described in Secs.~\ref{subsubsec:Trolley_Other_Systematic_Effects} and~\ref{sec:transients}.

The 378 fixed probes mounted above and below the storage region to continuously track the field drift are synchronized with the trolley measurements during each mapping run. Because trolley runs interrupt muon data taking, the detailed field mapping is only performed approximately every three days, driven by the fixed probes' field tracking capability. The fixed probes provide information about the field drift during the muon data taking periods between trolley maps. Four or six probes (see Fig.~\ref{fig:trolley_layout_basic}) are installed  at 72 azimuthal locations, called stations, regularly spaced around the storage ring, allowing continuous monitoring of the magnetic field at each azimuthal station. The fixed probe \acp{FID} are read out through 20 multiplexers and mixed down to about \SI{50}{\kilo\hertz} and digitized. A computer controls the read sequence, including the probe selection and the recording of the digitized waveforms.
The synchronization of the trolley measurements to the fixed probes and the subsequent field tracking are discussed in Sec.~\ref{sec:Interpolation}.

The magnetic field DAQ serves as an access point for controlling individual field measurement systems. These include fixed probes, trolley control, trolley readout, calibration probe control, power supply feedback, surface coil settings, and environmental fluxgate sensors. These systems are each managed by custom front ends that run asynchronously and communicates with a common \ac{DAQ} core. The field \ac{DAQ} uses standalone hardware that runs independently from the detector \ac{DAQ}, which controls the rest of the Muon \gm Experiment. The field \ac{DAQ} collects data whenever the magnet is powered and runs decoupled from the main DAQ for the calorimeters, trackers, pulsed injection systems, and other hardware. The field and main \acp{DAQ} used a common \SI{10}{\mega\hertz} time reference disciplined by a Rb-clock and a \acl{GPS}, allowing measurement time stamps to be correlated with data from the detector \ac{DAQ} with high precision.

\subsection{Magnetic Field Analysis}
\label{subsec:Introduction_Magnetic_Field_Analysis}

The data are analyzed to extract \opprimetildeatTexp as one input for the calculation of \amu. The evaluation of the trolley and fixed probe data is based on multipole and Cartesian moments described in Sec.~\ref{subsubsec:Introduction_Trolley_Multipole_Expansion}. They form the basis for various steps in the overall analysis, which is outlined in Sec.~\ref{subsubsec:Introduction_Analysis_Flow}. Throughout the rest of this paper, we provide the details of these analysis steps and their implementation. For many steps, there were two or three parallel analysis implementations by independent teams that cross checked each other and refined systematic uncertainties. We highlight important analysis differences between the independent teams in Sec.~\ref{subsubsec:Introduction_Multiple_Analysis_Approaches}.

\subsubsection{Multipole and Cartesian Moments}
\label{subsubsec:Introduction_Trolley_Multipole_Expansion}

The \ac{NMR} probes measure the magnitude of the magnetic field, $\norm{\BB} = \sqrt{B^2_x+B^2_y+B^2_z}$, and are often referred to as ``scalar magnetometers.'' Due to the design of the magnet and the shimming, the magnetic field is predominantly in the $y$ direction, i.e., $B^{}_x,\ B^{}_z \ll B^{}_y$. The difference between the \ac{NMR} measurement of $\norm{\BB}$ and the field component in the $y$ direction can be approximated to first order as
\begin{linenomath}\begin{equation}
		\norm{\BB} - B^{}_y \approx \frac{B^2_x+B^2_z}{2 B^{}_y}.
\end{equation}\end{linenomath}
During the shimming procedure, measurements of the radial and longitudinal components, $B^{}_x$ and $B^{}_z$, were performed at $\approx 100$ azimuthal locations. The azimuthally averaged radial field was determined to be $B^{}_x/\norm{\BB} <40$~\ac{ppm} during \RunOne with the applied \ac{SCC} settings, and the measurement of the average longitudinal field was consistent with zero. Local variations in the longitudinal component were typically $B^{}_z/\norm{\BB} <\SI{100}{ppm}$ with respect to $\norm{\BB}$, leading to $(\norm{\BB} - B^{}_y)/ \norm{\BB} = \mathcal{O}(\SI{10}{ppb})$. Therefore, it is well-justified (at our desired accuracy) to replace $\norm{\BB}$ with $B^{}_y$ and focus on its extraction from the data. From here forward, we will use the convention $B = \norm{\BB}$ and make the approximation $B \approx B^{}_y$.

The measurements from the trolley and fixed probes represent the field magnitudes $B(x,y,\phi=\phi^{}_k)$ at an azimuthal slice $\phi^{}_k$. We can extract the field's spatial dependence in these \ac{2D} slices in terms of moments $\mi$ of the magnetic field. For the trolley probe geometry, the parametrization of $B$ in a slice comes from the general solution to the source-free Laplace equation for the scalar potential in polar coordinates $(r,\ \theta)$,
\begin{linenomath}\begin{equation}
		B \approx B^{}_y = A^{}_0 + \sum_{n=1} \left(\frac{r}{r^{}_0}\right)^n [A^{}_n \cos(n \theta) + B^{}_n \sin(n \theta)],
		\label{eq:mult_By}
\end{equation}\end{linenomath}
where, here and in Table~\ref{tab:mults_in_Cart} only, $r = \sqrt{x^2+y^2}$ is the in-slice radius from the center of the muon orbit and $r^{}_0=\SI{4.5}{\centi\meter}$ is a normalization to the outer edge of the muon storage region. The $A^{}_n$ and $B^{}_n$ parameters are the multipole strengths, also known as the normal and skew multipoles, respectively. These names are often written as ``normal/skew (2$n$+2)-pole,'' such as the ``normal 2-pole (normal dipole),'' ``skew 4-pole (skew quadrupole)," or ``normal 6-pole (normal sextupole).'' The 17 trolley measurements from a given azimuthal slice are transformed into the multipole basis defined by Eq.~\eqref{eq:mult_By}.

The fixed probe geometry for both the four- and six-probe stations (see Fig.~\ref{fig:trolley_layout_basic}) are symmetric in a Cartesian coordinate system and are therefore parameterized as Cartesian field moments, which are analogous to the multipole moments. These Cartesian moments are the $x$ and $y$ derivatives of $B^{}_y$ evaluated at $x=y=0$. These moments are also normalized to $r_0$ in analogy with the multipole moments. The fixed probe measurements are used to make discrete estimates of the moments by calculating sums and differences of the measurements.

Table~\ref{tab:mults_in_Cart} summarizes the moments $\mi$ in terms of the trolley multipole moments and the fixed probe Cartesian moments. Only six (four) moments can be calculated at a six-probe (four-probe) station as indicated in the Cartesian moment columns. Given the discrete positions of the fixed probes, it is possible to estimate the values of these moments at the center of the storage region in terms of the multipole strengths defined in Eq.~\eqref{eq:mult_By}, implying that the fixed probes can be used to track the lower-order moments up to $\msix$ in the time between the trolley maps. In practice, we only use the fixed probes to track the first five moments due to the high uncertainty associated with the sixth moment and its relative unimportance in the final result.

\begin{table*}[tb]
	\centering
\begin{tabular}{lccccc}\hline\hline
		&\rule{0pt}{1em} Trolley & \multicolumn{4}{c}{Fixed probe stations} \\ \cline{3-6}
		Moment (common name) & multipole & \rule{0pt}{1em}Cartesian & Multipole& \multicolumn{2}{c}{Cartesian moment} \\ \cline{5-6}
		& $B_y(r,~\theta)$ & derivative & $B_y(x,~y)$& 6-probe station & 4-probe station \\
		\hline
		\rule{0pt}{1em}$\mone$ (normal dipole) & $A_0$ & $B_y$ & $A_0$ & $A_0$ & $A_0$ \\
		$\mtwo$ (normal quadrupole) & $A_1 \frac{r}{r_0} \cos(\theta)$ & $\pdiff{B_y}{x}$ & $A_1 \frac{1}{r_0} x$ & $\frac{A_1}{r_0}$ & $\frac{A_1}{r_0}$ \\
		$\mthree$ (skew quadrupole) & $B_1 \frac{r}{r_0} \sin(\theta)$ & $\pdiff{B_y}{y}$ & $B_1 \frac{1}{r_0} y$ & $\frac{B_1}{r_0}$ & $\frac{B_1}{r_0}$ \\
		$\mfour$ (skew sextupole) & $B_2 \left(\frac{r}{r_0}\right)^2 \sin(2 \theta)$ & $\pdiff{^2B_y}{x\partial y}$ & $2 B_2 \left(\frac{1}{r_0}\right)^2 xy$ & $\frac{2B_2}{r_0^2}$ & $\frac{2B_2}{r_0^2}$ \\
		$\mfive$ (normal sextupole) & $A_2 \left(\frac{r}{r_0}\right)^2 \cos(2 \theta)$ & $\pdiff{^2B_y}{x^2}$ & $2 A_2 \left(\frac{1}{r_0}\right)^2 (x^2 - y^2)$ & $\frac{2A_2}{r_0^2}$ & - \\
		$\msix$ (skew octupole) & $B_3 \left(\frac{r}{r_0}\right)^3 \cos(3 \theta)$ & $\pdiff{^3B_y}{x^2\partial y}$ & (Unused) & (Unused) &  \\
		$\mseven$ (normal octupole) & $A_3 \left(\frac{r}{r_0}\right)^3 \sin(3 \theta)$ & $\vdots$ &  &  &  \\
		\multicolumn{1}{c}{$\vdots$} & $\vdots$ & $\vdots$ & & &  \\
		\hline\hline
	\end{tabular}
	\caption{The first measurable moments for both the multipole and Cartesian basis. The parameters $A_n$ and $B_n$ are the multipole strengths for the normal and skew moments, respectively, defined in Eq.~\eqref{eq:mult_By}. Here, $r = \sqrt{x^2+y^2}$. In this experiment $r_0 = 4.5\,\si{\cm}$, a scale set by the radius of the collimated muon beam. Notice that evaluating these moments at $(0,0)$ recovers the multipole strengths, creating a relationship between the Cartesian and multipole moments.}
	\label{tab:mults_in_Cart}
\end{table*}
 
\subsubsection{Analysis Flow}
\label{subsubsec:Introduction_Analysis_Flow}

\tikzstyle{block} = [rectangle, draw, fill=white, text width=0.165\textwidth, text centered, rounded corners, minimum height=4em, font=\footnotesize]
\tikzstyle{line} = [draw, -latex']

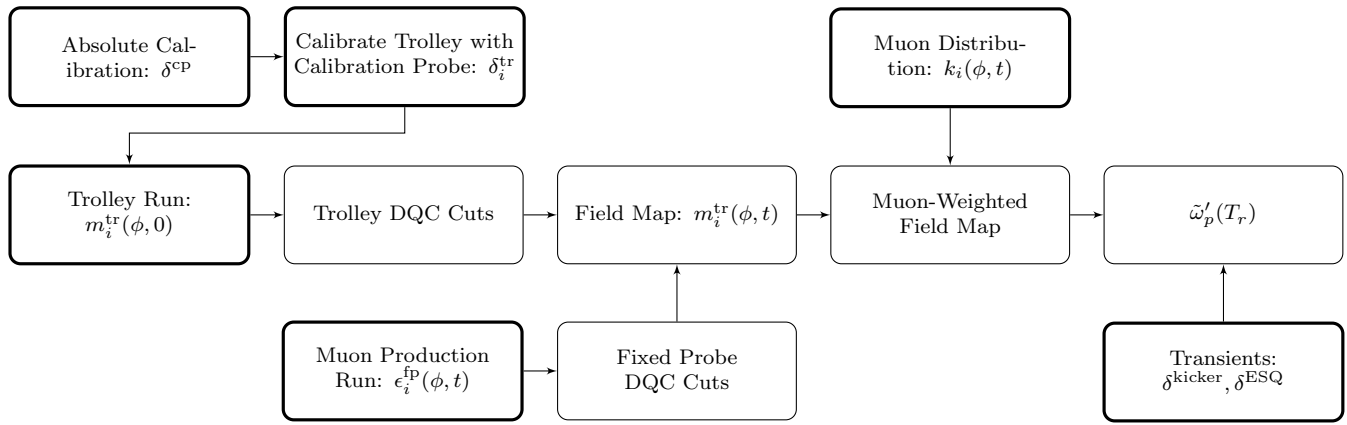
\begin{figure*}
	\centering
	\begin{tikzpicture}[node distance = 0.3 in and 0.175in]
		\node [block, very thick, fill=white, text width=0.165\textwidth, font=\footnotesize] (pp_cal) {Calibrate Trolley with Calibration Probe: $\delta_{i}^{\tr}$};
		\node [block, very thick, fill=white, text width=0.165\textwidth, font=\footnotesize, left =of pp_cal] (abs_cal) {Absolute Calibration: $\delta^{\pp}$};
		\node [block, fill=white, text width=0.165\textwidth, font=\footnotesize, very thick, below =of abs_cal] (trolley_run) {Trolley Run: $m_{i}^{\tr}(\phi,0)$};
		\node [block, fill=white, text width=0.165\textwidth, font=\footnotesize, right =of trolley_run] (tr_DQC) {Trolley DQC Cuts};
		\node [block, fill=white, text width=0.165\textwidth, font=\footnotesize, right =of tr_DQC] (field_map) {Field Map: $m_{i}^{\tr}(\phi,t)$};
		\node [block, fill=white, text width=0.165\textwidth, font=\footnotesize, below =of field_map] (fp_DQC) {Fixed Probe DQC Cuts};
		\node [block, fill=white, text width=0.165\textwidth, font=\footnotesize, very thick, left =of fp_DQC] (prod_run) {Muon Production Run: $\epsilon_{i}^{\fp}(\phi,t)$};
		\node [block, fill=white, text width=0.165\textwidth, font=\footnotesize, right =of field_map] (muon_weighted) {Muon-Weighted Field Map};
		\node [block, fill=white, text width=0.165\textwidth, font=\footnotesize, very thick, above =of muon_weighted] (muon_dist) {Muon Distribution: $k_i(\phi,t)$}; 
		\node [block, fill=white, text width=0.165\textwidth, font=\footnotesize, right =of muon_weighted] (B_tilde) {\opprimetildeatTexp};
		\node [block, fill=white, text width=0.165\textwidth, font=\footnotesize, very thick, below =of B_tilde] (transients) {Transients: $\delta^\mathrm{kicker}, \delta^\mathrm{ESQ}$};
		
		\path [line] (pp_cal) |- ++(0,-0.4in) -| (trolley_run);
		\path [line] (trolley_run) -- (tr_DQC);
		\path [line] (tr_DQC) -- (field_map);
		\path [line] (prod_run) -- (fp_DQC);
		\path [line] (fp_DQC) -- (field_map);
		\path [line] (muon_dist) -- (muon_weighted);
		\path [line] (field_map) -- (muon_weighted);
		\path [line] (muon_weighted) -- (B_tilde);
		\path [line] (abs_cal) -- (pp_cal);
		\path [line] (transients) -- (B_tilde);
	\end{tikzpicture}
	\caption{A flow chart of the field analysis showing the calibration chain through the data processing. The muon distribution is an input that is external to the field analysis, and is required to calculate the muon-weighted field average. Bold items show input measurements to the analysis. Not shown is the \ac{NMR} frequency extraction step required for each of the field measurements.}
	\label{flow:analysis}
\end{figure*}

The first step in the magnetic-field analysis represented in Fig.~\ref{flow:analysis} is the extraction of \ac{FID} parameters such as the frequency, amplitude, and length from all \ac{NMR} measurements, described in Sec.~\ref{subsec:NMR_Frequency_Extraction}. Data quality cuts are applied on these extracted parameters to discard \ac{FID} waveforms that correspond to instrument failures or severe field instabilities. A brief overview of these cuts is given in Sec.~\ref{subsec:Data_Quality_Control}.

Throughout this section and the rest of this paper, we use the symbol $\delta$ to refer to systematic and statistical effects. The uses of these symbols represent both corrections and uncertainties from the effect in question.

In Eq.~\eqref{eq:amu}, $\opprimetildeatTexp$ is the average frequency that would be measured by a spherical water sample at the calibration reference temperature $T_{r}=\SI{34.7}{\celsius}$ in the same position as the detected muons. This shielded proton frequency is related to the calibration probe frequency $\omega^{\pp}$ through a set of corrections, denoted by $\delta^{\pp}(T_{r})$ that account for the probe materials, effects due to sample shape and susceptibility, temperature, and other probe related effects: 
\begin{linenomath}\begin{equation}
		\opprime (T_{r})=\omega^{\pp} \left[1+\delta^{\pp}(T_{r})\right].
		\label{eq:w_p}
\end{equation}\end{linenomath}
The determination of $\delta^{\pp}(T_{r})$ is the absolute calibration step in Fig.~\ref{flow:analysis} and was mainly performed in a dedicated calibration setup including a solenoid magnet as discussed in Sec.~\ref{sec:shielded-proton-correction}.

The calibration probe is then used to calibrate the trolley probes, detailed in Sec.~\ref{sec:pp-trly-calib}. This step determines the relationship between each trolley probe $n$ and the shielded proton frequency via a calibration constant
\begin{linenomath}\begin{equation}
		\opprime (T_{r}) =\omega^{\tr}_n \left(1+\delta^{tr}_n \right).
		\label{eq:trolley_cals}
\end{equation}\end{linenomath}
Since the moments $m^{}_i$ are linear combinations of trolley probe measurements $\omega^{\tr}_n$, we can generalize to
\begin{linenomath}\begin{equation}
        m'^{}_{p,i}=\mtr_i\left (1+\delta^{\tr}_i\right ).
		\label{eq:w_pp}
\end{equation}\end{linenomath}
Details of the trolley map analysis step are given in Sec.~\ref{sec:Trolley_Data_Analysis}.

The fixed probe field moments $m_i^{\fp}$ are synchronized to the trolley field moments when the trolley passes each fixed probe station at a specific time $t=0$. A first-order Taylor expansion of the trolley moments in terms of the fixed probe moment yields
\begin{linenomath}\begin{eqnarray}
		\mtr_i(\phi,t) & = & \mtr_i(\phi,0) +  \sum_j J_{ij}(\phi) \left[ m^{\fp}_j(t)-m^{\fp}_j(0)\right ] \nonumber \\
		& & + \epsilon_{i}^{\ho}(\phi,t),
		\label{eq:FPTracking}
\end{eqnarray}\end{linenomath}
where the subscripts $i$ and $j$ indicate specific field moments. The Jacobian $J_{ij}(\phi)=\frac{\partial \mtr_i(\phi)}{\partial m^{\fp}_j}$ relates small changes in fixed probe moments to small changes in the trolley moments for each station (indicated by the $\phi$ dependence) and $\epsilon^{\ho}_i(\phi, t)$ represents the effects of higher-order moments that the fixed probes cannot track. Note that $\epsilon^{\ho}_i(\phi, 0)\equiv 0$. Because $\epsilon^{\ho}_i(\phi, t)$ cannot be tracked due to the limited number of fixed probes in a station, we model it as a random walk and include its effect only as an uncertainty. The full procedure for synchronizing and tracking the field with the fixed probes is discussed in Sec.~\ref{sec:Interpolation}. We can rewrite Eq.~\eqref{eq:FPTracking} as
\begin{linenomath}\begin{equation}
		\mtr_i(\phi,t)=\mtr_i(\phi,0) + \epsilon_i^{\fp}(\phi,t) + \epsilon^{\ho}_i(\phi, t),\label{eq:FPTracking2}
\end{equation}\end{linenomath}
where
\begin{linenomath}\begin{equation}
		\epsilon^{\fp}_i(\phi, t)= \sum_j J_{ij}(\phi) \left [m^{\fp}_j(t)-m^{\fp}_j(0)\right ].
		\label{eq:delta_fp}
\end{equation}\end{linenomath}

Assuming that the trolley calibrations ($\delta^{\tr}_n$) do not change over time, we can combine the fixed probe tracking, trolley maps, trolley calibration, and calibration probe corrections to:
\begin{linenomath}\begin{eqnarray}
		m^{\prime}_{p,i} (\phi,t,\Tr) & = & \left[ \mtr_i(\phi,0) + \epsilon_i^{\fp}(\phi,t) +\epsilon^{\ho}_i(\phi,t) \right] \nonumber \\
		& & \times \left( 1 + \delta_i^{\tr} \right),
		\label{eq:Combined}
\end{eqnarray}\end{linenomath}
with the field moment index $i$. Note that in this equation, $\delta^{\pp}(T)$ is absorbed into $\delta_i^{\tr}$ through Eq.~\eqref{eq:w_pp}. These moments are then weighted by the muon distribution in space and time and averaged over time $t$ and azimuth $\phi$ to determine $\opprimetildeatTexp$, as described in Sec.~\ref{sec:MuonWeighting}.
 
\subsubsection{Multiple Analysis Approaches}
\label{subsubsec:Introduction_Multiple_Analysis_Approaches}
For several of the key analysis steps described above, the analysis was performed by at least two independent teams in order to provide important cross checks and test different algorithms against each other. Comparison of the parallel analyses often found a high degree of consistency. In cases where noticeable differences were identified, a detailed comparison of the approaches allowed us to develop and implement improved algorithms. Sections~\ref{sec:Data_Extraction}--\ref{sec:MuonWeighting} present the final analysis that led to the reported result for the measurement of \opprimetildeatTexp. Here, we highlight a few of the notable differences between the different trolley calibration and field tracking algorithms. The details associated with these differences will be explained in the analysis sections of the paper.

Three individual analyzers performed the trolley calibration analysis (see Sec. ~\ref{sec:pp-trly-calib}) for our \RunOne data set with the following main differences:
\begin{itemize}
	\item One analysis used a zero-crossing counting method for the frequency extraction of the calibration probe, while the other two used the Hilbert transform method (see Sec.~\ref{subsec:NMR_Frequency_Extraction}).
	\item The calibration analysis in \RunOne had to correct both the normal long-term drift of the magnetic field due to slow changes in the magnet and a field oscillation with an amplitude of about \SI{20}{ppb} and a period of 2\,min. The three analyzers chose different approaches for selecting and treating the fixed probe data used to correct the calibration and trolley probe measurements.
	\item The analysis needed to account for uncertainties associated with gradients in the magnetic field that coupled to the error in the relative positioning of the probes. The determination of local field gradients was based on polynomial fits to local maps, and each analyzer chose fits with different orders and ranges.
\end{itemize}
All cross-checks showed consistency between the three analyses at the 10-ppb level for the probe calibration offsets. 

For synchronizing the trolley and fixed probes and the subsequent tracking (see Sec.~\ref{sec:Interpolation}), two independent analyses~\cite{Tewsley-Booth:phd,Osofsky:phd} were implemented with the following major differences:
\begin{itemize}
	\item Three fixed probe stations located in regions with large field gradients exhibited significantly more noise than typical. One analysis replaced the measurements from these stations with the average of the stations' nearest neighbors. The other analysis relied on long averaging times to improve resolutions.
	\item The trolley and fixed probes were read out at 2 and \SI{0.7}{\hertz} respectively and were not simultaneous. One analysis worked with these original asynchronous times while the second interpolated both to produce a time series at 1-s intervals.
	\item During synchronization between the trolley and a given fixed probe station, the fixed probe station was tied to a local azimuthal average of trolley measurements when the trolley was closest to that station. One analysis used about $\pm$ \SI{2.5}{\degree} of trolley measurements for each station, while the other analysis only used $\pm$ \SI{1}{\degree} with a secondary synchronization to take into account the unused parts of the trolley maps.
	\item While the trolley is near a fixed probe station, its magnetization distorts the local field measured by that station. This ``trolley footprint'' window is vetoed in the fixed probe data when the trolley is nearby. The analyses differed in the implementation of the veto window, the interpolation across the missing data, and the usage of other fixed probe stations to account for short-term field fluctuations.
\end{itemize}
A blind analysis comparison campaign focused on these differences to understand each choice's impact on the final results. The treatment of the poor-resolution stations was the dominant contribution to the difference. The two analyses differed by maximally \SI{30}{ppb} over a field tracking time interval of about three days and only by \SI{1.5}{ppb} after averaging over the entire tracking period. 
     \graphicspath{{./dataextraction/figures/}}

\section{Data Extraction and Preparation} \label{sec:Data_Extraction}

\subsection{NMR Frequency Extraction} \label{subsec:NMR_Frequency_Extraction}

The \ac{NMR} technique generates \acp{FID}, which are the signals measured in the probe coil due to the precessing magnetization across the sample. The finite size of the sample combined with a nonuniform magnetic field affects the evolution of the frequency and signal amplitude during the FID. Therefore, it is critical to develop algorithms that determine the relationship between the frequency evolution and $B$ and to understand features associated with the observed signal that stem from the nonuniformities in the magnetic field. The following is a summary of frequency extraction and its related uncertainties. Further details can be found in \cite{Hong:2020}.

In the first step of the data analysis, the frequency and other characteristics including the \ac{FID} length and amplitude are extracted from the digitized waveforms of the calibration, fixed, and trolley probes.
\begin{figure}
        \includegraphics[width=3.375in]{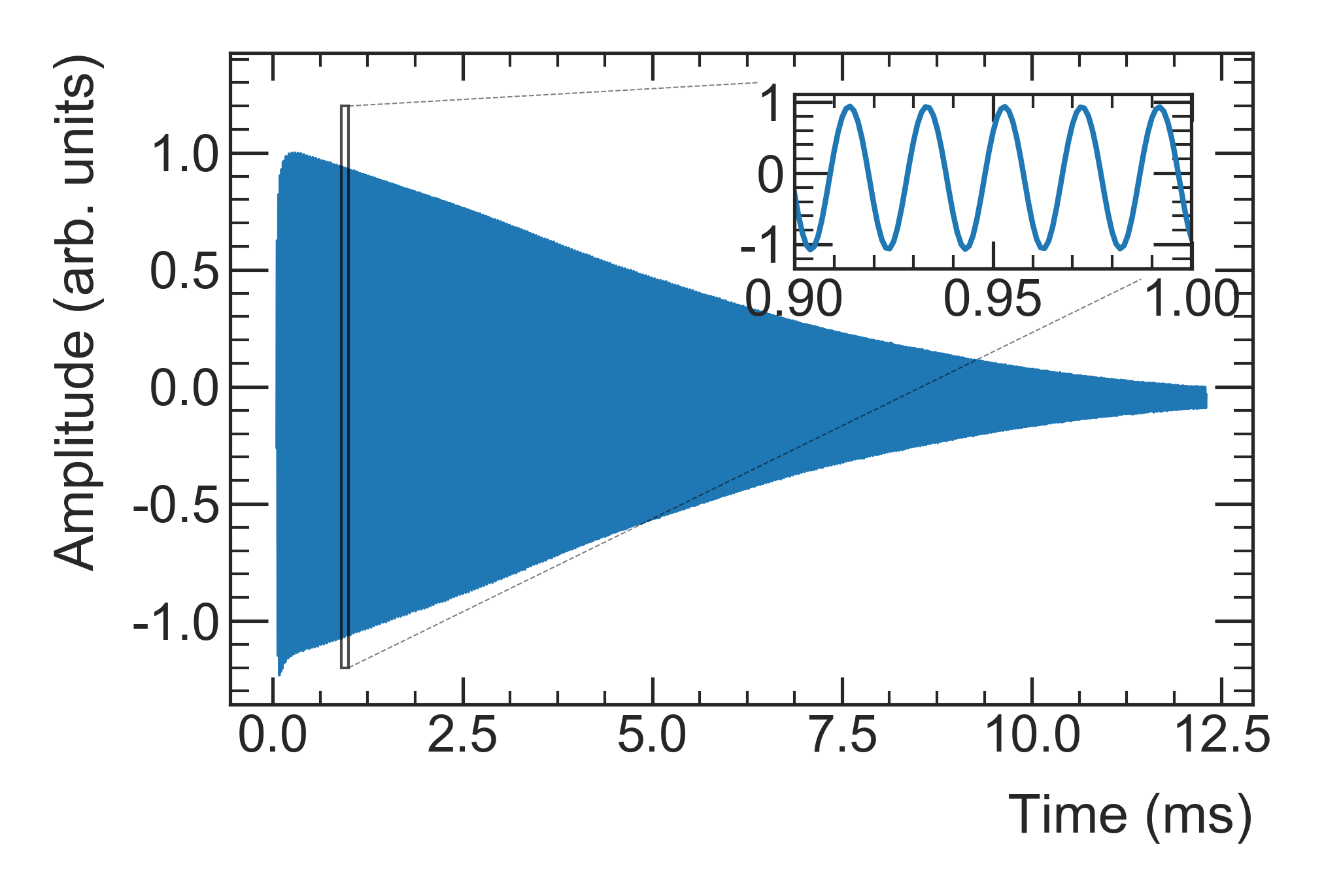}
	\caption{A typical free induction decay (FID) from a trolley probe. The zoomed inset shows the periodic behavior that is used to measure $\omega(t)$.}
	\label{fig:typical_fid}
\end{figure}
A typical \ac{FID} signal is shown in Fig.~\ref{fig:typical_fid}. Two algorithms were used to analyze these signals: 
\begin{itemize}
	\item For trolley and fixed probes, the main frequency extraction algorithm extracts the phase function $\Phi(t)=\tan^{-1}(f(t)/\mathcal{H}(f(t)))$ from the discrete Hilbert transform $\mathcal{H}(f(t))$ of the \ac{FID} signal $f(t)$. To mitigate effects of a time-varying baseline, finite \ac{FID} length, and sampling period, we apply time- and frequency-domain filters to the extraction of $\Phi(t)$.
	\item For the calibration probe, an alternative extraction of the phase function $\Phi(t)$ uses an iterative baseline subtraction and identification of zero-crossing times in the oscillatory \ac{FID} signal, which correspond to a phase advance of $\pi$.
\end{itemize}

The initial frequency of the \ac{NMR} signal, $\omega(t=0)$, is related to the phase function by $\omega(0)= \frac{d\Phi}{dt}(0)$ \cite{Cowan:1996}. A polynomial fit is used to extract $\omega(0)$, shown in Fig.~\ref{fig:phase_fit}. The truncation order (up to fifth order) and range of the fit (roughly 40\% of the FID length\footnote{The FID length is defined as the time when the envelope’s amplitude falls below $1/e$ of the initial amplitude.}) were chosen to optimize the combined statistical and systematic uncertainties. While a lower truncation order and longer fit range generally reduce the statistical uncertainty, the non-linear terms of $\Phi(t)$ increase the systematic uncertainty.

\begin{figure}[ht]
	\centering
	\includegraphics[width=3.375in]{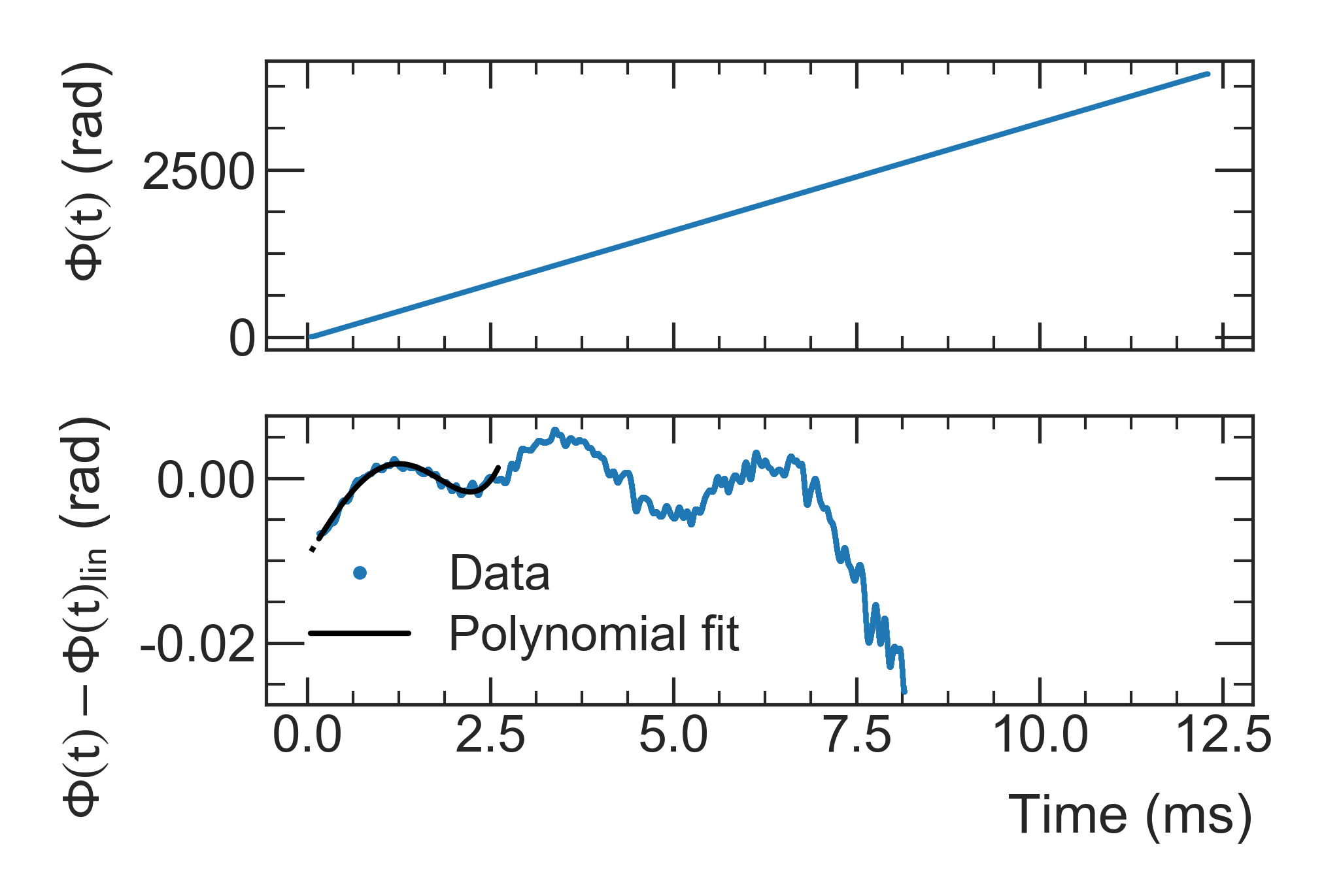}
    \caption{The upper panel shows $\Phi(t)$ as extracted from the Hilbert transform of the signal in Fig.~\ref{fig:typical_fid}. The blue points in the lower panel are the difference between $\Phi(t)$ and a linear fit $\Phi(t)^{}_\mathrm{lin}$ to $\Phi(t)$. The black line is a polynominal fit to these residuals, the dotted part shows the extrapolation outside the fit range to $t=0$.}
	\label{fig:phase_fit}
\end{figure}

We developed the phase-template method for fixed and trolley probes, which reduces the effect from static, non-linear terms by subtracting an initial phase template, $\Phi^{}_0(t)$ from each $\Phi(t)$. In the case of the trolley, only static effects extracted in an optimized field are subtracted by $\Phi^{}_0(t)$. The fixed probes generally observe small frequency changes due to temporal field changes; the non-linear terms in $\Phi(t)$ change less than the linear term  $\frac{d\Phi(t)}{dt}$, measurement-to-measurement.

The systematic and statistical effects related to the frequency extraction were extensively studied using simulated and real \acp{FID}, real noise waveforms recorded in the magnetic field without initiating the \ac{NMR} sequence, and waveforms recorded with the regular \ac{NMR} sequence without the main magnetic field present. The following main uncertainties were identified:
\begin{itemize}
	\item The systematic fit uncertainty $\epsilon_f$ is dependent on the frequency extraction algorithm and quantifies the difference between the fitted value and the true $\omega(0)$. It originates from approximating the phase function with a truncated polynomial or from artifacts of the applied filter.
	\item The intrinsic systematic uncertainty from the simulation $\epsilon_i$ is the difference between the extracted $\omega(0)$ and the frequency $\omega^{}_0$ corresponding to the magnetic field at the center of the probe. This uncertainty is driven by the probe geometry and the magnetic-field inhomogeneity during the measurement and was independent of our choice of algorithm.
	\item The statistical uncertainty $\delta \omega^{}(0)$ is caused by the noise in the \ac{FID} waveform. It is determined from the standard deviation of the fit values for several \acp{FID} measured in the same magnetic field.
\end{itemize}

The determination of these uncertainties was performed for the calibration and trolley probes and will be reported in Secs.~\ref{subsec:SystematicsFid} and~\ref{subsubsec:Trolley_Systematics_FID}, respectively. For the fixed probes, the systematic uncertainties are absorbed in the synchronization step with the trolley, and statistical uncertainties are negligible due to long averaging times.

\subsection{Data Quality Control}\label{subsec:Data_Quality_Control}
In preparation for the determination of \opprimetildeatTexp described in the following sections, data quality selection was performed to only include field measurements where the magnetic field changed slowly with respect to the measurement period. All analyses apply common data quality selections that fall into the following two categories: 
\begin{itemize}
	\item Event Level Effects: Data quality flags were introduced at the individual \ac{FID} level (see Appendix~\ref{subsec:Data_Quality_Control_Instrument_Failures}) to identify intermittent measurement failures. These flags are based on the \ac{FID} parameters; cut thresholds were determined based on identifying outliers from the distributions of these parameters over a short period.
	\item Global Effects: Several types of magnetic-field instabilities were identified over the course of \RunOne. The two main causes of these instabilities were sudden magnet coil movements that generated abrupt changes in the magnetic field and failures in the fixed probe electronics crates that drove erroneous changes in the feedback system (see Appendix~\ref{subsec:Data_Quality_Control_Severe_Field_Instabilities}).  Data analysis is vetoed for $\SI{\pm2}{\minute}$ around these easily identifiable abrupt changes. Dedicated studies showed that the field tracking outside the veto window is uncompromised. 
\end{itemize}

The \ac{FID} quality cuts are applied to only the \opprimetildeatTexp analysis and not the \oa analysis because omitting individual \acp{FID} has negligible effects on the field tracking and the final determination of \amu. However, during periods with magnetic-field instabilities the field is not reliably tracked. Therefore, these periods must be excluded from the \oa analysis. Additional veto windows were applied to all analyses roughly every two hours during the 12-s-long transitions in the \ac{DAQ}, during which no field data are recorded.

\subsection{Run-1 Datasets}\label{subsec:Dataextraction_Run1_Datasets}
During the \RunOne data taking period, experimental conditions were varied in each of the pulsed high voltage systems, the \ac{ESQ} and the fast kicker. The \RunOne data are grouped into four distinct subsets according to the \ac{ESQ} and kicker high voltages as shown in Table~\ref{tab:datasets}. For the analysis of \oa, periods with different set points are analyzed individually, and separate beam-dynamics corrections are applied \cite{E989SRBDpaper}. The magnetic-field analysis produces a separate result for each of these data subsets in Sec.~\ref{sec:final_results} for $\opprimetildeatTexp$.

\begin{table}[htb]
	\centering
	\begin{tabular}{ccc}
		\hline\hline
		\rule{0pt}{1em} \RunOne  &  ESQ & Kicker \\
		data subset &  (kV)  &  (kV)  \\
		\hline
		\RunOneA    & 18.3 & 130    \\
	        \RunOneB    & 20.4 & 137    \\
		\RunOneC    & 20.4 & 130    \\
		\RunOneD    & 18.3 & 125    \\
		\hline\hline
	\end{tabular}
	\caption{Summary of the \RunOne data subsets. The different voltages on the beam-injection systems impact the \oa analysis, and the magnetic-field analysis is grouped accordingly.}\label{tab:datasets}
\end{table}
    \section{The Calibration Probe} \label{sec:shielded-proton-correction}

To determine \amu as written in Eq.~\eqref{eq:amu}, a well-characterized \ac{NMR} standard is required. For that purpose, the calibration probe with a cylindrical high-purity water sample was constructed. Its material perturbations were characterized so that its measured Larmor frequencies can be corrected to those expected of a shielded proton in a spherical water sample $\opprime(\Tr=\SI{34.7}{\celsius})$ with high accuracy and precision. The local magnetic field is then obtained using Eq.~\eqref{eq:Btoomega}. The calibration of this probe is transferred to each of the 17 trolley probes, compensating for the trolley probes' material effects and differences in diamagnetic shielding.

\subsection{Systematic Effects}
\label{sec:pp_syst}

A set of corrections, described below, are required to relate the \ac{NMR} frequencies measured by the calibration probe $\opmeas(T)$ to $\opprime(\Tr)$ via\footnote{In principle, the corrections would be multiplicative but we use the approximation $(1+\delta^{}_1)\cdot(1+\delta^{}_2) \approx (1+\delta^{}_1 + \delta^{}_2)$ because the corrections $\delta^{}_1$ and $\delta^{}_2$ are $\mathcal{O}(ppm)$ or less and the term $\delta^{}_1 \cdot \delta^{}_2$ is hence negligible.}
\begin{align} \label{eqn:omega-p-meas}  
   \opprimeatTexp = \opmeas(T) \times \left[ \right. & 1 + \delta^T(\Tr-T) \nonumber \\
    & \left. + \delta^{b}\left( \mathrm{H_2O}, T \right) + \delta^{t} \right], 
\end{align}
where $\delta^T$ corrects for the temperature dependence of the diamagnetic shielding of H$_2$O between the temperature $T$ of the measurement and the chosen reference temperature $\Tr = \SI{34.7}{\celsius}$ \cite{Phillips:1977, Neronov:2014, Petley:1984}; $\delta^{b}$ is a correction dependent on water magnetic susceptibility and sample shape; and $\delta^{t}$ is the sum of corrections for the probe materials and other effects related to the probe. The probe temperature $T$, typically close to \SI{26}{\celsius}, was measured to \SI{0.5}{\celsius} with a PT-1000 sensor installed in the probe near the sample.

The correction $\delta^{t}$ consists of several contributions:
\begin{equation} 
	\delta^{t} = \delta^{s} + \delta^{p} + \delta^\mathrm{RD} + \delta^{d}. \label{eqn:delta-t}
\end{equation}
Here, $\delta^{s}$ denotes the correction for effects due to the probe materials, the probe's angular orientation about its long axis, the pitch angle relative to the field axis, and the magnetic images it induces in the surrounding magnet's iron. We split this term into two parts $\delta^{s} = \delta^{s,\ \mathrm{intr}} + \delta^{s,\ \mathrm{config}}$. Here, $\delta^{s,\ \mathrm{intr}}$ corrects for the effects that are intrinsic to the probe and $\delta^{s,\ \mathrm{config}}$ corrects for the specific probe configuration when used in the experiment at Fermilab. The correction $\delta^{p}$ is due to the water sample and the water sample holder and $\delta^{\mathrm{RD}}$ is the contribution from radiation damping \cite{Vlassenbroek:1995}, an effect where the \ac{NMR}-induced signal in the \ac{RF} coil affects the proton spin precession. Finally, $\delta^{d}$ is the proton dipolar field perturbation \cite{Jeener:1995}.

\subsubsection{Intrinsic Effects: \texorpdfstring{$\delta^{b}\left(\mathrm{H_2O}\right),\ \delta^{s,\ \mathrm{intr}},\ \delta^{p},\ \delta^\mathrm{RD},\ \delta^{d}$}{delta\^{}T, delta\^{}s,intr, delta\^{}p, delta\^{}RD, delta\^{}d}}
\label{sec:pp_intrinsic_systematics}
Intrinsic systematic effects in the calibration probe are terms that affect the probe's measured frequency independent of its environment. These corrections and uncertainties were measured at \ac{ANL} in a dedicated \ac{MRI} solenoid and include the bulk magnetization and several of the material perturbations.

A correction due to the bulk magnetic susceptibility $\delta^{b}$ is required because the calibration probe uses a cylindrical water sample perpendicular to the field, not a spherical sample.  The magnetization of the water molecules in one location of the sample perturbs the field at other locations, and the magnitude depends on the shape and volume susceptibility of the \ac{NMR} sample. In SI units:
\begin{equation}
	\delta^{b}\left(\mathrm{H_2O}, T\right) = \left( \varepsilon -  \frac{1}{3}\right) \chi \left(\mathrm{H_2O}, T\right), 
\label{eq:susc}
\end{equation}
where $\varepsilon$ is the shape factor of the sample.
For a sphere $\varepsilon = 1/3$ so the field perturbation from this effect would vanish, whereas $\varepsilon= 1/2$ for an infinite cylinder perpendicular to the field ~\cite{Osborn:1945,Durrant:2002,Hoffman:2005}. 

The recommended value for the volume magnetic susceptibility of water $\chi \left(\mathrm{H_2O}, T=\SI{20}{\celsius}\right) = -9.032\times 10^{-6}$ was measured at temperature of $T=\SI{20}{\celsius}$~\cite{Schenck:1996}. A comparison with an additional measurement taken at an unknown temperature, $\chi \left(\mathrm{H_2O}\right) = -9.060(3) \times 10^{-6}$~\cite{Blott:1993} is used to estimate an uncertainty of $3\times 10^{-8}$. 
The measured, small, temperature dependence of the magnetic susceptibility \cite{Philo:1980} is used to determine the magnetic susceptibility of water at an experimental measurement temperature $T$.

The intrinsic probe correction $\delta^{s,\ \mathrm{intr}}$ was measured in the \ac{MRI} magnet by removing the 5-mm-diameter cylindrical water sample and measuring the field shift caused by the remaining calibration probe materials, when a test probe was inserted inside the calibration probe. The dependence on the probe's roll and pitch\footnote{The roll is the angle of the rotation around the probe’s long axis and the pitch is the long axis’ angle with respect to horizontal.} was measured. 

For the estimation of $\delta^p$, ASTM type-1 water from different vendors was utilized, and degassed and non-degassed water samples were examined. A variety of additional tests were performed in which the glass water sample tube was rotated, and different sample tubes were used.  No systematic shifts were observed within an uncertainty of \SI{2}{ppb}. The $\delta^{\mathrm{RD}}$ term was estimated by varying the magnetization tip angle and detuning the probe's resonant circuit. No relevant effects larger than \SI{3}{ppb} were observed, consistent with expectations \cite{Vlassenbroek:1995}. The value for $\delta^{d}$ is based on estimates for the specific probe geometry described in Sec.~\ref{subsec:Introduction_Measuring_Magnetic_Field}, and the effect is estimated to be less than \SI{2.5}{ppb} \cite{Jeener:1995}.

The results for all of the terms described in this section are shown in Table~\ref{tab:pp-pert-results} and $\delta^{s}$ is evaluated for the configuration used at \ac{FNAL} as described in the next section.

\subsubsection{Configuration Effects: \texorpdfstring{$\delta^{s,\ \mathrm{config}}$}{delta\^{}s,config}}
The configuration specific $\delta^{s,\ \mathrm{config}}$ accounts for four additional corrections, which arise when the calibration probe is used in the storage ring magnet at \ac{FNAL}. First, new materials were added to support the probe whose field perturbation must be determined: an aluminum holder clamped around the probe, a long aluminum rod used to move the probe into the measurement region, and a new SMA connector and cable. The perturbations of the aluminum holder, SMA connector, and cable were measured in the \ac{MRI} solenoid, and were consistent with expectations based on the volumes, distances from the \ac{NMR} sample, and magnetic susceptibilities of the materials.

Second, when inserted between the iron magnet poles, magnetic images of the magnetized components of the probe perturb the field at the water sample. The image effects were measured in the \ac{MRI} solenoid by observing the field perturbation from the calibration probe on a test probe located one image distance (\SI{18}{\centi\meter}) away, and were consistent with calculations. The total correction $\delta^{s}$ including the probe, holder, and rod and their images was also measured directly in the storage volume, and was consistent with the measurements performed with the \ac{ANL} solenoid. The effect of the rod could not be verified in the solenoid, but the measurement result in the storage ring magnet was consistent with expectations.

Third, when installed on the long rod, the long axis of the probe is not exactly perpendicular to the field so its pitch angle is nonzero. The probe angle with respect to the vertical field was measured using a camera and plate with fiducial markings, and found to be offset by $\SI{0.7}{\degree}$. The material effects for a probe pitched at $\SI{2.5}{\degree}$ were measured at \ac{ANL} and scaled linearly, yielding a difference of $\SI{4(4)}{ppb}$ with respect to a probe aligned with the field.

\begin{table*}[!hbt]\centering
	\small{\begin{tabular}{lccc}
		\hline\hline
		Quantity & Symbol & Correction (ppb) & Uncertainty (ppb) \\
		\hline
		Bulk Magnetic Susceptibility  & $\delta^{b} \left(\mathrm{H_2O},T\right)$ & -1505.9 to -1505.6 & 6 \\T Dependence of Diamagnetic Shielding & $\delta^T(\Tr-T)$ & -99.1 to -86.0 & 5 \\
Intrinsic and Configuration-Specific Probe Effects      & $\delta^{s}$ & 15.2 & 12 \\
		Water Sample & $\delta^{p}$ & 0 & 2 \\
		Radiation Damping             & $\delta^{\mathrm{RD}}$ & 0 & 3 \\
		Proton Dipolar Field          & $\delta^{d}$ & 0 & 2 \\ \hline
Total                         &             & -1589.8 to -1576.4  & 15 \\
		\hline\hline
	\end{tabular}}
	\caption{The calibration probe corrections due to effects described in the text. The temperature-dependent entries were evaluated for the calibration probe's temperatures ranging from $T=\SI{25.13}{\celsius}$ to \SI{26.4}{\celsius} during the calibration of the trolley probes. Positive values indicate that the field measurements are increased to correct a given effect.}
	\label{tab:pp-pert-results}
\end{table*}

The fourth correction arises because the material perturbation measurements involve the probe displacing air, which is paramagnetic due to the molecular oxygen, whereas the probe displaces vacuum when used during the calibration procedure. This vacuum shift is effectively the magnetic perturbation due to a volume of air in the shape of the probe, estimated as $\SI{-2(2)}{ppb}$.

\subsubsection{Correcting the Measurement to the Shielded Proton Frequency: \texorpdfstring{$\delta^T(\Tr-T)$}{delta\^{}T(T\_r-T)}}

To extract the shielded-proton precession frequency from calibration probe measurements, we solve Eq.~\eqref{eqn:omega-p-meas}, applying all corrections. With $\delta^T(\Tr-T) =  \SI{-10.36(30)}{ppb/\celsius} \times (\Tr-T)$ and calibration probe temperatures of around \SI{26}{\celsius}, the typical value for this correction was $\delta^T \approx \SI{90}{ppb}$. These shielded-proton frequencies are then transferred to the trolley via a detailed calibration program, which we discuss in Sec.~\ref{sec:pp-trly-calib}.

\subsection{Cross Checks with Spherical Water Sample and \texorpdfstring{\hethree}{He-3}}

The difference between cylindrical and spherical samples was verified by comparing cylindrical calibration probe frequencies with those of the spherical sample probe used in the \ac{BNL} E821 experiment \cite{Fei:1997sd}. The measurements were taken in the stable homogeneous field of an \ac{MRI} magnet at \SI{1.45}{\tesla} at \ac{ANL}. The measured difference $\SI{1514(15)}{ppb}$ agrees with expectations from Eq.~\eqref{eq:susc}, with the uncertainty dominated by the asphericity of the \ac{BNL} water sample. To account for the finite length of our water sample, a small correction of 0.02\% was applied to the shape factor $\varepsilon=1/2$ of an infinte cylinder~\cite{Hoffman:2005}.

As a cross check with considerably different systematics, a \hethree~probe described in~\cite{Farooq2020,Farooq:phd} was also compared with the \ac{BNL} spherical water probe.  After correcting the \ac{BNL} probe to $\SI{25}{\celsius}$ and for material effects, the ratio of \hethree~to spherical probe frequencies was measured to be $0.761\,786\,139 (29) (\SI{38}{ppb})$. This result agrees with a previous measurement \cite{Flowers:1993} of the ratio of frequencies from \hethree~and water in a spherical sample,
\begin{equation*}
	\frac{\mu^{}_h(\hethree)}{\mu'_p} = -0.761\,786\,1313 (33) (\SI{4.3}{ppb}).
\end{equation*}
The cylindrical calibration probe was therefore calibrated to \hethree\ indirectly through the \ac{BNL} spherical probe, effectively validating the calibration probe to \SI{10(38)}{ppb}.
    \graphicspath{{./pp-trly-calib/figures/}}

\section{Trolley Calibration} \label{sec:pp-trly-calib}

The field measured by each of the 17 trolley probes is a combination of the storage ring magnetic field and additional perturbations introduced by the NMR probes, their sample shape, and surrounding magnetized materials in the trolley. The trolley probe calibration procedure described in this section provides a set of offsets $\delta_{j}^{\tr}(\Tr)$ (see Eq.~\eqref{eq:trolley_cals}), used to correct the measured frequency of probe $j$ to the shielded proton frequency at $\Tr=\SI{34.7}{\celsius}$.
The offsets are due primarily to differences in diamagnetic shielding of protons in water versus petroleum jelly, sample shape, and magnetic perturbations from magnetization of the materials used in the NMR probes and trolley body.
This procedure allows the trolley frequency maps to be converted into maps of the magnetic field in the storage volume. 
The trolley calibration constants are extracted from the difference of trolley probe frequencies $\omega^{\tr}_{j}$ and calibration probe measurements corrected to the shielded proton frequency $\opprimej\xspace(\Tr)$, with the two probes swapped into the same position. Remaining misalignments and magnetic footprints of the calibration probe on the trolley and vice versa during the actual calibration measurement lead to procedure specific corrections $\delta^\mathrm{align}_{j}$ and $\delta^{\fp}_{j}$. The $\omega^{\tr}_{j}$ in Eq.~\eqref{eq:trolley_cals} have to be expressed in terms of the actual measured trolley frequencies $\omega^{\tr,\ \mathrm{meas}}_{j}$ via $\omega^{\tr}_{j} = \omega^{\tr,\ \mathrm{meas}}_{j} \left( 1 - \delta^\mathrm{align}_{j} - \delta^{\fp}_{j}\right)$.  The difference in trolley probe temperature between calibration and trolley field mapping is taken into account in the trolley map analysis (see Sec.~\ref{subsec:Trolley_Systematic_Effects}). The trolley calibration constants are extracted as
\begin{align}\label{eq:calibration_offsets}
    \delta^{\tr}_{j}(\Tr) = & \frac{\opprimej(\Tr) - \omega^{\tr,\ \mathrm{meas}}_{j}}{\omega^{\tr,\ \mathrm{meas}}_{j}}  + \delta^\mathrm{align}_{j} + \delta^{\fp}_{j}.
\end{align}

The calibration procedure described in Sec.~\ref{sec:calib-proc} was performed for all 17 trolley probes. The full campaign took about two weeks to complete, meaning it was not feasible to repeat the procedure often. For the \RunOne analysis, the calibration campaign was performed during the \ac{FNAL} accelerator summer shutdown following the production period. The calibration of the central probe was performed multiple times as a cross check. We have performed four calibration campaigns, associated with each annual running period, and preliminary analyses of the \RunTwo and \RunThree calibration data show good consistency with the \RunOne results discussed here.

\subsection{Calibration Procedure}
\label{sec:calib-proc}
Each trolley probe was calibrated with the following procedure:
\begin{enumerate}
    \item The calibration probe (Sec.~\ref{sec:shielded-proton-correction}) was mounted on a translation stage in the vacuum chamber. The translation stage allowed the calibration probe to be moved to each trolley probe position at a specific azimuthal location. 
	\item The \ac{SCC} and a set of local azimuthal coils were used to impose known, large field gradients in the calibration region, allowing precision determination of the two probes' positions.
    \item The field was shimmed locally with the \ac{SCC} based on a local field map by the calibration probe.
    \item The trolley and calibration probe were rapidly swapped back and forth into the same position. Several measurements were taken with each probe in this calibration position.
	\item Nearby fixed probes tracked the magnetic-field drift during the calibration procedure.
\end{enumerate}

To determine the probe's position, we imposed large gradients in all three directions using the \ac{SCC} and azimuthal coils to colocate the calibration probe and the target trolley probe $j$. The difference of the field with and without these large gradients uniquely determined the probe position. This procedure allowed the position to be determined with a precision of typically \SI{0.5}{\milli\meter}. 

With the large external gradients turned off, remaining spatial field gradients in the storage region will couple to small position offsets between the probes. To minimize this systematic uncertainty, the field in the vicinity of a target trolley probe was mapped using the calibration probe and shimmed locally with the \ac{SCC} and a set of azimuthal coils to reduce local field gradients to less than \SI{30}{\nano\tesla/\milli\meter} (\SI{21}{ppb/\milli\meter}). The calibration probe mapped the residual field gradients so we could correct any errors from the remaining misalignment between the probes.

The magnetic field in the muon storage region drifts over time. We used the power supply feedback to suppress this drift and monitored the remaining magnetic-field drift using the fixed probes. Repeated ``rapid swaps" between the trolley and the calibration probe help mitigate the effects of long-term drifts in ``ABA"-style measurements \cite{Swanson:2010}. Measurements were taken with the trolley at the calibration location for \SI{30}{\second}, then the trolley was retracted upstream azimuthally by $\approx\SI{4}{\degree} \simeq\SI{50}{\centi\meter}$. The calibration probe was then moved into the calibration location and we took measurements for \SI{30}{\second}. We repeated this sequence at least 4 times per probe and up to 10 times for some probes.

\subsection{Analysis}
\label{sec:calib-analysis}

To extract $\delta_j^{\tr}$ for a trolley probe $j$ via Eq.~\eqref{eq:trolley_cals}, the data taken during the rapid swapping are analyzed as discussed in Sec.~\ref{sec:rapid-swap}. Since both probes cannot be placed exactly at the same position when they are moved into the measurement position, the analysis must also account for the small, relative position misalignments of the trolley probe $j$ and the calibration probe. This analysis is described in Sec.~\ref{sec:misalignment-correction}.

\subsubsection{Rapid Swapping Analysis}
\label{sec:rapid-swap}

From the ABA$\ldots$ series of measurements, the A and B measurements are interpolated to common times, which allows us to correct for linear drifts that occured while the two probes were being swapped. In these measurements, the drift rate was up to $\SI{100}{ppb/\hour}$.

During \RunOne, we observed an oscillation in the magnetic field with an amplitude of \SIrange{10}{20}{ppb} and a period of \SI{2}{\minute}, which is shorter than the measurement and swapping periods of $\sim$\SI{10}{\minute}. Therefore, the ABA method does not remove this oscillation. However, the oscillation is not a localized effect in the calibration region but coherent around the entire ring and it can be removed using data from the fixed probes. The shape of the oscillation is shown in Fig.~\ref{fig:fxpr-osc}, where the slow field drift has already been corrected. Table~\ref{tab:fcc-results} shows the statistical uncertainty from this procedure for all trolley probes.

\begin{figure}[!hbt]
	\centering
	\includegraphics[width=0.98\linewidth]{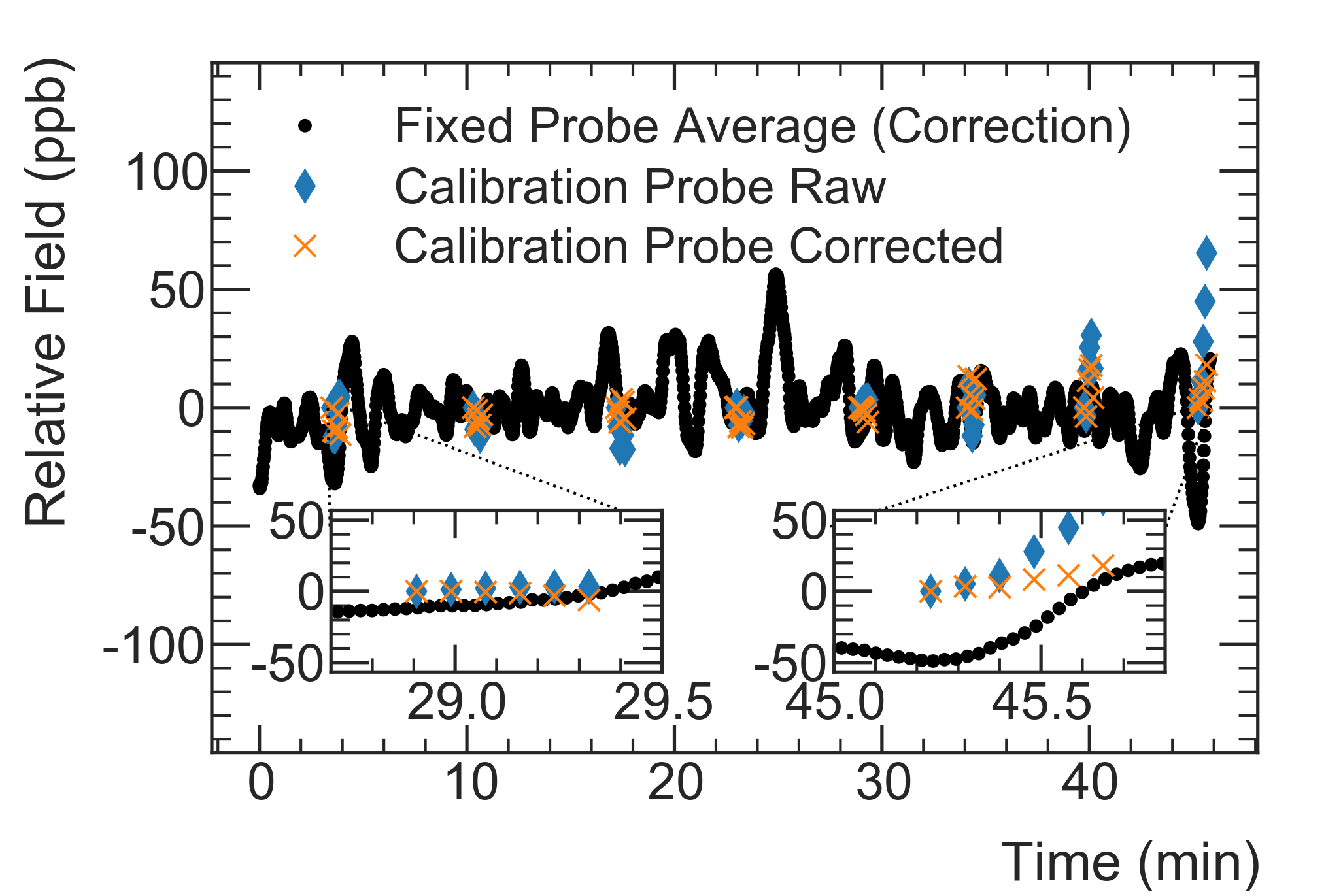}
    \caption{Oscillatory signal as measured by the fixed probes (black dots), where linear drift corrections have been applied. The calibration probe data before (blue diamonds) and after (orange crosses) the correction is overlaid to show the typical size of corrections.}
	\label{fig:fxpr-osc}
\end{figure}

\subsubsection{Misalignment Correction: \texorpdfstring{$\delta^\mathrm{align}_{j}$}{delta\^{}align,j}}
\label{sec:misalignment-correction} 

The difference between the frequencies with and without imposed gradients are calculated using an ABA method. The drift-corrected differences are called $\Delta \omega_{j,q}^\mathrm{tr}$, where $q$ ranges over $x$, $y$, and $\phi$ and $j$ indicates the probe number. The transverse gradients were fitted across the 17 trolley probes. For the azimuthal gradient, the trolley was moved azimuthally through the calibration region in $\approx \SI{0.5}{\centi\meter}$ steps.

\begin{figure}[!hbt]
	\centering

	\includegraphics[width=0.98\linewidth]{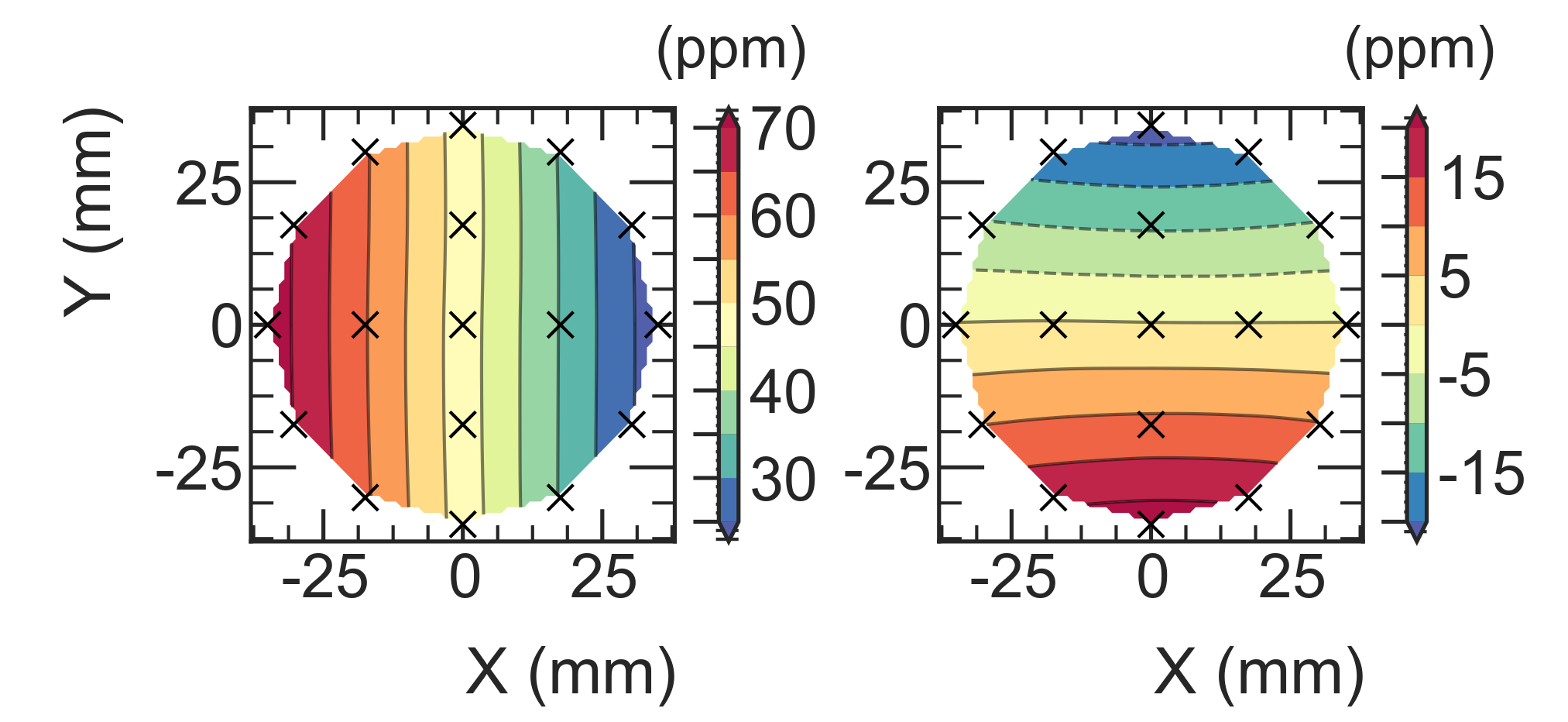}
    \caption{$\Delta \omega_{j,x}^\mathrm{tr}$ (left) and $\Delta \omega_{j,y}^\mathrm{tr}$ (right) measured by the trolley. The graphs were fitted to a two-dimensional polynomial to extract the large imposed field gradient. These gradients are used to uniquely identify a probe's location in the $xy$ plane.}
	\label{fig:transverse_dB}
\end{figure}

Figure~\ref{fig:transverse_dB} shows the field gradients used to locate each trolley probe in the $x$ and $y$ directions. The combination of the two uniquely determines each probe's $xy$ position. This uniqueness can be extended to $xyz$ by including the azimuthal gradient measurement. From these measurements, we obtain the strength of the imposed gradients $\pdiff{\omega^{\mathrm{grad}}_q}{q}$. The calibration probe was moved in the field with the same imposed gradient to find the location where its $\Delta \omega'^{}_{\mathrm{cp},q}$ values matched the trolley's $\Delta \omega_{j,q}^\mathrm{tr}$. In practice, the calibration probe's position was iterated until $|\Delta \omega'^{}_{\mathrm{cp},q}-\Delta \omega_{j,q}^\mathrm{tr}|/2\pi \le \SI{20}{\hertz}$ (\SI{324}{ppb}) for $x$ and $y$ directions and $\le \SI{5}{\hertz}$ (\SI{81}{ppb}) for the $\phi$ direction, corresponding to a position alignment better than \SI{0.5}{\milli\meter}. The remaining difference determines the two probes' misalignment $s^q$. Using the measured gradients, the misalignment in each direction can be extracted via:
\begin{equation}\label{eq:misalignment}
    s_{j}^q = \left( \Delta \omega'^{}_{\mathrm{cp},q} - \Delta \omega_{j,q}^\mathrm{tr} \right) \left/ \pdiff{\omega^{\mathrm{grad}}_q}{q} \right. .
\end{equation}

Prior to the rapid swapping, the local field inhomogeneities around each trolley probe's position were mapped with the calibration probe. Additional imposed fields generated with the \ac{SCC} and the azimuthal coils reduced the local gradients to less than \SI{21}{ppb/\milli\meter} (\SI{30}{\nano\tesla/\milli\meter}). The calibration probe was used to map the residual local shimmed field $\omega^{}_\mathrm{local}$. The misalignment between the target trolley probe and the calibration probe together with the local gradients created an error that is corrected since we measure both the misalignment and gradient. The misalignment correction is then
\begin{equation*}
    \delta^\mathrm{align}_{j} = \boldsymbol{\grad} \omega^{}_\mathrm{local} \cdot \Bs_{j}.
\end{equation*}

To minimize the time between the rapid swaps, the measurements of $\Delta \omega'^{}_{\mathrm{cp},q}$ and $\Delta \omega_{j,q}^\mathrm{tr}$ were only performed prior to the first swap. While the calibration probe can be placed into the same position repeatedly with a precision of $<\SI{0.1}{\milli\meter}$, the trolley positioning during the rapid swapping was based on the less precise encoder readings. This results in position variations of $\mathcal{O}(\SI{1}{\milli\meter})$. To correct for this position variation, the more precise bar code position was used offline to determine an additional azimuthal position offset $\delta^{\mathrm{tr}}_{j,s}$, for each placement of the trolley during the swap $s$ in the sequence. This leads to a modification of Eq.~\eqref{eq:misalignment} for $q=\phi$:
\begin{equation}\label{eq:misalignment_withbarcode}
    s_{j,s}^\phi = \left( \Delta \omega'^{}_\mathrm{{cp},\phi} - \Delta \omega_{j,\phi}^\mathrm{tr} \right) \left/ \pdiff{\omega^{\mathrm{grad}}_\phi}{\phi} \right. + \delta^{\mathrm{tr}}_{j,s}.
\end{equation}

\subsection{Systematic Effects}
\label{sec:calib-systematic} 

Multiple systematic uncertainties arise from the trolley calibration procedure. 
They comprise a statistical component from the rapid swapping in $\omega^{\prime}_{p,j}(\Tr) - \omega^{\tr}_{j}$ and systematic uncertainties arising from the analysis of the \acp{FID}, from the misalignment, and the remaining magnetic footprints of the probes.

\subsubsection{Frequency Extraction Uncertainty: \texorpdfstring{$\epsilon^{}_{i},\ \epsilon^{}_{f}$}{epsilon, epsilon\_f}}
\label{subsec:SystematicsFid}

The calibration constants are based on a zero-crossing algorithm for the frequency extraction of the calibration probe and the Hilbert transform algorithm for the trolley. The systematic fit uncertainty ($\epsilon^{}_{f}$) and the intrinsic systematic uncertainty ($\epsilon^{}_{i}$) (see Sec.~\ref{subsec:NMR_Frequency_Extraction}) are estimated based on simulated \acp{FID}. The large gradients required to colocate the probes produce large field nonuniformities over the probe samples. Thus systematic effects from frequency extraction are larger for these measurements than in the well-shimmed field during the rapid swapping. The full calibration procedure was compared with an independent analysis utilizing the Hilbert transform for the calibration probe frequency extraction. The results agreed within the stated uncertainties.

\subsubsection{Position Misalignment Uncertainty: \texorpdfstring{$\delta^\mathrm{align}_{j}$}{delta\^{}align}}
\label{sec:SystematicsPosition}

The determination of the position misalignment is based on imposing additional large gradients $\pdiff{\omega^{}_\mathrm{q-grad}}{q}$ with the \ac{SCC} and the azimuthal coils. These large gradients degrade the field uniformity and result in larger systematic effects from \ac{FID} frequency extraction. However, the same gradients in the denominator of Eqs.~\eqref{eq:misalignment} and~\eqref{eq:misalignment_withbarcode} suppress the effect of the frequency uncertainty on the actual misalignment, leading to a misalignment uncertainty of less than \SI{0.4}{\milli\meter}.

A set of local measurements of the shimmed field $\omega_\mathrm{local}$ in the vicinity of the probe's location is used to evaluate the local gradient $\grad \omega_\mathrm{local}$ at the actual position of the probe. A lack of knowledge of higher-order and cross-term derivatives in this local field map causes systematic effects in this evaluation.
The residual field was only measured at two positions along some directions for some probes, hence not constraining second- and higher-order gradients along this axis. For those probes and directions, the largest observed gradient is used to estimate an upper limit for the uncertainty of \SI{3}{ppb/mm}, which then couples to the misalignments $s^x_j$ and $s^y_j$.
The azimuthal direction was not mapped for all probes. The observed variations in gradient of up to \SI{22}{ppb/mm} are used as an uncertainty, which couples to the azimuthal misalignment $s^\phi_j$. The resulting uncertainties range of \SIrange{0}{13}{ppb}.

No second-order cross-terms (e.g., $\pdiff{^2 \omega}{x \partial y}$) were explicitly measured. They are estimated from quadratic terms measured along the $x$ and $y$ directions. The cross terms are assumed to be smaller than two times the largest quadratic derivative along the $x$ and $y$ axes ($\pdiff{^2}{x^2}$, $\pdiff{^2}{y^2}$).  
The largest uncertainty generated by the cross term is $\epsilon^{}_\mathrm{cross} = \SI{12}{ppb/\milli\meter^2}$, leading to an uncertainty of $\epsilon^{}_\mathrm{cross}\Delta q^{}_i \Delta q^{}_j$ for $i \neq j$ in a range of 0 to \SI{8}{ppb}.

\subsubsection{Trolley and Calibration Probe Magnetic Footprints}
\label{sec:footprints}
During the calibration probe measurements in the rapid swapping procedure the trolley was azimuthally retracted by $\sim\SI{4}{\degree}$. The calibration probe was used to measure the remaining magnetic footprint of the trolley \emph{in situ} as a function of relative trolley position in a range from \SIrange{3}{100}{\degree}. No perturbations are observed for relative distances larger than $\sim\SI{25}{\degree}$. The probe-independent correction due to the perturbation from the magnetic footprint of the trolley retracted by $\sim\SI{4}{\degree}$ is $\delta_{{fp,tr}}=\SI{40(8)}{ppb}$.

During the trolley measurements the calibration probe is retracted radially inwards. The material of the probe itself and its aluminum fixture perturb the field at the location of the trolley probes slightly. The size of the resulting corrections $\delta^{\fp,\,\pp}_{j}$ ranges from 2 to \SI{7}{ppb} depending on the trolley probe location and the uncertainties were in the range of 1 to \SI{6}{ppb}. Table~\ref{tab:fcc-results} lists the associated uncertainties of the total footprint correction $\delta^{\fp}_{j} = \delta^{\fp,\,\tr}_{j} - \delta^{\fp,\,\cp}_{j}$ for all trolley probes.

\subsection{Results}
\label{sec:calib-results} 

The final calibration coefficients $\delta^{\tr}_j(\Tr)$ were determined via Eq.~\eqref{eq:calibration_offsets} and are shown Table~\ref{tab:fcc-results} along with the statistical and systematic uncertainties described above. The total uncertainty also includes the uncertainty of $15\,$ppb from the corrections to $\opprimej(\SI{34.7}{\celsius})$ from Table~\ref{tab:pp-pert-results}.

\begin{table*}[!hbt]
    \centering
    \begin{tabular}{ccccccc}
        \hline \hline
        Probe &   $\delta^{\tr}_j(\Tr)$  &   Statistical Uncertainty & \multicolumn{3}{c}{Systematic Uncertainties}  &   Total \\\cline{4-6}
        &                               &               & Misalignment & Freq. Extr. & Footprint &  \\  
        & [\si{ppb}] & [\si{ppb}] & [\si{ppb}] & [\si{ppb}] & [\si{ppb}] & [\si{ppb}]\\
        \hline
        \rule{0pt}{1em}
        1   &  1470  &   6   &  27 & 6   & 9  & 33 \\
        2   &  1363  &   11  &  3  & 7   & 9  & 22 \\
        3   &  1538  &   9   &  29 & 11  & 8  & 36 \\
        4   &  1392  &   4   &  11 & 2   & 9  & 21 \\
        5   &  1504  &   8   &  3  & 2   & 9  & 20 \\
        6   &  1719  &   7   &  4  & 13  & 8  & 23 \\
        7   &  1888  &   16  &  4  & 19  & 8  & 30 \\
        8   &  1236  &   10  &  6  & 7   & 8  & 22 \\
        9   &  1352  &   4   &  18 & 8   & 8  & 27 \\
        10  &  389   &   22  &  2  & 11  & 8  & 30 \\
        11  &  2873  &   4   &  21 & 18  & 8  & 32 \\
        12  &  1794  &   7   &  15 & 17  & 8  & 29 \\
        13  &  1989  &   34  &  14 & 22  & 9  & 46 \\
        14  &  1248  &   9   &  13 & 21  & 9  & 32 \\
        15  &  1211  &   17  &  15 & 10  & 10 & 31 \\
        16  &  329   &   7   &  40 & 18  & 9  & 48 \\
        17  &  2786  &   20  &  14 & 22  & 9  & 37 \\
        \hline \hline
    \end{tabular}
    \caption{Calibration coefficients $\delta_j^{\tr}(\Tr)$ and their statistical and systematic uncertainties. The total uncertainties in the last column are the quadrature sum of the statistical and systematic uncertainties listed and the uncertainty of \SI{15}{ppb} for the corrections to $\opprimej(\Tr)$ from Table~\ref{tab:pp-pert-results}.}\label{tab:fcc-results}
\end{table*}

While most probes have total uncertainties of about 20 to \SI{30}{ppb}, a few of the probes on the outer circle have total uncertainties as large as \SI{48}{ppb}, which is driven by large field nonuniformities for probes located nearest to the trolley rails and the iron pole pieces. Many measurements of the field gradient at the outer probes were performed after the main calibration campaign to determine the misalignment, and the drift of the azimuthal gradient contributes significantly to the systematic uncertainty.

    \graphicspath{{./trolley/figures/}}

\section{Trolley Field Mapping}
\label{sec:Trolley_Data_Analysis}

The determination of \opprimetildeatTexp requires precision measurement of the field in the region in which the muons are stored. However, continuous field measurements with \ac{NMR} in the storage region would physically interfere with the muons. The trolley provides detailed frequency maps over the entire storage region. We determined the azimuthally averaged field with a precision of 30 \ac{ppb}. Critically, the trolley is also retracted from the storage region during muon injection periods. While mapping, the set of probes in the fixed probe station are synchronized to the trolley probes. Trolley runs take about four hours in total to execute and are performed approximately every three days to minimize interruptions to muon data taking. The fixed probes continuously track field drifts between the trolley runs. Therefore, we have occasional precise measurements of the field in the storage volume that are interpolated with continuous, less precise measurements. This section covers the analysis of the trolley frequency maps and the corresponding systematic corrections and uncertainties. The relationship between the trolley map and the fixed probe measurements is discussed in Sec.~\ref{sec:Interpolation}.

\subsection{Trolley Maps: \texorpdfstring{$\omega_{j}^{\tr}(\phi,0)$}{omega\^{}tr(phi,0)}}
\label{subsec:Trolley_Magnetic_Field_Maps}

\begin{figure}[!hbt]
    \centering
    \includegraphics[width=3.375in]{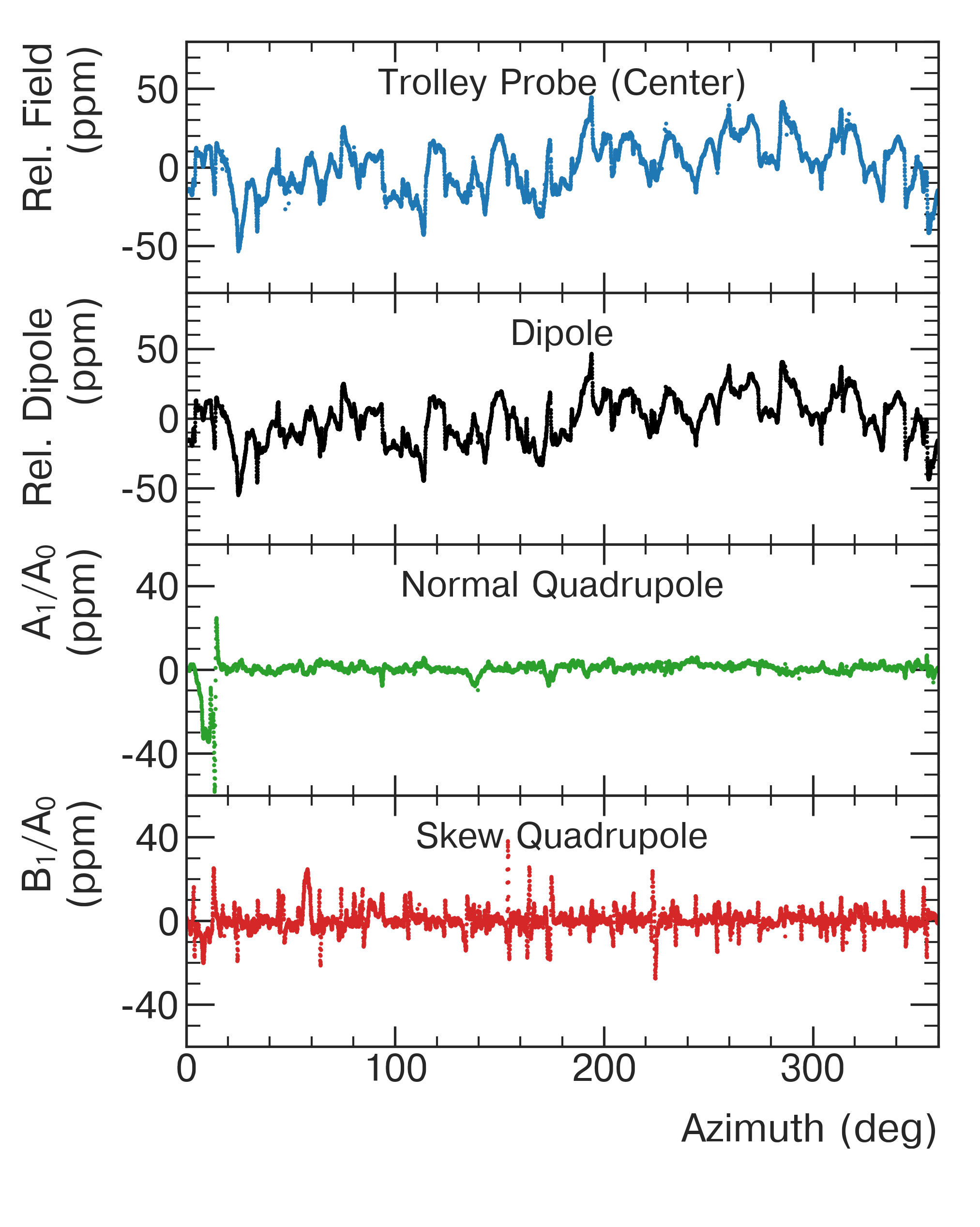}
    \caption{A typical field map from a trolley run (25th of April 2018, approximately at 3 a.m.). On top, the raw, relative frequency $(\omega_{j}(\phi,0)-\avg{\omega_{j}})/\avg{\omega_{j}}$ for the central trolley probe, $j=1$.
    The lower three plots show the corresponding lowest-order multipoles, dipole (black), normal quadrupole (green), and skew quadrupole (red), as a function of azimuth. The dipole distribution has an RMS of \SI{16}{ppm} with a peak-to-peak variation of \SI{101}{ppm}.}
    \label{fig:Trolley_FieldMap}
\end{figure}

\begin{figure}[!hbt]
    \centering
    \includegraphics[width=3.375in,valign=m]{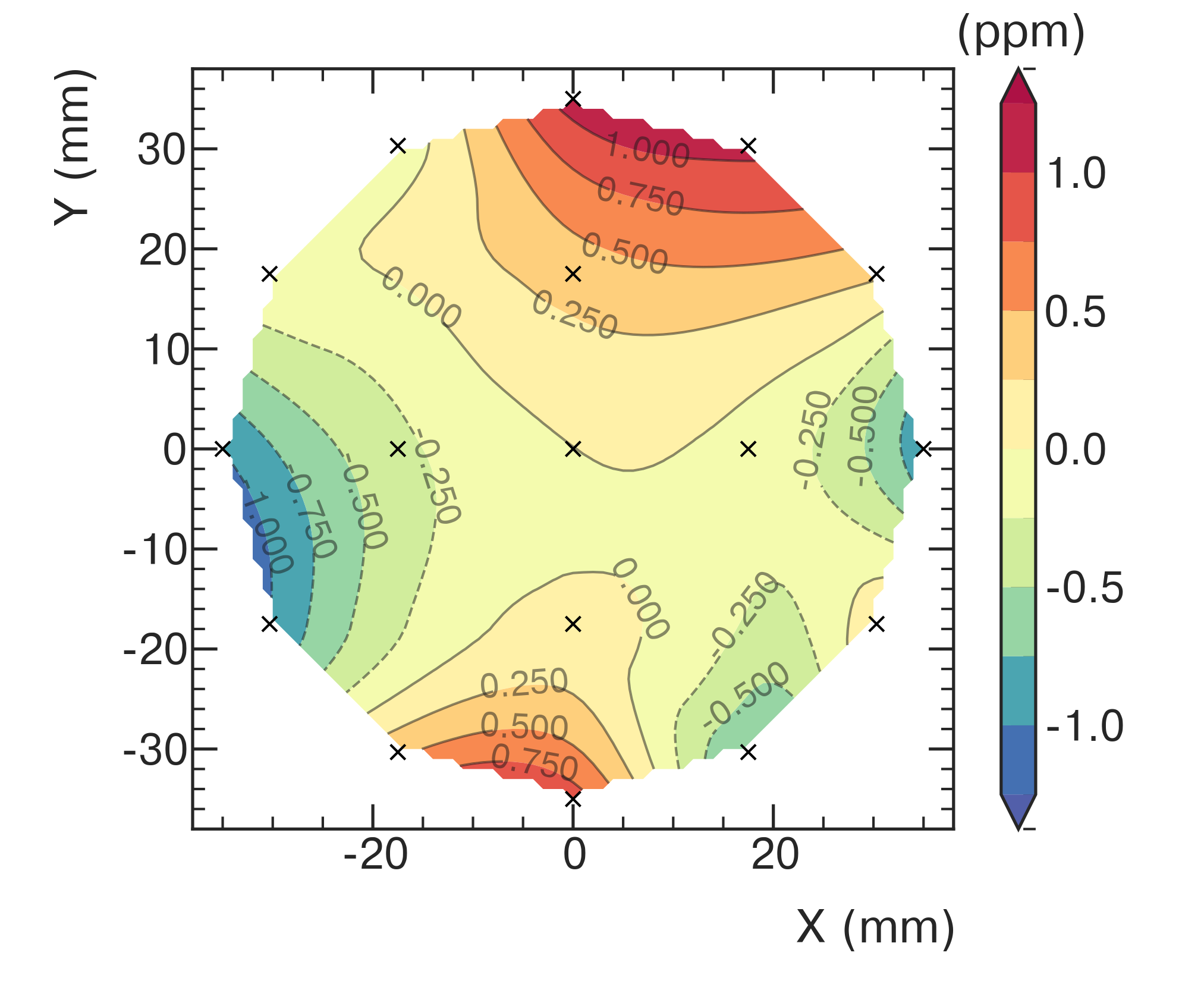}
    \caption{Variations in the azimuthally averaged, relative frequency $(\omega^{\tr}_{j}(\phi,0)-\avg{\omega^{\tr}_{j}})/\avg{\omega^{\tr}_{j}}$, for the central probe ($j=1$). The locations of the 17 trolley probes are indicated by (x). Their raw frequencies are averaged and the field variations are interpolated.}
    \label{fig:Trolley_FieldMap_Avg}
\end{figure}

The trolley moment maps $m_i^{\tr}(\phi,t=0)$ in Eq.~\eqref{eq:Combined} are extracted from the frequency maps $\omega_{j}^{\tr}(\phi,0)$ that are directly measured in the continuous trolley runs by the 17 probes (index $j$). Note that the part of a trolley run, that generates the baseline trolley maps, takes about an hour. Therefore, calling a trolley run time $t=0$ is a notational convenience. The finite duration of the trolley run is taken into account in Sec.~\ref{sec:Interpolation}. Figure~\ref{fig:Trolley_FieldMap} shows the results from a typical trolley run. The top panel shows the raw, relative frequency $(\omega^{\tr}_{j}(\phi,0)-\avg{\omega^{\tr}_{j}})/\avg{\omega^{\tr}_{j}}$ for the central probe ($j=1$), where $\avg{\omega^{\tr}_{j}}$ is the azimuthal average frequency of that probe. The bottom three plots show the extracted multipole moments $m_{i}^{\tr}(\phi,0)$ for the dipole ($i=1$), normal quadrupole ($i=2$), and skew quadrupole ($i=3$), normalized to the dipole moment.

\begin{table}
  \centering
  \begin{tabular}{llrr} \hline \hline
    \multicolumn{2}{l}{Normalized moment strength} & \multicolumn{1}{c}{normal} & \multicolumn{1}{c}{skew} \\
    & & \multicolumn{1}{c}{[ppb]} & \multicolumn{1}{c}{[ppb]} \\
    \hline
    \rule{0pt}{1em}Dipole & $(A^{}_0/A^{}_0,\ -)$ &  1\,000\,000\,000 & - \\
    Quadrupole & $(A^{}_1/A^{}_0,\ B^{}_1/A^{}_0)$&   300  &   399\\
    Sextupole & $(A^{}_2/A^{}_0,\ B^{}_2/A^{}_0)$  & -1\,247 &   395\\
    Octupole & $(A^{}_3/A^{}_0,\ B^{}_3/A^{}_0)$  &   14 &   273\\
    Decupole & $(A^{}_4/A^{}_0,\ B^{}_4/A^{}_0)$  &   39 & -1\,319\\
    Dodecupole & $(A^{}_5/A^{}_0,\ B^{}_5/A^{}_0)$ & -756 &  -187\\
    Tetradecupole & $(A^{}_6/A^{}_0,\ B^{}_6/A^{}_0)$ & -1\,067 &    -0\\
    \hline
    \hline
  \end{tabular}
    \caption{Strength of each moment from the fit in Fig.~\ref{fig:Trolley_FieldMap_Avg} with $r^{}_0=\SI{4.5}{\centi\meter}$ and normalized to the strength of the dipole $A^{}_0$ (see Eq.~\eqref{eq:mult_By} and Table~\ref{tab:mults_in_Cart}), and averaged over azimuth.} \label{tab:azi_avg_moments}
\end{table}

An azimuthally averaged relative frequency distribution for a typical trolley run is shown in Fig.~\ref{fig:Trolley_FieldMap_Avg}. The corresponding azimuthally averaged quadrupole moments are \SI{6.7}{ppb/\milli\meter} (normal) and \SI{8.87}{ppb/\milli\meter} (skew), and higher-order moments are shown in Table~\ref{tab:azi_avg_moments}.

The frequency maps are an integral part of the magnetic field tracking described in Sec.~\ref{sec:Interpolation}. Specifically, they provide precise baseline measurements of the field, which are interpolated using the fixed probes. The trolley maps are averaged over \SI{5}{\degree} of azimuth into 72 bins that correspond to each fixed probe station. The edges of the bins are defined by the midpoints between adjacent fixed probe stations. Azimuthal averages of the 72 bins are used for the systematic uncertainty evaluation (Sec.~\ref{subsec:Trolley_Systematic_Effects}), but do not enter directly into the determination of \opprimetildeatTexp, which is mainly based on fixed probe data and a synchronization of each station to the trolley data as discussed in Sec.~\ref{sec:Interpolation}.

\subsection{Systematic Effects: \texorpdfstring{$\delta_{i}^{\tr,\ \mathrm{syst}}$}{delta\^{}tr,syst}}
\label{subsec:Trolley_Systematic_Effects}

The final field moment maps $m_{i}^{\tr}(\phi,t=0)$ that enter in Eq.~\eqref{eq:Combined} can be derived from the measured maps via $m_{i}^{\tr}(\phi,0) = m_{i}^{\tr,\ \mathrm{meas}}(\phi) \left( 1 + \delta_{i}^{\tr,\ \mathrm{syst}} \right)$,
where $\delta_{i}^{\tr,\ \mathrm{syst}}$ is the sum of systematic corrections and their uncertainties caused by the following effects: 
\begin{itemize}
    \item $\delta_{i}^\mathrm{freq}$: frequency extraction biases and uncertainties from the \acp{FID} from the trolley \ac{NMR} probes,
    \item $\delta_{i}^\mathrm{motion}$: effects that are introduced by the continuous trolley motion and dominated by eddy currents in the trolley shell,
    \item $\delta_{i}^\mathrm{pos}$: corrections for transverse and azimuthal trolley position offsets,
    \item $\delta_{i}^\mathrm{temp}$: corrections due to the temperature of the trolley NMR probes during the field mapping,
    \item $\delta_{i}^\mathrm{multipole}$: field variations that are not described by the moments in Table~\ref{tab:azi_avg_moments},
    \item $\delta_{i}^\mathrm{config}$: differences in the experiment configuration during a trolley run from nominal muon storage conditions.
\end{itemize}

The uncertainties associated with the field maps in \RunOne are treated conservatively and combined as correlated uncertainties.
The following sections discuss these systematics in more depth.
An overview of their numerical values for both the correction and associated uncertainty is given in Table~\ref{tab:Trolley_Systematic_Effects_Overview}.

\begin{table*}[ht]
	\centering
	\begin{tabular}{lccccccc}
	  \hline
          \hline
	  Quantity & \multicolumn{2}{c}{Dipole} & \multicolumn{2}{c}{Normal Quadrupole} & \multicolumn{2}{c}{Skew Quadrupole} \\
		\quad $<\delta_{i}^{X}>$ & Corr. [ppb] & Unc. [ppb] & Corr. [ppb] & Unc. [ppb] & Corr. [ppb] & Unc. [ppb] \\
		\hline
		\rule{0pt}{1em}freq & & & & & &  \\
		\quad syst, fit &  $<1$ & 10 & 1 & 0 & 0 & 0  \\
		\quad stat &  0.0 & 0.1 & 0.0 & 0.2 & 0.0 & 0.2  \\
		\hline
		\rule{0pt}{1em} motion & -15 & 18 & 21 & 10 & -8 & 12 \\
		\hline
		position & & &  & & &  \\
		\quad transverse  & 0 & 12 & 0 & 27 & 0 & 4  \\
		\quad azimuthal  & 0 & 4 & 0 & 2 & 0 & 4 \\
		\hline
            	temperature & 0 & $15-27$ & - & - & - & -  \\
		\hline
		multipoles & 0 & 1 & 0 & 1 & 0 & 1  \\
		\hline
		config & & &  & & & & \\
		\quad garage & -5 & 22 & - & - & - & -  \\
		\quad collimators & $<1$ & $<1$ & - & - & - & - \\
		\quad ground loop & -2 & 0 & -2 & 0 & 3 & 0  \\
		\hline
		{\bf Total} & -21 & $36-43$ & 20 & 29 & -5 & 13  \\
		\hline
                \hline
	\end{tabular}
	\caption{Overview of all contributions to $<\delta_{i}^{\tr,\ \mathrm{syst}}>$. Uncertainties from different trolley regions are treated as correlated uncertainties, leading to a conservative uncertainty estimate. Ranges are specified for uncertainties that vary between data subsets; the range is defined by the minimum and maximum uncertainties from the \RunOneA through \RunOneD data subsets.}
	\label{tab:Trolley_Systematic_Effects_Overview}
\end{table*}

\subsubsection{Trolley Frequency Extraction: \texorpdfstring{$\delta_{i}^\mathrm{freq}$}{delta\^{}freq}}
\label{subsubsec:Trolley_Systematics_FID}
The uncertainty in the extracted \ac{NMR} frequency can be split into $\delta_{i}^{\mathrm{freq}}(\phi) = \delta_{i}^\mathrm{freq,\ stat}(\phi) + \delta_{i}^\mathrm{freq,\ syst}(\phi)$, the statistical uncertainty and the systematic uncertainty which combines the fit uncertainty $\epsilon_f$ and the intrinsic uncertainty $\epsilon_i$ (see Sec.~\ref{subsec:NMR_Frequency_Extraction}).
The systematic contribution is evaluated based on \ac{FID} simulation, taking into account the local field shape around the azimuth $\phi$ as described in Sec.~\ref{subsec:NMR_Frequency_Extraction}.
Systematic stop-and-go trolley runs collect frequency data while the trolley is stationary before being moved to the next position. These measurements are free of motion effects described below and are used to extract the probes' statistical resolution. The resulting uncertainties are statistically independent for each field map but sampled from the same underlying distribution.
The probe resolution is extracted from the variance of measurements taken over \SI{5}{\second} while the trolley is stationary; the field drift is negligible on this timescale.

\subsubsection{Trolley Motion: \texorpdfstring{$\delta_{i}^\mathrm{motion}$}{delta\^{}motion}}
\label{sec:Trolley_Systematics_Eddy}

\begin{figure}[!hbt]
  \centering
  \begin{tikzpicture}
      \node (img) {\includegraphics[width=3.375in]{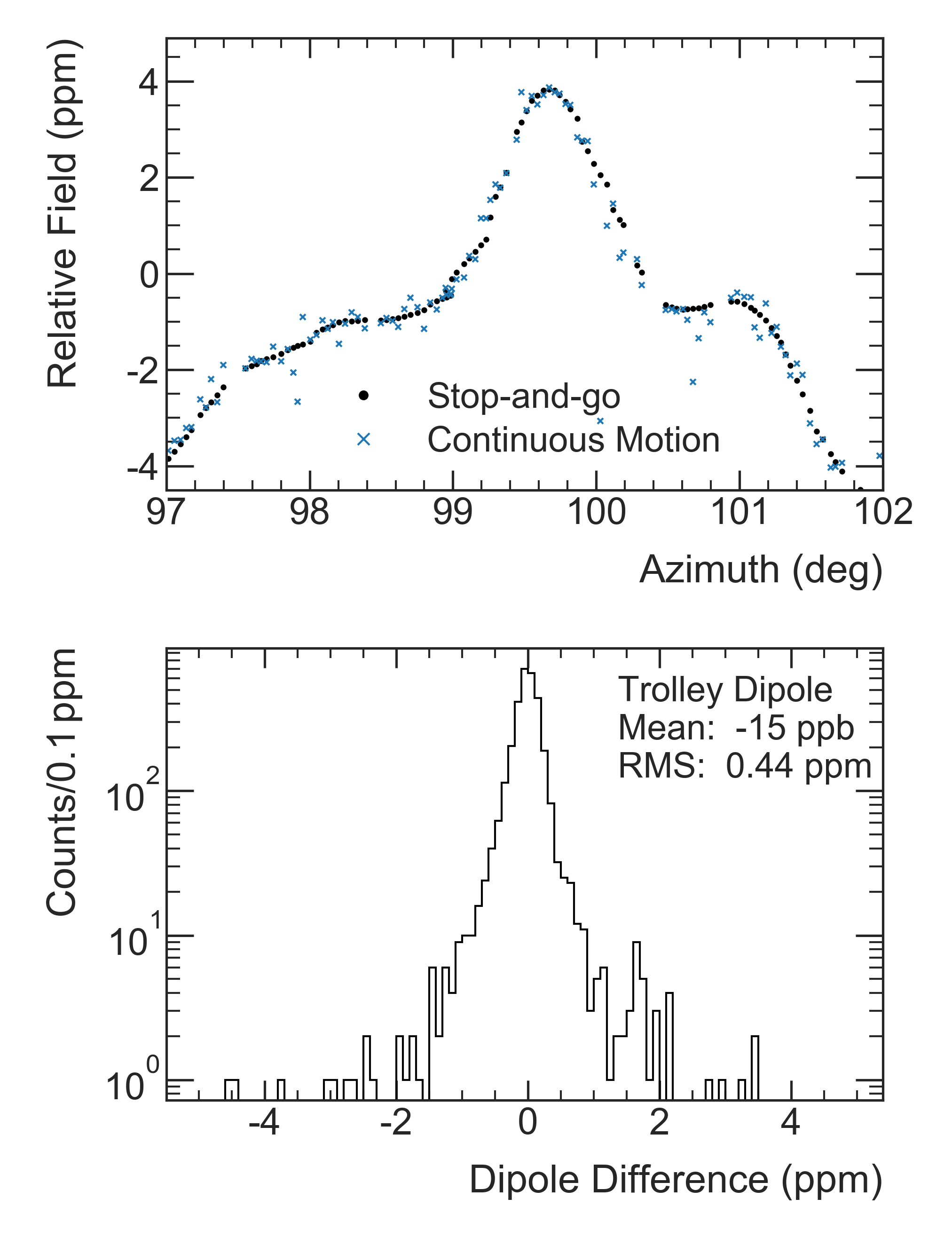}};
      \node [align=center,yshift=-0.35in, xshift=0.85in] at (img.north west){(a)};
      \node [align=center,yshift=-2.55in, xshift=0.85in] at (img.north west){(b)};
  \end{tikzpicture} 
 \caption{The difference between moving and static trolley measurements. 
   (a) Comparison of the frequencies measured in a selected azimuthal region for the normal trolley motion and the stop-and-go operation. (b) The distribution of the dipole differences between motion and static measurements over the full ring.}
 \label{fig:Trolley_Systematic_Effects_Eddy}
\end{figure}

The trolley movement through the nonuniform magnetic field generates eddy currents in the conducting components, most significantly the aluminum shell. These produce transient field variations that affect the trolley map leading to the correction $\delta_{i}^\mathrm{motion}$. It was determined in two ways: 1) comparison of the frequency measurements from two trolley-run modes, one with continuous motion (standard trolley run) and one in stop-and-go, and 2) the comparison of maps taken in the clockwise and counterclockwise directions.

Figure~\ref{fig:Trolley_Systematic_Effects_Eddy}a shows a comparison of the continuous and stop-and-go modes over a narrow azimuthal range. Taking the differences of these frequencies for each probe allows the construction of the azimuthally averaged differences in the field moments shown in Fig.~\ref{fig:Trolley_Systematic_Effects_Eddy}b for the dipole moment. The resolution for the moving trolley is two orders of magnitude worse than what is observed in the static situation.
Additionally, large eddy current spikes generate fluctuations of the measured trolley probe frequencies of up to \SI{20}{ppm} with decay constants on the order of \SI{100}{\milli\second}.
The statistical and systematic uncertainties are determined from the statistics-scaled RMS and dedicated studies that removed spikes from the maps, respectively.
The dipole moment correction is $\delta_{1}^\mathrm{motion} = -15(2)(17)$\,ppb.

\subsubsection{Trolley Position: \texorpdfstring{$\delta_{i}^\mathrm{position}$}{delta\^{}position}}
Extracting moments from trolley data requires knowledge of its position in $x$, $y$, and $\phi$ for each measurement.
The trolley's azimuthal position is determined from the bar code reader, and the uncertainty in the trolley's azimuthal location propagated into the uncertainty $\delta_{i}^\mathrm{pos,\ azi}$ in the field maps.
Position deviations in the transverse directions from the ideal circular muon orbit of radius \SI{7.112}{\meter} predominantly originate from the location and shapes of the rails, generating an uncertainty $\delta_{i}^\mathrm{pos,\ vert}$. The total trolley position uncertainty is $\delta_{i}^\mathrm{position} = \delta_{i}^\mathrm{pos,\ azi} + \delta_{i}^\mathrm{pos,\ vert}$.

\paragraph{Transverse Trolley Position: \texorpdfstring{$\delta_{i}^\mathrm{pos,\ vert}$}{delta\^{}pos,vert}}
The trolley rails have shape distortions with respect to their design curvature and limitations on the precision of their placement inside the vacuum chambers. Extensive rail surveillance data were collected prior to installation using laser tracking, and additional trolley motion verification was performed during installation. The vertical and radial offsets of the rails and their corresponding roll of the trolley are shown in Fig.~\ref{fig:rail_mis}. These data are analyzed to determine the trolley probes' vertical and radial displacements and any roll movement during trolley motion.

\begin{figure}[!hbt]
        \centering
        \includegraphics[width=0.48\textwidth]{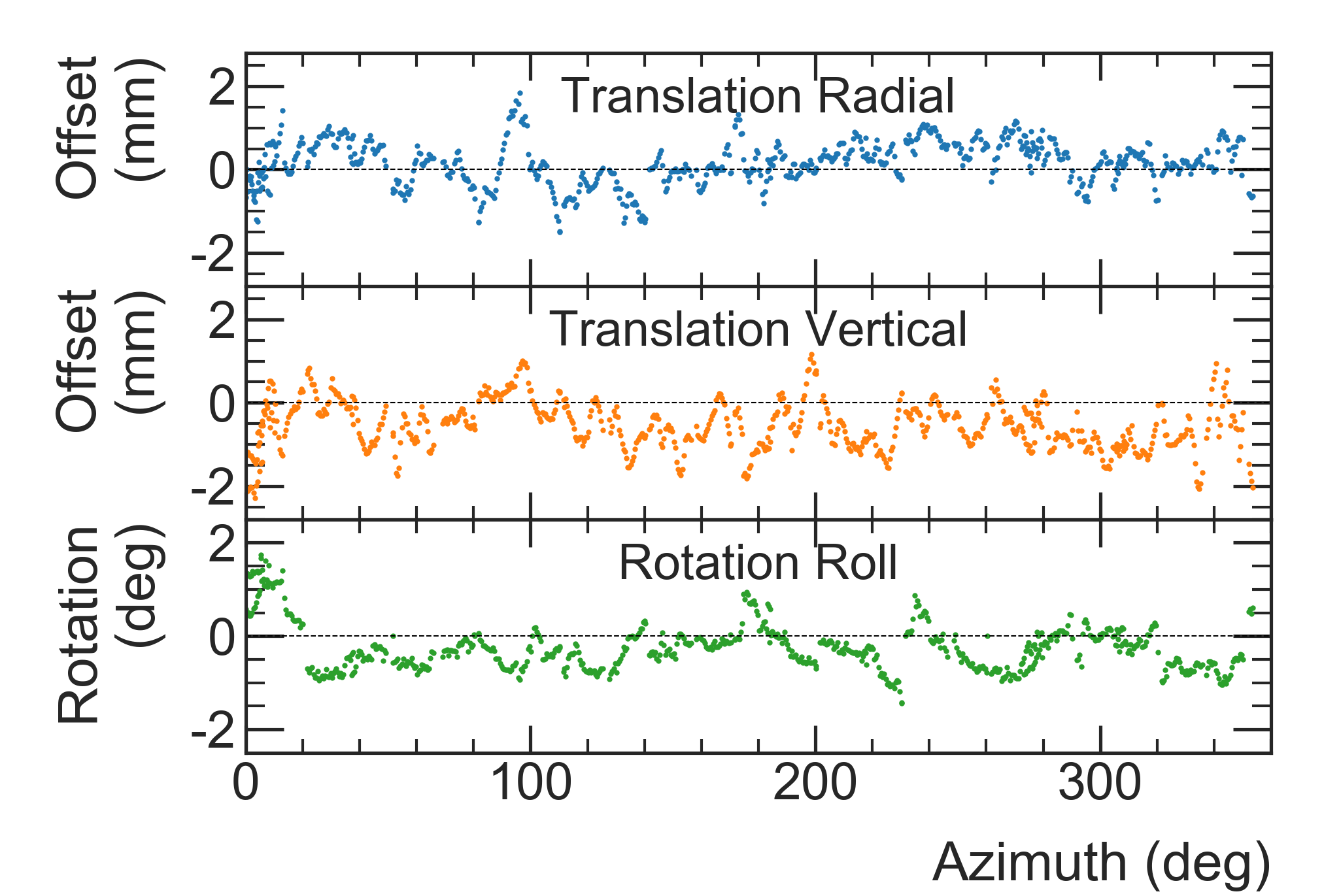}
        \caption{Surveyed offsets (top: radial with a mean of \SI{0.2}{\milli\meter} and RMS of \SI{0.5}{\milli\meter}, middle: vertical with a mean of \SI{-0.6}{\milli\meter} and RMS of \SI{0.6}{\milli\meter}) and rotation (bottom: roll with a mean of \SI{-0.3}{\degree} and a RMS of \SI{0.6}{\degree}) of the trolley rails.}
        \label{fig:rail_mis}
\end{figure}

The multipole moment extraction is performed without accounting for the positional distortions. By repeating the multipole fits with slightly different probe positions determined by including linearly interpolated displacement information at each azimuthal location, systematic uncertainties are determined for the azimuthally-averaged dipole (\SI{12}{ppb}), normal quadrupole (\SI{27}{ppb}), and skew quadrupole (\SI{4}{ppb}).

\paragraph{Azimuthal trolley position: \texorpdfstring{$\delta_{i}^\mathrm{pos,\ azi}$}{delta\^{}pos,azi}}
The bar code reader provides the azimuthal position via the recording of regular, 2-mm-wide alternating dark and bright marks etched into the vacuum chambers. The bar code reader is equipped with two sensor groups that are \SI{12}{\centi\meter} apart and record the same bar code patterns with a small time delay. In the \RunOne analysis, only one group is used to determine the azimuthal position, resolving about 80\% of the full azimuth. For the remaining 20\%, the position information is determined using less precise rotary encoders installed in the cable winding mechanism. The differences between reconstructed bar code positions for the two sensor groups determines the precision of the bar code reader to be \SI{0.2}{\milli\meter}.

Because there are small gaps between adjacent vacuum chambers and some regions that rely on the encoders, a conservative overall position resolution of \SI{2}{\milli\meter} is used. A random variation of the azimuthal trolley positions with a 2-mm-wide Gaussian distribution is used to estimate a systematic uncertainty of \SI{4}{ppb} on the average dipole field.

\subsubsection{Temperature Correction: \texorpdfstring{$\delta_{i}^\mathrm{temp}$}{delta\^{}temp}}
\label{subsubsec:Trolley_Temperature_Correction}

\begin{table*}[!htb]
    \centering
    \begin{tabular}{lccccc}
      \hline\hline
	    \rule{0pt}{1em}Data Subset & Temperature & \multicolumn{2}{c}{$<\delta_{\mathrm{dipole}}^{\mathrm{temp}}(T_{\mathrm{run}})>$} & \multicolumn{2}{c}{$<\delta_{\mathrm{dipole}}^{\mathrm{temp}}(\Delta T)>$}\\
	    & [C] & Corr. [ppb] & Unc. [ppb] & Corr. [ppb] & Unc. [ppb] \\
	    \hline
		\rule{0pt}{1em}\RunOneA & 27.44 & 0 & 27 & 0 & 4 \\
		\RunOneB & 27.99 & 0 & 25 & 0 & 4 \\
		\RunOneC & 28.79 & 0 & 20 & 0 & 4 \\ 
		\RunOneD & 30.06 & 0 & 14 & 0 & 4 \\
		\hline\hline
    \end{tabular}
    \caption{The mean temperature and uncertainties per data subset. The different trolley runs are linearly interpolated weighted by the number of decay muons.}
    \label{tab:Trolley_Systematic_Effects_Temp}
\end{table*}

The temperature of the trolley probes increases during operation due to the trolley's electronics' power dissipation. The precession frequency produced by these \ac{NMR} probes has a temperature dependency of \SI{0(5)}{ppb/\celsius} which was measured with a dedicated setup in the stable and homogeneous solenoid at \ac{ANL}. 
Because the trolley temperature during the field mapping runs differed from the temperature during the trolley calibration, a run-specific uncertainty $\delta_{i}^\mathrm{temp}(T_{\mathrm{run}})$ is applied.

The mean temperatures of all trolley runs, linearly interpolated and weighted by the corresponding number of decay muons, are grouped into data subsets shown in Table~\ref{tab:Trolley_Systematic_Effects_Temp}.
The temperature also varies during the one-hour duration of a trolley run and adds an additional uncertainty. Temperature changes on the order of \SIrange{1.59}{1.70}{\celsius} are observed during the trolley runs. 
This corresponds to assigned systematic uncertainty $\delta_{i}^\mathrm{temp}(\Delta T)$ of \SI{4}{ppb}.
The data-subset-specific systematic uncertainty from the temperature is the sum of these two parts: $\delta_{i}^\mathrm{temp} = \delta_{i}^\mathrm{temp}(T_{\mathrm{run}}) + \delta_{i}^\mathrm{temp}(\Delta T)$.

\subsubsection{Other Systematic Corrections: \texorpdfstring{$\delta_{i}^\mathrm{config}$, $\delta_{i}^\mathrm{multipoles}$}{delta\^{}config, delta\^{}multipoles}}
\label{subsubsec:Trolley_Other_Systematic_Effects}
Other systematic effects include those that arise from the experiment's different configuration during field mapping compared to muon data taking.
The configuration differences during the trolley measurement generate three systematic contributions from (1) the change in the configuration of the garage, (2) the change in the orientation of the beam collimators, and (3) an electrical ground loop. All of these effects are constant for all trolley runs. An additional systematic is caused because the truncated moment expansion does not completely describe the magnetic field.
The trolley is unable to measure higher-order moments accurately, leading to an uncertainty $\delta_{i}^\mathrm{multipoles}$. A 3D fit framework has been developed in \cite{Bodwin:2019} to describe the field maps in toroidal harmonics that obey the Laplace equation. Their framework was also used in Muon g-2 to study the influence of field components not captured by the used multipoles.

The trolley was moved radially in and out of the storage region by a sliding rail section and only measured the magnetic field when this segment of the rails was inserted. However, the segment of the rails was retracted during muon injection. The magnetization of this rail section changed the magnetic field during the trolley measurement in a way that the muons do not experience. A similar systematic effect is caused by three copper collimators\footnote{The experiment is equipped with five collimators but in \RunOne only three of them were used.}. The collimators are retracted during field mapping measurements to prevent interference with the trolley's motion, but inserted during muon injection. 

Corrections and uncertainties are determined for both the garage and collimator effects by modeling their magnetization and estimating the two configurations' differences. Additionally, the effect from the garage was measured by the fixed probe system.  The systematic effects are \SI{-5(22)}{ppb} for the garage and less than \SI{1(1)}{ppb} for the collimators.

Over a small azimuthal extent of $\approx\SI{5}{\degree}$, the trolley shell makes contact with the grounded kicker plates. This provides an additional ground path for the return current of the trolley power, which normally flows through the coaxial cable connected to the trolley. The imbalance in current paths generates a small magnetic field and affects the trolley probes and all fixed probe stations between the trolley and the end of the coaxial cable at the trolley drive.
Dedicated measurements that broke the ground loop showed systematic shifts for the azimuthally-averaged dipole (\SI{-2}{ppb}), normal quadrupole (\SI{-2}{ppb}), and skew quadrupole (\SI{3}{ppb}). The electrical contact causing the ground loop effect has since been corrected for future datasets.

\graphicspath{{./interpolation/figures/}}

\interfootnotelinepenalty=10000

\section{Magnetic Field Tracking}
\label{sec:Interpolation}

Changes of \opprimetilde between trolley map measurements are predominantly due to changes of the magnetization and geometry of the magnet’s ferromagnetic components and may include hysteresis. We track \opprimetilde with the 72 fixed probe stations, each containing four or six probes mounted outside the vacuum chambers (see Fig.~\ref{fig:trolley_layout_basic}). The procedure of synchronizing the fixed probes during the trolley run and tracking certain moments accounts for the changes of \opprimetilde during muon storage, up to uncertainties that are discussed in Sec.~\ref{subsec:calib_drift}.

The tracking procedure incorporates the following main steps, which will be described in more detail in Secs.~\ref{subsec:interp} and~\ref{subsec:interpolation_systematics}: 
\begin{enumerate}
	\item After the application of all data quality cuts (see Sec.~\ref{subsec:Data_Quality_Control}) and the trolley probe calibration offsets, all \ac{NMR} measurements obtained from the trolley mapping and fixed probes are converted into 2D moments according to the prescription in Sec.~\ref{subsubsec:Introduction_Trolley_Multipole_Expansion}.
	\item Because the magnetization of the trolley's materials and eddy currents in its shell distort a fixed probe station's local field, algorithms are applied to remove this magnetic footprint from the fixed probe measurements. 
	\item The four or five moments tracked at each four- or six-probe station shown in Table~\ref{tab:mults_in_Cart} are synchronized to the moments measured by the trolley during a trolley run using a Jacobian described below (and in Appendix~\ref{appendix:jacobian}).
	\item The field's evolution is interpolated by tracking the changes in the fixed probe measurements from the baseline measured during a trolley run $\Bmtrst(0)$.
	\item Corrections are added to the interpolated field map for systematic sources such as temperature variations, magnetic configuration changes, trolley systematic effects, and fast field transients.
\end{enumerate}

The tracking procedure combined with the calibration probe corrections from Sec.~\ref{sec:shielded-proton-correction} provide the field moments $m^{\prime}_{p,i} (\phi_k,t,T) = \mtr_i(\phi_k,t) \left( 1 + \delta_i^{\tr} \right)$ [see also Eq.~\eqref{eq:Combined}], where $\phi_k$ refers to the azimuthal locations of the $k$th fixed probe station. As a reminder, $\mtr_i(\phi,t)$ and $\mfp_j(\phi,t)$ denote the $i$th multipole (trolley) and $j$th Cartesian (fixed probe) moment, respectively. Additionally, $\Bmtr(\phi,t)$ and $\Bmfp(\phi,t)$ denote vectors in the vector space of moments in a slice of azimuth $\phi$ and time $t$. These vectors are, in principle, elements of $\mathbb{R}^{17}$ (trolley), $\mathbb{R}^6$ (six-probe stations), or $\mathbb{R}^4$ (four-probe stations). However, in practice, we truncate the trolley and six-probe stations to only $\mathbb{R}^{9}$ and $\mathbb{R}^{5}$ due to the large uncertainties in the tracking of the higher-order moments. The effect of this truncation is negligible because the influence of the higher-order moments on the average magnetic field is suppressed when the field is weighted by the muon distribution, discussed in Sec.~\ref{sec:MuonWeighting}.

For a specific station $s$ at $\phi=\phi_{\st}$, the field moments are $\Bmtrst(t)=\Bmtr(\phi_{\st},t)$ and $\Bmfpst(t)=\Bmfp(\phi_{\st},t)$. In practice, $\Bmtrst(0)$ is averaged over $\sim\SI{5}{\degree}$ of azimuth and $\Bmfpst(0)$ is averaged over the amount of time it takes the trolley to traverse that azimuth, about \SI{40}{\second}. With this notation and neglecting the untrackable higher order moments $\epsilon_{\st}^\ho(t)$ for now, Eq.~\eqref{eq:FPTracking} from Sec.~\ref{subsubsec:Introduction_Analysis_Flow} becomes 
\begin{linenomath}\begin{align}
		\Bmtrst(t) &= \Bmtrst(0) + \BJst \cdot \left[ \Bmfpst(t) - \Bmfpst(0) \right],
		\label{eq:Taylor_form}
\end{align}\end{linenomath}
where $t=0$ is the synchronization time during the trolley run for that particular station. $\BJst$ is the Jacobian with elements $\Jstij=\pdiff{ \mtr_i (\phi^{}_{\st})}{\mfp_j (\phi^{}_{\st})}$. The Jacobian matrix is $9\times5$ for the six-probe stations and $9\times 4$ for the four-probe stations. Because the fixed probes can only track lower-order moments, $\Jstij = 0$ for $i \ge 6$ ($i \ge 5$ for the four-probe stations). For moments that are measurable by the trolley but not the six-probe stations, we linearly interpolate between the two trolley runs. The moment $\mfive$, which can be tracked by a six-probe station but not a four-probe station, is estimated in four-probe stations to be the average of $\mfive$ from the nearest neighbors (which are always six-probe stations). This approximation is mathematically equivalent to increasing the weight of six-probe stations that neighbor four-probe stations.

When considering the azimuthal average over the full storage ring, we sum over the stations weighted by their azimuthal spacing $W_{\st} = \Delta \phi_\st / 2\pi$,
\begin{linenomath}\small\begin{equation}
		\avg{\Bmtr(t)}_\phi = \sum_{\st} W_{\st} \left\{  \Bmtrst(0) + \BJst \cdot \left[ \Bmfpst(t) - \Bmfpst(0) \right] \right\}. \label{eq:Taylor_form_azi}
\end{equation}\normalsize\end{linenomath}

Equation~\eqref{eq:Taylor_form_azi} has four quantities of interest: $\Bmtrst(0)$, $\Bmfpst(0)$, $\Bmfpst(t)$, and the Jacobian matrix $\BJst$. The baseline measurements $\Bmtrst(0)$ and $\Bmfpst(0)$ for each fixed probe station are measured simultaneously during a trolley run. Trolley measurements are grouped according to the closest fixed probe station ($\sim \pm \SI{2.5}{\degree}$ around a fixed probe station), establishing $t=0$ for each station and synchronizing the two sets of probes. From the fixed probe stations' measurements, $\Bmfpst(t)$ is calculated for times between the two trolley runs. The Jacobian matrix is determined analytically from each fixed probe station's geometry. Details of the explicit Jacobians for the general six- and four-probe stations and some stations with special geometry are given in Appendix~\ref{appendix:jacobian}.

\subsection{Tracking Analysis}
\label{subsec:interp}

The tracking analysis has five primary steps outlined above. This section addresses the first four, which are needed as inputs to Eq.~\eqref{eq:Taylor_form}. The final step is to determine systematic corrections and uncertainties and is covered in detail in Sec.~\ref{subsec:interpolation_systematics}. 

\subsubsection{Data Preparation}

Before beginning the tracking analysis, the data quality selection described in Sec.~\ref{subsec:Data_Quality_Control} is performed. Then, the trolley calibration offsets described in Sec.~\ref{sec:pp-trly-calib} are added to the frequency measurements from the trolley as shown in Eq.~\eqref{eq:w_p}. 
The trolley and fixed probe NMR measurements are converted into the multipole moment and the Cartesian moment bases, respectively. During trolley runs, there are $\approx$9000 sets of moments for the trolley and each of the 72 fixed probe stations; during the muon production runs, there are 72 sets of moments every \SI{1.4}{\second} between each pair of trolley runs.

\subsubsection{Trolley Footprint Replacement}

As the trolley is pulled past fixed probe station $s$, the trolley's magnetization and eddy currents in its shell perturb the field at the location of the station's probes.
This perturbation, the ``trolley footprint", needs to be removed from the fixed probe data before performing the time averaging of the station moments to calculate $\Bmfpst(0)$. All measurements from the fixed probes are vetoed during the time when the trolley was close enough to influence the station measurably ($\pm\SI{12.5}{\degree}$ of azimuth about the fixed probe location, approximately $\SI{200}{\second}$). The vetoed data are replaced by an estimate of the fixed probe moments' unperturbed values. We use data from outside the veto window when the trolley is sufficiently far away not to perturb the measurements and interpolate over the vetoed data points. The interpolating function is a model of the local drift of the station during the veto window. The local model is a fifth order polynomial fit to the unperturbed data from the station. It is corrected with data measured over the rest of the ring to account for global field transients that would otherwise be missed in the veto region. An example of a footprint replacement for one of the fixed probe stations is shown in Fig.~\ref{fig:footprint_replacement}. A subset of the interpolated points are used to calculate the fixed probe baseline moments $\Bmfpst(0)$.

\begin{figure}
	\centering
	\includegraphics[width=0.95\linewidth]{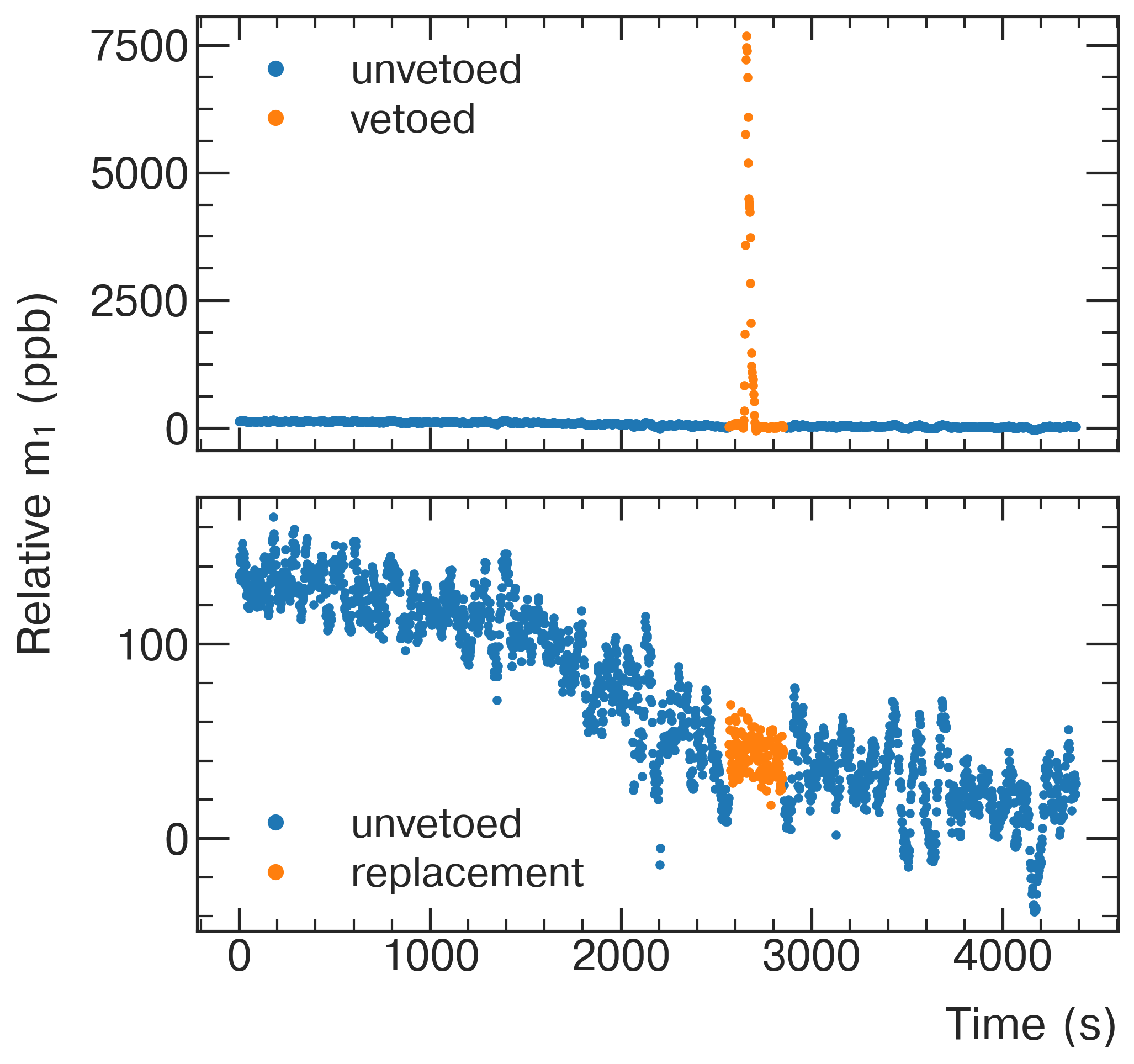}
	\caption{The trolley footprint as seen by the fixed probe station, and the replacement data after vetoing the perturbed region.}
	\label{fig:footprint_replacement}\end{figure}

\subsubsection{Synchronization and Tracking}
\label{subsubsec:sync}

In the synchronization step, we find the trolley and fixed probe baseline ($t=0$) moments $\Bmtrst(0)$ and $\Bmfpst(0)$ in Eq.~\eqref{eq:Taylor_form} for each of the 72 fixed probe stations. The trolley measurements closest to each station, $\sim\SI{5}{\degree}$ per station, are averaged with a weight determined by the azimuthal step-size of each trolley measurement. These azimuthal averages are the baseline trolley moments $\Bmtrst(0)$ for that station. During the time interval when the trolley is closest to the station, the footprint-corrected fixed probe measurements are averaged over time to calculate the baseline fixed probe moments $\Bmfpst(0)$.

Equation~\eqref{eq:Taylor_form} can be rearranged to group baseline terms together
\begin{linenomath}\begin{align}
		\Bmtrst(t) &= \BJst \cdot \Bmfpst(t) + [\Bmtrst(0) - \BJst \cdot \Bmfpst(0)] \nonumber \\
		&= \BJst \cdot \Bmfpst(t) + \Bc_{\st}(0).\label{eq:sync_form}
\end{align}\end{linenomath}
In this form, $\Bc_{\st}(0)$ is a synchronization constant for station $s$, measured entirely during the trolley run. During \RunOne, there was one production trolley pair that only had a trolley one before the period due to magnet issues preventing us from bookending the period. However, trolley runs bookend most production data sets, so two synchronization constants can be calculated for those data sets, one from each adjacent trolley run. In general the two values are not equal, implying that the synchronization drifted over the course of the production period. This effect is the ``tracking error." Because the goal is to track the field between one trolley run at time $t=0$ and the next at $t=T$, we replace $\Bc_{\st}$ in Eq.~\eqref{eq:sync_form}  with a time-dependent form $\Bc_{\st}(t)$. With no additional information about this term between times $t=0$ and $t=T$, we express $\Bc_{\st}(t)$ as a linear interpolation from $\Bc_{\st}(0)$ to $\Bc_{\st}(T)$. The time-dependent synchronization is
\begin{linenomath}\begin{align}
		\Bmtrst(t) =& \BJst \cdot \Bmfpst(t) + \Bc_{\st}(t) \nonumber \\
		=& \BJst \cdot \Bmfpst(t) \nonumber\\
		&+ \left[ \Bc_{\st}(0) + \frac{\Bc_{\st}(T) - \Bc_{\st}(0)}{T} t \right] + \mathbf{\Delta}_{\st}(t),
		\label{eq:sync_form_time}
\end{align}\end{linenomath}
where $\mathbf{\Delta}_{\st}(t)$ is the non-linear component of the drift of moments that the fixed probes cannot track, which leads to the tracking error. This term is the leading source of uncertainty in the field tracking analysis. The process described in Eq.~\eqref{eq:sync_form_time} is called ``backward interpolation" because it involves correcting for drift from the first (``forward") synchronization by interpolating backward in time from the second (see Fig.~\ref{fig:backward_interp} for an example of the effect). Long stationary trolley runs to measure the tracking error rate suggest that it follows the statistics of a random walk with known initial and final constraints; we approximate $\mathbf{\Delta}_{\st}(t)$ as a Brownian bridge \cite{BrownianMotion2008}. The distribution of the differences $\Bc_{\st}(T) - \Bc_{\st}(0)$ from 11 trolley pairs is used to parameterize the rate of the random walk for each station, and therefore estimate the uncertainty on our time averages of $\Bmtrst$ from not knowing the functional form of $\mathbf{\Delta}_{\st}(t)$. With the time-dependent tracking error, the Jacobians, and the fixed probe measurements, Eq.~\eqref{eq:sync_form_time} is evaluated to determine each station's $\Bmtrst(t)$. This quantity is an estimate of what the trolley would measure at station $s$ at time $t$. The set of all 72 $\Bmtrst(t)$ for a given time constitutes our field map at time $t$.

\begin{figure}
	\centering
	\includegraphics[width=0.95\linewidth]{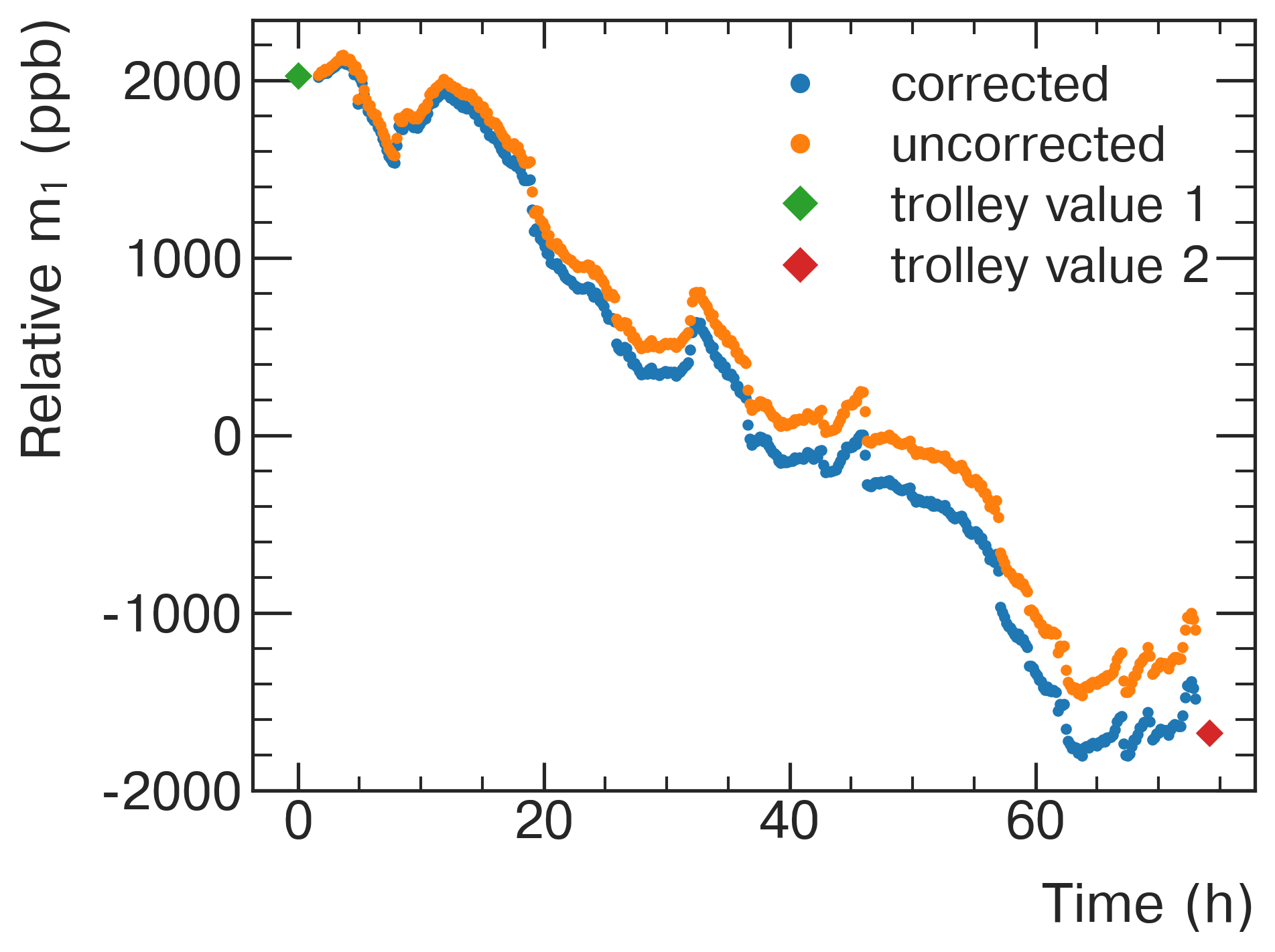}
	\caption{Before the backward correction, the uncorrected tracking curve (light) can disagree with the measurement from the second trolley run. After the correction (dark), $\Bmtrst(t)$ is equal to the corresponding trolley measurements $\Bmtrst(0)$ (left diamond) and $\Bmtrst(T=\SI{74}{\hour})$ (right diamond) at both bookending trolley runs. Note that this plot shows only a single station.}
	\label{fig:backward_interp}
\end{figure}

Figure~\ref{fig:run1_track} shows the results of the tracking analysis over all four major data subsets in \RunOne. The dipole and the normal quadrupole are each averaged over azimuth and shown as a function of time. The dipole trend generally behaves smoothly and the drift is understood to be caused by the selection of probes used in the stabilizing power supply feedback algorithm. The normal quadrupole term is sensitive to temperature variations and exhibits a diurnal structure in addition to slow drifts. The moments are used as inputs in the muon weighting in Sec.\ref{sec:MuonWeighting}.

\begin{figure}
	\includegraphics[width=0.95\linewidth]{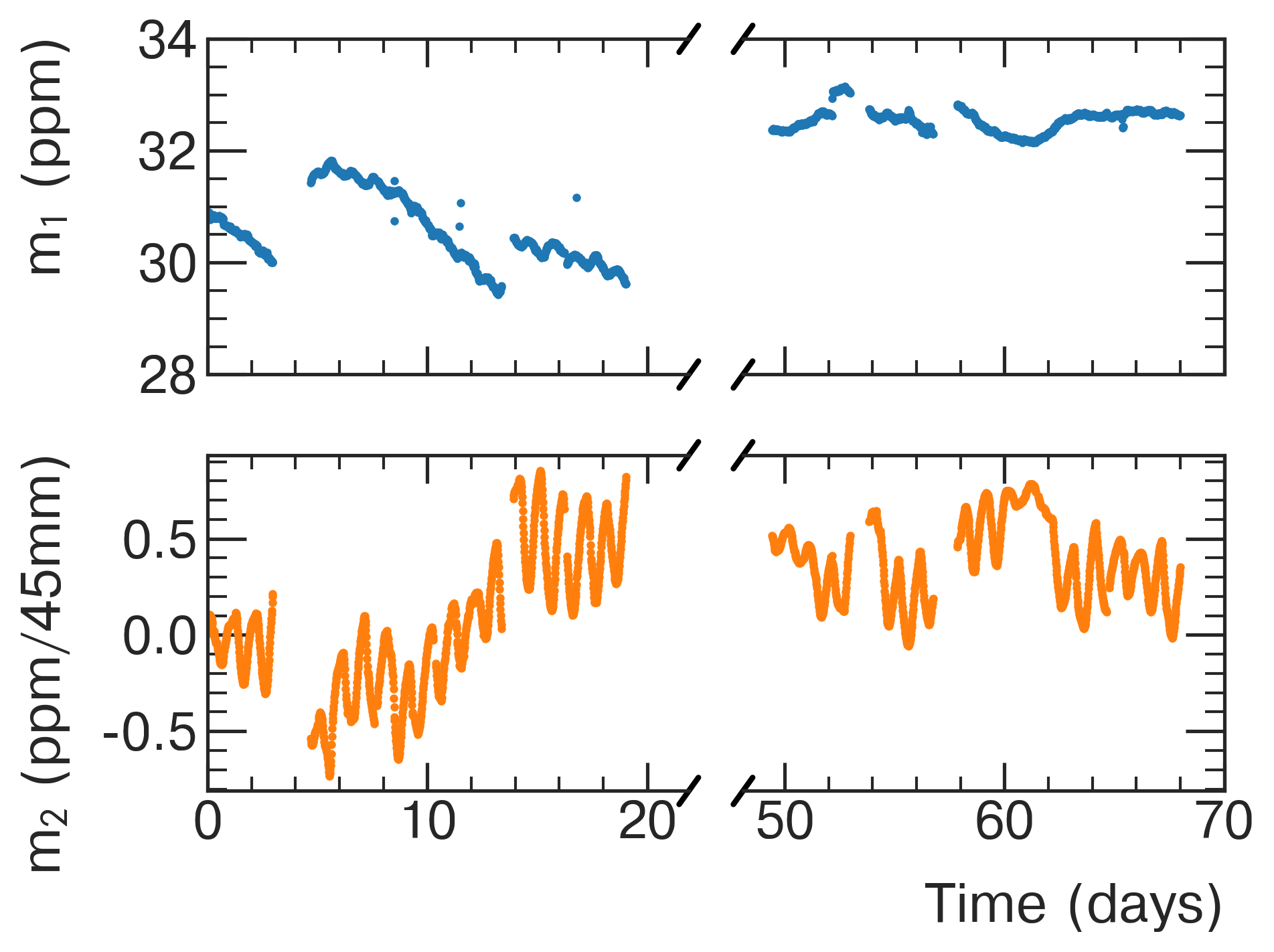}
	\caption{The first two moments (dipole and normal quadrupole), each azimuthally averaged, tracked over the four major data subsets in \RunOne. The average field $\mone$ is reported as \si{ppm} away from a reference value of \SI{61.79}{\mega\hertz}, so the value is $(f - \SI{61.79}{\mega\hertz})/\SI{61.79}{\mega\hertz}$. The moment $\mtwo$ is also reported as \si{ppm} of \SI{61.79}{\mega\hertz}, but with a central value of \SI{0}{\hertz}. The major discontinuities coincide with magnet ramps and major configuration changes.}
	\label{fig:run1_track}
\end{figure}

\subsection{Systematic Effects}
\label{subsec:interpolation_systematics}
The systematic uncertainties are presented by the term they enter in Eq.~\eqref{eq:Taylor_form}. The shown values of the corrections and uncertainties are for the azimuthally and time-averaged field (the averaging procedure is discussed in Sec.~\ref{sec:MuonWeighting}). The trolley baseline systematics are discussed in Sec.~\ref{subsec:Trolley_Systematic_Effects} and shown in Table~\ref{tab:Trolley_Systematic_Effects_Overview}. Fixed probe baseline and fixed probe run systematic uncertainties refer to uncertainties on the relevant terms in Eq.~\eqref{eq:Taylor_form}. An overview of their numerical values for both the correction and associated uncertainty is given in Table~\ref{tab:interp_systematics}. The tracking error systematic refers to the uncertainty related to the untrackable term $\mathbf{\Delta}_{\st}(t)$ in Eq.~\eqref{eq:sync_form_time}.

\begin{table}  
	\centering       
	\begin{tabular}{lcc}
		\hline \hline
		\rule{0pt}{1em}	Quantity & Corr. [ppb] & Unc. [ppb] \\
\hline
		$\mfp(0)$ &  &   \\
		\quad trolley footprint & 0 & 8  \\
\hline
		$\mfp(t)$ & &   \\
		\quad fixed probe resolution & 0 & 1 \\
		\hline
		$\mathbf{\Delta}_{\st}(t)$ & 0 & 22--43  \\
		\hline \hline
	\end{tabular}
	\caption{The systematic corrections and uncertainties from the field tracking analysis. The uncertainties are categorized by where they enter in Eq.~\eqref{eq:Taylor_form}. The trolley baseline systematics can be found in Table~\ref{tab:Trolley_Systematic_Effects_Overview}.}
	\label{tab:interp_systematics}
\end{table}

\subsubsection{Fixed Probe Baseline Systematic Effects: \texorpdfstring{$\Bmfpst(0)$}{m\^{}fp\_st(0)}}

The fixed probe baseline systematic is driven by the trolley footprint replacement and short ($\sim$\SI{1}{\hour}) averaging times of the fixed probe noise during the trolley run. We estimate these effects by implementing the same footprint-replacement algorithm used during the trolley run on fixed probe data in which the trolley is not present. The fixed probe baseline calculated from the replacement data can then be directly compared to a baseline calculated from the measured value. This process can be repeated for all the fixed probe stations and all moments over many sample data sets. The resulting uncertainty is \SI{8}{ppb}.

\subsubsection{Fixed Probe Run Systematic Effects: \texorpdfstring{$\Bmfpst(t)$}{m\^{}fp\_st(t)}}

The primary source of uncertainty on the fixed probe measurement is caused by the measurement noise on the fixed probes. Here, noise is defined as the standard deviation of a measurement over times short enough for field drift to be negligible. Despite some fixed probes being quite noisy from measurement to measurement, over very long averaging times (3 days) the contribution to the uncertainty on the azimuthal average is reduced to under \SI{1}{ppb}.

\subsubsection{Tracking Error Systematic Effects: \texorpdfstring{$\mathbf{\Delta}_{\st}(t)$}{Delta\_st(t)}}
\label{subsec:calib_drift}
The dominant source of uncertainty in the field tracking comes from the tracking error between the trolley and fixed probes, discussed in Sec.~\ref{subsubsec:sync}. This drift is parameterized by the difference in the synchronization constants $\Bc_{\st}$ from Eq.~\eqref{eq:sync_form} between trolley runs, $\Bc_{\st}(T) - \Bc_{\st}(0)$ and is modeled as a Brownian bridge. Its uncertainty is derived analytically, using the equations for the variance and covariance of points in a Brownian bridge process \cite{BrownianMotion2008}. One time period during Run-1d did not have a trolley run after the muon data period due to the magnet's safety monitoring systems triggering a ramp down, so the tracking error for that period is instead modeled as a random walk. Because each trolley baseline is corrected for temperature (see Sec.~\ref{subsubsec:Trolley_Temperature_Correction}), this model also accounts for temperature drift in the fixed probes that influence their frequency measurements.

To average $N$ measurements $\Bx$ with normalized weights $\Ba$, we need to know the $N \times N$ covariance matrix $\BSigma$. Then the average of the measurements is $\Ba \cdot \Bx$, and the variance of the average is $\Ba \cdot \BSigma \cdot \Ba$. In our case, the weights $\Ba$ are related to the number of muons in the storage ring at a given time (described in detail in Sec.~\ref{sec:MuonWeighting}). The expectation value of a random walk or Brownian bridge is zero, so there is no correction associated with the tracking error. However, the variance of either process is not zero. For a Brownian bridge between times $0$ and $T$, the covariance between any two times during the process $t_1 \le t_2$ is
\begin{equation}
	\sigma(t_1, t_2) = M \frac{(T-t_2)t_1}{T},
	\label{eq:bb_unc}
\end{equation}
where $M$ parametrizes the rate of the process. The value peaks at $t_1 = t_2 = T/2$, showing that the variance of the Brownian bridge is largest in the middle of the process and decreases to zero at either bound. We use this functional form to construct the covariance matrix for all the measurements between adjacent trolley runs, and then use that matrix to calculate the variance on the average of the measurements described above; the tracking error uncertainty is the square root of the variance on the average. The same process is repeated for the unbookended data period, except the drift is modeled as a random walk instead of a Brownian bridge. The covariance of a random walk is
\begin{equation}
	\sigma(t_1, t_2) = M t_1
	\label{eq:rw_unc}
\end{equation}
for $t_1 \le t_2$ and the same $M$ as above. The variance during a random walk increases linearly in time.

To use either Eq.~\eqref{eq:bb_unc} or~\eqref{eq:rw_unc}, we must have an estimate of the parameter $M$. As alluded to above, $M$ is estimated by considering the differences $\Bc_{\st}(T) - \Bc_{\st}(0)$ for each trolley run pair. These differences for each of the $72 \times 9$ station-(trolley) moment combinations can be interpreted as sampling the random walk space and can be normalized by the square root of the time between the measurement for each trolley pair, which varies from 54 to \SI{88}{\hour}. For each station-moment combination, the RMS of the normalized samples is taken as an estimate of the random walk rate. The azimuthal average of the random walk rate is calculated for each moment, taking into account correlations between adjacent stations using an autocorrelation function of the differences $\Bc_{\st}(t=T) - \Bc_{\st}(t=0)$ over $s$. It is then used in the equations above to calculate the covariance matrix $\BSigma$.

Any two separate random walks (or Brownian bridges) are uncorrelated with each other. When we average multiple trolley pairs within each data subset, the tracking error uncertainties become smaller. The more trolley pairs averaged in a single data subset, the lower the uncertainty will be for that subset. Therefore, despite the uncertainty being $\sim$\SI{40}{ppb} for a single trolley pair (or \SI{73}{ppb} for the period with no closing trolley run), the uncertainties for the four data subsets are significantly lower, 22 to \SI{43}{ppb}. The values of the synchronization uncertainty for each subset are shown in Table~\ref{tab:bb_unc_runs}.

\begin{table}
	\begin{tabular}{lcc}
		\hline \hline Data subset & Number of & Tracking error $\Delta$ \\
                & trolley pairs & (ppb) \\ \hline
		\RunOneA & 1 & 43 \\
		\RunOneB & 2 & 34 \\
		\RunOneC & 3 & 25 \\
		\RunOneD & 5 & 22 \\
		\hline \hline
	\end{tabular}
	\caption{The tracking error uncertainty for each data subset in \RunOne. Note that the uncertainty decreases with the number of trolley pairs in the subrun.}
	\label{tab:bb_unc_runs}
\end{table}
    \section{The Muon-Weighted Magnetic Field}
\label{sec:MuonWeighting}
\graphicspath{{./beam-conv/figures/}}

The average magnetic field experienced by the muons as they precess in the storage ring is expressed in terms of $\tilde \omega_p '$ (see Eq.~\eqref{eq:Btoomega}). It is determined by weighting the frequency maps with the muon distribution and averaging over space and time.
The quantities needed for this determination are the muon distribution as a function of space and time, $\rho^{\mu}(r,y,\phi,t)$ and calibrated, interpolated frequency maps $\opprime(r,y,\phi,t)$ that represent the field in the $ry\phi$ basis (see Fig.~\ref{fig:coordinates}).
Over a time interval $[0,T]$, with a muon distribution bounded radially and vertically, the resulting muon-weighted magnetic field, expressed in terms of the shielded proton precession frequency is,
	\begin{linenomath}\small\begin{equation}
		\tilde \omega_p' = \frac{\int_0^T\dd t \int_0^{2\pi}\dd\phi \int_{r_1}^{r_2}\dd r \int_{-y_0}^{y_0}\dd y ~ r \rho^\mu(r,y,\phi,t) \omega_p'(r,y,\phi,t)}{\int_0^T\dd t \int_0^{2\pi}\dd\phi \int_{r_1}^{r_2}\dd r \int_{-y_0}^{y_0}\dd y ~ r \rho^\mu(r,y,\phi,t)}.
	\label{eq:op_tilde}
	\end{equation}\normalsize\end{linenomath}

\subsection{Time Averaging}
\label{subsec:intrafill}

Before evaluating the integral over $t$ in Eq.~\eqref{eq:op_tilde}, we consider the relevant timescales involved. The storage time of a muon injection (the intrafill time) is on the order of hundreds of microseconds and is considered in-depth in Appendix~\ref{app:muon_dist}. On the submillisecond timescale, the magnetic field can be considered constant (see Sec.~\ref{sec:transients} for small corrections to this assumption). On the timescale of tens of seconds, the magnetic field drifts, but the muons' spatial distribution remains constant, except for fluctuations in the total number of muon decays detected. The calorimeter data acquisition produces data binned on this timescale, allowing us to track the number of muon decays detected. On the timescale of hours, the trackers sum the muons' spatial distribution information, generating distributions such as the one shown in Fig.~\ref{fig:sample_muon_dist}. On the timescale of days, driven by the time between trolley runs, we produce a value for \opprimetilde for each trolley pair, and then combine the results from multiple trolley pairs into four data subsets, \RunOneA--d. These timescales are summarized in Table~\ref{tab:timescales}

\begin{figure}
	\includegraphics[width=.96\columnwidth]{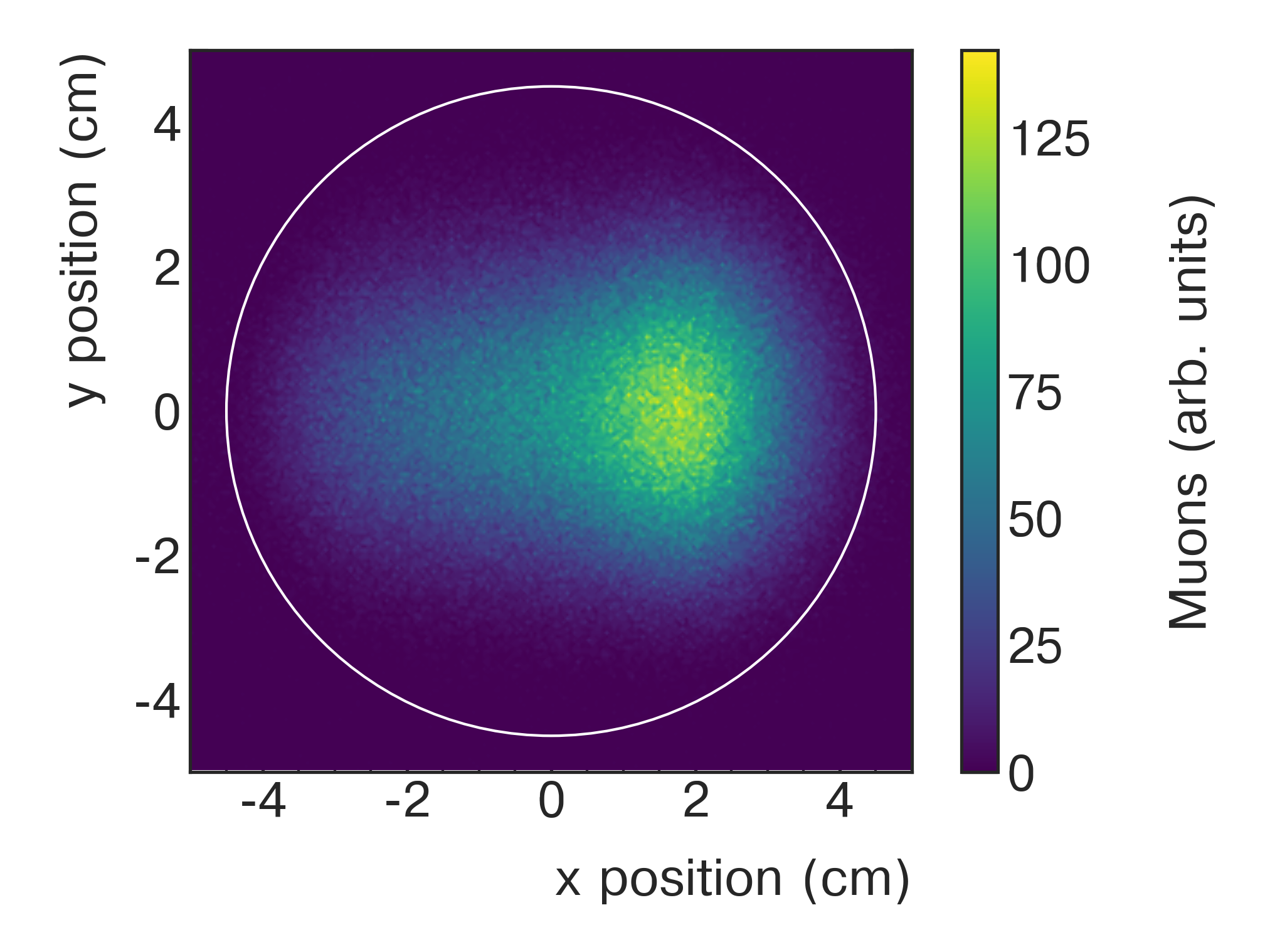}
	\caption{A typical example of the muon distribution measured by the trackers after integrating for several hours. This distribution is used to weight the field map.}
	\label{fig:sample_muon_dist}     
\end{figure}

\begin{table*}\centering
	\begin{tabular}{lll}
		\hline \hline
		Name & Duration & Usage \\
		\hline
		\rule{0pt}{1em}Intrafill & $\sim \SI{700}{\micro\second}$ & Storage time for each muon injection \\
		Magnetic-Field Measurements & $\sim \SI{1}{\second}$ & Measurements of magnetic field \\
		Calorimeter Bins & $\sim \SI{10}{\second}$ & Tracking field drift, muon distribution time-dependence \\
		Tracker Bins & $\sim \SI{3}{\hour}$ & Tracking drift of muon distribution\\
		Trolley Pairs & $\sim \SI{3}{\day}$ & Resynchronization of the fixed probes by the trolley\\
		Data Set & 1-5 trolley pairs & Combination of $\omega_a$ and $\tilde \omega_p'$ \\
		\hline \hline
	\end{tabular}
	\caption{The six relevant timescales used in the muon-weighted averaging and magnetic-field analysis. Each time scale is averaged over and then binned into the next highest scale. This procedure is repeated up to the data set level.}
	\label{tab:timescales}
\end{table*}

Each trolley run pair is broken down into the same time bins as the tracker data. These bins, indicated by index $q$, span the time intervals bounded by $s^{}_{q} \le t \le u^{}_{q}$, where $u^{}_{q} - s^{}_{q} \approx \SI{3}{\hour}$. Equation~(\ref{eq:op_tilde}) is evaluated assuming that the muons' spatial distribution is constant, but the overall number varies. 
Essentially, $\rho^{\mu}_{}$ is factored into a time-dependent and a time-independent part,
\begin{linenomath}\begin{equation}
		\rho^{\mu}(r,y,\phi,t) = N(t) \sigma^{\mu}(r,y,\phi).
		\label{eq:muon_dist_factor}
\end{equation}\end{linenomath}
The time-averaged field for each time bin $q$ is the average of the field weighted by the number of muon decays detected in that bin, 
\begin{linenomath}\begin{equation}
	\opprimeq(r,y,\phi) = \frac{\integral{\opprime(r,y,\phi, t)N(t)}{t}{s^{}_{q}}{u^{}_{q}}}{\integral{N(t)}{t}{s^{}_{q}}{u^{}_{q}}},
	\label{eq:muon_time_integral}
\end{equation}\end{linenomath}
where the subscript $q$ indicates the average of the quantity in bin $q$.

The decay positrons detected in the calorimeter are used as a proxy measurement for the number of muons in the storage region $N(t)$. These data are available in the intermediate time bins (approximately \SI{10}{\second}) and are integrated for each tracker bin $q$. Figure~\ref{fig:muon_avg_time} shows a typical detected muon decay time series and the dipole field over a 60-h time interval.

\begin{figure}
	\includegraphics[width=.96\columnwidth]{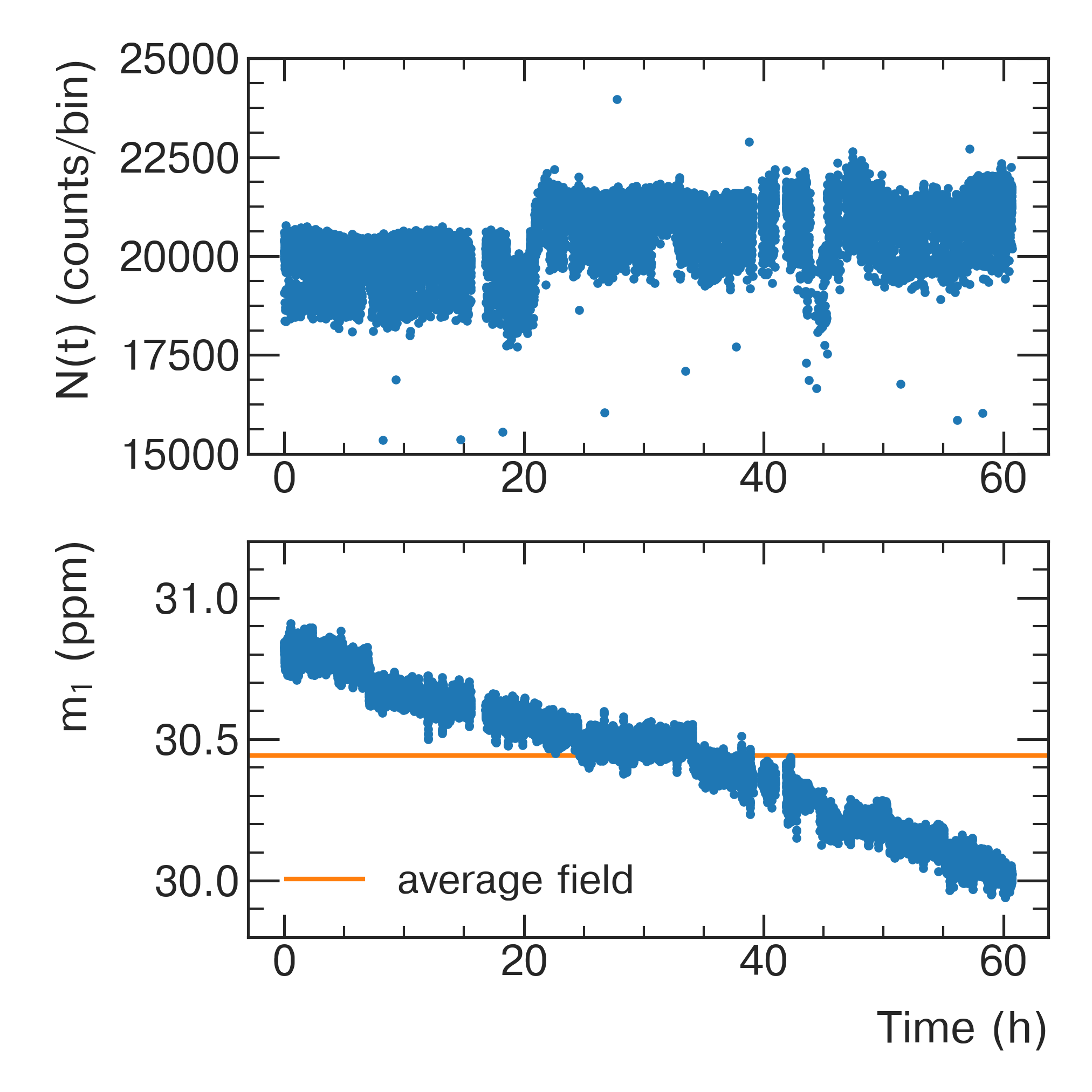}\caption{The number of muons integrated per bin over a typical trolley run pair and the azimuthally averaged dipole field over the same time. The number of integrated muons, represented by the number of observed decay positrons, is used to weight the field when evaluating the integral over time in Eq.~\eqref{eq:op_tilde}.}
	\label{fig:muon_avg_time}
\end{figure}

Using Eqs.~\eqref{eq:muon_dist_factor} and~\eqref{eq:muon_time_integral}, we can write Eq.~\eqref{eq:op_tilde} for an individual bin $q$:
\begin{linenomath}\begin{equation}
		\tilde \omega_{p,q}^{'} = \frac{\integral{}{\phi}{0}{2\pi} \integral{}{r}{r_1}{r_2} \integral{r \sigmamuq(r,y,\phi) \opprimeq(r,y,\phi)}{y}{-y_0}{y_0}}{\integral{}{\phi}{0}{2\pi} \integral{}{r}{r_1}{r_2} \integral{r \sigmamuq(r,y,\phi)}{y}{-y_0}{y_0}}.
		\label{eq:op_volume}
\end{equation}\end{linenomath}
Note that $\sigmamuq$ is the density determined by the trackers and beam dynamics, and the finest binning we have for the muon distribution. The muon distribution is reconstructed from tracker profiles and propagated to other azimuthal locations using beam dynamics simulations. After the three spatial integrals in Eq.~\eqref{eq:op_volume} are evaluated for each of the tracker bins $q$ (see next Sec.~\ref{subsec:spatial_avg}), the resulting set of $\opprimetildeq$ are averaged together, weighted by the total number of detected positrons in each bin, to determine $\opprimetilde$ over the trolley run pair interval $[0,T]$,
\begin{equation}
	\opprimetilde = \frac{\sum_q N_q \opprimetildeq}{\sum_q N_q},
	\label{eq:final_time_avg}
\end{equation}
with $N_q = \integral{N(t)}{t}{s^{}_{q}}{u^{}_{q}}$.

\subsection{Spatial Averaging}
\label{subsec:spatial_avg}

The spatial averaging procedure described here is performed for each time bin $q$ described above. The result of the spatial averaging is $\tilde \omega_{p,q}^{'}$ for each time bin, used as input for Eq.~\eqref{eq:final_time_avg}. The azimuthal part of the integral is broken down into azimuthal bins, indexed by $j$, set by the spacing between the fixed probe stations. Bin $j$ is defined by bounds $\eta^{}_{j} \le \phi \le \psi^{}_{j}$ with $\psi^{}_{j} - \eta^{}_{j} \approx \SI{2\pi/72}{\radian}$. We average the muon distribution within each azimuthal bin and use that average value $\sigmamuqj(r,y)$ for all positions in the bin, so the azimuthal portion of the spatial integral is
\begin{linenomath}\begin{equation}
	\frac{1}{\psi^{}_{j} - \eta^{}_{j}}\integral{\sigmamuq\opprimeq(r, y, \phi)}{\phi}{\eta^{}_{j}}{\psi^{}_{j}} = \sigmamuqj(r, y) \opprimeqj,
\end{equation}\end{linenomath}
with
\begin{linenomath}\begin{equation}
	\opprimeqj(r,y) = \frac{1}{\psi^{}_{j} - \eta^{}_{j}} \integral{\opprimeq(r, y, \phi)}{\phi}{\eta^{}_{j}}{\psi^{}_{j}}
\end{equation}\end{linenomath}
is the frequency map azimuthally averaged over a given fixed probe station. As before, the subscript $j$ indicates that the quantity has been averaged over azimuthal bin $j$ (and the subscript $q$ continues to mean the quantity is averaged over time bin $q$). The azimuthal average in these bins is the natural product of the field tracking described in Sec.~\ref{sec:Interpolation}. The full azimuthal integral is then just the sum over $j$ for all 72 stations, weighted by each station's azimuthal extent $\frac{\psi^{}_{j} - \eta^{}_{j}}{2\pi}$.

The two-dimensional integral of $r$ and $y$ is also performed per azimuthal bin. Assuming that the muon distribution is independent of azimuth within a bin $j$, the integral can be written
\begin{linenomath}\begin{equation}
		\opprimetildeqj = \frac{\integral{}{r}{r_1}{r_2}\integral{r \sigmamuqj(r,y) \opprimeqj(r,y)}{y}{-y_0}{y_0}}{\integral{}{r}{r_1}{r_2}\integral{r \sigmamuqj(r,y)}{y}{-y_0}{y_0}}.
		\label{eq:2d_int}
\end{equation}\end{linenomath}

The magnetic field is parametrized with moments, discussed in Sec.~\ref{subsubsec:Introduction_Trolley_Multipole_Expansion}, so the field $\opprimeqj(r,y)$ can be written
\begin{linenomath}\begin{equation}
		\opprimeqj(r,y) = \sum_i m^{}_{i,q,j} f^{}_{i}(r,y),
		\label{eq:moment_sum}
\end{equation}\end{linenomath}
where the functions $f^{}_{i}(r,y)$ encode the spatial dependence of the moments, $i$, in Eq.~\eqref{eq:mult_By} (shown explicitly in Table~\ref{tab:mults_in_Cart}) and the $m^{}_{i,q,j}$ are the moment strengths averaged in bins $q$ and $j$. The sum runs over all of the tracked moments. Combining Eqs.~\eqref{eq:2d_int} and Eq.~\eqref{eq:moment_sum} yields
\begin{linenomath}\begin{align}
    \opprimetildeqj &= \sum_i m^{}_{i,q,j} \frac{\integral{}{r}{r_1}{r_2} \integral{r \sigmamuqj(r,y) f^{}_i(r,y)}{y}{-y_0}{y_0}}{\integral{}{r}{r_1}{r_2}\integral{r \sigmamuqj(r,y)}{y}{-y_0}{y_0}} \nonumber\\
		&= \sum_i m^{}_{i,q,j} k^{}_{i,q,j},
		\label{eq:2d_int_param}
\end{align}\end{linenomath}
with
\begin{linenomath}\begin{equation}
    k^{}_{i,q,j} = \frac{\integral{}{r}{r_1}{r_2} \integral{r \sigmamuqj(r,y) f_i(r,y)}{y}{-y_0}{y_0}}{\integral{}{r}{r_1}{r_2}\integral{r \sigmamuqj(r,y)}{y}{-y_0}{y_0}}.
		\label{eq:beam_param}
\end{equation}\end{linenomath}
These $k$ parameters are calculated for each azimuthal bin $j$ in each time bin $q$. The values in the time and azimuthal bins are combined as described above, yielding the value of $\opprimetilde$ for the full trolley run pair. The average field experienced by the muons in a given dataset is
\begin{linenomath}\begin{equation}
		\opprimetilde = \avg{\sum_i k^{}_{i,q,j} m^{}_{i,q,j}}^{}_{q,j},
		\label{eq:sum_km}
\end{equation}\end{linenomath}
where the index $i$ is summed over all moments and the brackets indicate the quantity is averaged over the time bins $q$ and azimuthal bins $j$. The averaging is performed as prescribed in Eq.~\eqref{eq:final_time_avg}. Sample values of the $k^{}_i$ parameters (averaged over all bins) are shown in Fig.~\ref{fig:k_i_plot}. Note that $k^{}_1 = 1$ analytically.

\begin{figure}
	\includegraphics[width=.96\columnwidth]{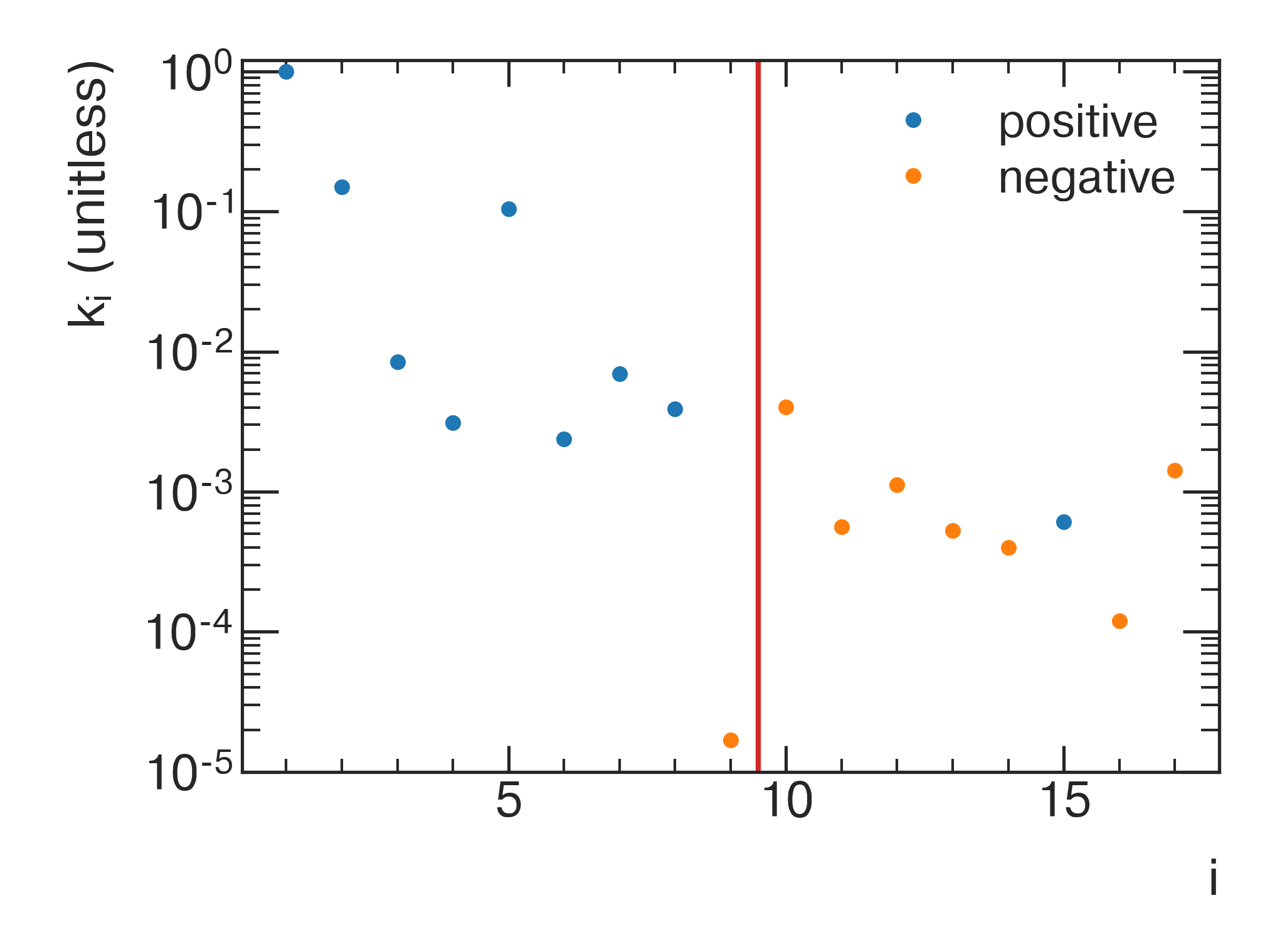}
	\caption{The amplitudes of the muon beam moments $k^{}_i$ decrease as $i$ increases. The moments with positive (negative) amplitudes are shown in dark (light). This decrease implies that the effect of higher-order moments on the average field is suppressed by the muon distribution. The muon distribution can be thought of as a low-pass filter on the moments of the field. The vertical line show the truncation order. All moments (and beam parameters) to the right of the vertical line are truncated.}
	\label{fig:k_i_plot}
\end{figure}
 
\subsection{Systematic Effects}\label{sec:weighting-systematics}

Table~\ref{tab:muon_weight_simp} shows the corrections and uncertainties related to the muon beam distribution. Because the field is highly uniform, \opprimetilde is dominated by the dipole field contribution. The corresponding uncertainties, $\delta \opprimetilde$ can be grouped into terms that include uncertainties in the field moment ($k_i \delta m_i$) and terms that include uncertainties in the muon beam moments ($m_i \delta k_i$). The beam and field moments are uncorrelated and thus no cross terms contribute to the overall uncertainty for \opprimetilde.

\begin{table}[ht]
	\centering
	\begin{tabular}{lcc}
		\hline \hline
		Contribution & Correction [ppb] & Uncertainty [ppb]\\
		\hline
		\multicolumn{3}{c}{Uncorrelated Uncertainties} \\
		$\delta^{\textrm{in-fill}}$ &-1 -- -4& 0 \\
		\hline
		\multicolumn{3}{c}{Correlated Uncertainties} \\  
		$\delta^{\textrm{tracker~x}}$ & 0 & 1 -- 9 \\
		$\delta^{\textrm{tracker~y}}$ & 0 & 7 -- 19 \\
		$\delta^{\textrm{tracker~accept}}$ & 0 & 1 -- 2 \\
		$\delta^{\textrm{y cod B-rad}}$ & 0 & 2 -- 3 \\
		$\delta^{\textrm{cod ESQ}}$ & 1 -- 2 & 4 -- 5 \\
		$\delta^{\textrm{calo accept}}$ & 0 & 0 -- 3 \\
		\hline
		\rule{0pt}{1em}Total & -3 -- 1 & 11 -- 20 \\
		\hline \hline
	\end{tabular}
	\caption{The contributions to the uncertainty on \opprimetilde from $m_i \delta k_i$ terms are shown. Ranges are specified when the corrections or uncertainties vary across the four \RunOne data subsets.}
	\label{tab:muon_weight_simp}
\end{table}

The systematic effects from the field moment uncertainties have been described in detail in the previous sections on the frequency extraction (Sec.~\ref{subsec:NMR_Frequency_Extraction}), probe calibration (Secs.~\ref{sec:shielded-proton-correction} and~\ref{sec:calib-analysis}), trolley measurements (Sec.~\ref{sec:Trolley_Data_Analysis}), and the magnetic field tracking (Sec.~\ref{sec:Interpolation}). When weighted by the muon distribution, terms correlated across measurements contribute 50-55~ppb per data subset. Additional terms are uncorrelated for each trolley pair and generate contributions between 22-43~ppb for the four \RunOne data subsets. These results are collected in the final uncertainty table in Sec.~\ref{sec:final_results}.

The remaining systematic effects from the muon distribution uncertainties $\delta k_i$ are due to the uncertainty in the muon decay position reconstruction from the trackers and the uncertainty from beam dynamics simulations used to propagate the tracker profiles to other azimuthal locations. The general process for estimating these systematic uncertainties is to introduce reasonable perturbations to the distributions before calculating the $k_i$. The resulting variation of the beam parameters is used to estimate the uncertainty on the muon weighting.

\subsubsection{Muon Tracker Systematics: \texorpdfstring{$\delta^\mathrm{tracker}$}{delta\^{}tracker}}
The trackers used to measure the muon distribution are affected by several sources of uncertainty estimated by simulation, including misalignment of the physical devices, their resolution, and their spatial acceptance based on the decay position of the parent muon. For each variation, the measured muon distribution is modified based on the uncertainty of the parameter being studied. New $k_i$ parameters are determined for the resulting muon distribution.
The resulting variation of the beam parameters is used to estimate the uncertainty on the muon weighting.

Uncertainty in the tracker alignment leads to a $\pm\SI{0.6}{\milli\meter}$ horizontal and vertical position uncertainty in the measured muon distribution. The vertical position uncertainty couples to the skew quadrupole resulting in an uncertainty $\delta^\mathrm{tracker~y}$. For the different conditions of the four data subsets, this uncertainty was typically 7 to \SI{19}{ppb}. A similar procedure was followed to estimate the uncertainty from the trackers' horizontal alignment, resulting in $\delta^\mathrm{tracker~x} =$1 to \SI{9}{ppb}. The trackers' spatial acceptance uncertainty results in $\delta^\mathrm{tracker~accept} =$ 1 to \SI{2}{ppb}.

\subsubsection{Closed Orbit Distortion: \texorpdfstring{$\delta^\mathrm{cod}$}{delta\^{}cod}}
Several effects can distort the muon beam's closed orbit away from its ideal orbit, leading to an azimuth-dependent mean position of the muon distribution. This azimuthal dependence on the beam can couple to azimuthally dependent variations in the field gradients, leading to a shift in \opprimetilde. The dominant \ac{COD} contribution is the lowest-order Fourier component of the dipole moment vs azimuth, and is included in the standard muon distributions used for muon weighting. Additional distortions lead to corrections and uncertainties. The presence of a radial mean field with azimuthal variation would cause a vertical \ac{COD}; a misalignment of the \ac{ESQ} plates causes both a radial and vertical \ac{COD} by steering the beam. The discrete structure of the \acp{ESQ}, as well as higher-order Fourier terms, also cause small distortions. Corrections and uncertainties due to \ac{COD} effects are evaluated by generating a distribution of possible \acp{COD} based on each error source, shifting the muon distribution in each azimuthal bin, and calculating the resulting distribution of $k_i$ parameters which is used to propagate the uncertainty. The radial and vertical \acp{COD} due to \ac{ESQ} plate misalignment contribute corrections of $\simeq \SI{2}{ppb}$ with uncertainties of $\delta_{i,\textrm{cod ESQ}}=$ 2 to \SI{4}{ppb}. An additional uncertainty is attributed to the beam distortions generated by the radial component of the magnetic field and its uncertainties, $\delta_{\textrm{y cod B-rad}}=$ 2 to \SI{3}{ppb}. 

\subsubsection{Calorimeter Acceptance: \texorpdfstring{$\delta^\mathrm{calo~accept}$}{delta\^{}calo accept}}
The muon distributions used for muon weighting represent the true muon distribution in the ring. A subset of these muons enter the $\omega_{a}$ analysis according to the spatially varying calorimeter acceptance. Each muon in this subset has a different probability for its decay positron to be detected by a calorimeter and also experiences a different magnetic field along its trajectory. A set of muon distributions representing this subset is generated using spatial weighting based on calorimeter acceptance as a function of muon beam trajectories. The $k_i$ parameters are calculated for this set of muon distributions and used to evaluate the resulting uncertainty. A maximum uncertainty $\delta^{\textrm{calo~accept}}=\SI{3}{ppb}$ is identified.

\subsubsection{In-fill Time Dependence: \texorpdfstring{$\delta^\mathrm{in-fill}$}{delta\^{}cod}}
The spatial muon distribution is approximated as constant over time in the fill. However, during \RunOne, it was changing during the fill due to instabilities in the \ac{ESQ} system \cite{E989SRBDpaper}; this problem was fixed before \RunTwo. This leads to a time dependence of the muon-weighted field over each muon beam pulse.

Time-binned azimuthally averaged muon distributions are used to calculate the $k_i$ and the corresponding muon-weighted field as a function of time in the fill. The resulting time dependence, approximated by a linear fit, leads to a correction to the muon-weighted field $\delta^{\textrm{in-fill}}<\SI{4}{ppb}$.

\subsection{Results}\label{sec:weighting-results}

The muon distribution is highly symmetric but slightly outside the magic radius around the storage ring, leading to low values for the $k^{}_i$ parameters for $i>1$. Because the beam is not centered, the leading order, nondipole terms couple to the normal quadrupole and normal sextupole moments of the field, and are $k_2 \sim 0.15$ and $k_5 \sim 0.09$. All of the other parameters are at least a factor of 10 lower (see Fig.~\ref{fig:k_i_plot}). The low-$k$ values combined with the low values of the higher-order field moments mean that the effect on the average field experienced by the muons from their distribution over the nonuniform part of the field is small. The largest effect comes from the normal sextupole ($\sim \SI{8}{\Hz},~\sim \SI{128}{\ppb}$), which is larger than the effect of the normal quadrupole due to dedicated shimming efforts to reduce the normal quadrupole around the ring. The net difference between the average field and the dipole field is of the same order.

    \graphicspath{{./transients/figures/}}

\section{Fast Transient Fields}\label{sec:transients}

Two time-dependent, \si{\micro\second}-timescale magnetic fields are induced by the pulsed magnetic and electric fields from the kicker and \acp{ESQ} that are synchronized with each muon fill. These transient magnetic fields are not present during the trolley runs and must be included as corrections to \opprimetilde. The fixed probe system measures the field at intervals of 1.2 to \SI{1.4}{\second}, typically asynchronously with respect to muon injection. The fast transient fields change on much shorter timescales. Additionally, the skin depth effect in the aluminum vacuum chamber walls shields the fixed probes from both of these transients, which originate in the muon storage region. For these reasons, both transients required unique measurement solutions.

The kicker transient was studied with two dedicated fast magnetometers for the current experiment. The transient associated with the \acp{ESQ} was discovered in studies of correlations of the fixed probe measurements with the muon injection. A set of NMR probes was developed to measure the \ac{ESQ} transient.

An additional systematic uncertainty is assigned to transient fields associated with the booster ring near the muon campus at \ac{FNAL}. By synchronizing the field measurement systems to the injection cycles with the pulsed systems turned off, we were able to apply an upper limit of \SI{7}{ppb} to any stray transient fields from the booster.

\subsection{Kicker Transient Fields}
\label{subsec:kickertransient}

A set of three kicker magnets reside in the storage ring vacuum chambers \cite{E989SRBDpaper}. The kickers reduce the \SI{1.45}{\tesla} field locally by roughly \SI{22}{\milli\tesla} for \SI{150}{\nano\second} to deflect the injected muons onto the stored orbit. This kick consists of a current pulse through three pairs of thin curved aluminum plates, each \SI{1.27}{\meter} long, that subtend an angle of \SI{62.5}{\degree} at a radius of \SI{4.5}{\centi\meter} in the $xy$ plane. The pulsed field induces eddy currents in the surrounding metal, leading to field perturbations in the storage volume during the times muons are stored. The fixed \ac{NMR} probes are shielded from this rapid transient field by the skin depth effect of the aluminum vacuum chambers and do not have the required measurement bandwidth.

\subsubsection{Measurement}

We built two Faraday magnetometers to measure this transient, one similar to the one used in E821 \cite{Efstathiadis:2003fs} and the other substantially improved against vibrations caused by the pulsing systems, which we are going to describe next. Such magnetometers exploit the rotation of the polarization angle $\theta$ of linear polarized light that occurs in almost any isotropic dielectric in a magnetic field $\BB$ parallel to the light propagation direction $\Delta\theta(t)=V B(t) L$. Here $L$ is the length, and $V$ is the Verdet constant of the dielectric.

The magnetometer (see Fig.~\ref{fig:kicker_transient_magnetometer}) fits between the kicker plates and was made without any metal. Light from a \SI{405}{\nano\meter} diode laser passed through a Faraday isolator into a multi-mode fiber. The fiber went through a vacuum flange to the magnetometer. The unpolarized light was collimated, polarized by a \ac{PBSC}, and then its plane of polarization was rotated by a half-wave plate. The light reflected off a \SI{45}{\degree} mirror, and then passed through two \ac{TGG} crystals, each \SI{5}{\milli\meter} in diameter with their \SI{14.5}{\milli\meter} long axis parallel to $B^{}_y$. The Verdet constant was measured to be $V(\SI{405}{\nano\meter})\approx \SI{450}{\radian/\tesla\cdot\meter}$. The beam was reflected and passed through another \ac{PBSC} which directed $s$- and $p$-polarized light to different return fibers. Time dependence in the magnetic field $\BB(t)$ changes the plane of polarization and the fraction of light entering each of the two return fibers. Typically, $\approx \SI{1}{\milli\watt}$ was detected in photodiodes attached to each fiber. The photocurrents were subtracted, amplified, and digitized, yielding a voltage signal of the form $V(B)=V^{}_{0}\cos\left(2VBL+\phi\right)$ where $\phi$ depends on the waveplate angle.

The magnetometer was calibrated in two steps. The magnet was ramped from full field, \SI{1.45}{\tesla} at \SI{5173}{\ampere}, to \SI{0}{\ampere} at a rate of \SI{-0.5}{\ampere/\second} while the magnetometer voltage was recorded as a function of magnet current $V(I)$. A Hall sensor was inserted near the magnet gap and recorded $B(I)$. From the two measurements, the sensitivity $\dd V/\dd B=(\dd V/\dd I)\times (\dd I/\dd B)$ was determined. Prior to calibration, the magnetometer was inserted between the kicker plates in the storage volume and the $\lambda$/2 waveplate adjusted to maximize the sensitivity $\dd V/\dd B$ at the full field. Constraints on the design made precise waveplate adjustment difficult and the actual maximum $\dd V/\dd B$ occurred at \SI{5124}{\ampere}, which was used for subsequent measurements. The calibration value extracted at this current was 
\begin{align}
	\diff{V}{B} &=(\SI{12.5}{\milli\volt/\ampere}) \times (\SI{1}{\ampere}/\SI{183}{\micro\tesla}) \nonumber \\
	&= \SI{68.3(7)}{\milli\volt/\milli\tesla}. \nonumber
\end{align}
Because the magnetometer baseline voltage depends on the laser current and the coupling efficiency into the incident fiber, the data were scaled to the voltage observed during calibration. This correction was less than \SI{7}{\percent}.

In addition to showing the expected kicker pulse, the signals showed a repeatable pattern of oscillations in the few \si{\kilo\hertz} range that grew after each kicker pulse and spanned $\pm$\SI{1}{\micro\tesla}. This pattern is thought to be due to vibrations in the cages holding the kicker plates that jostled the magnetometer, causing variations in the detected light. The system was run at magnet currents of 4841 and \SI{4326}{\ampere} to reduce this background. Here, $\dd V/\dd B\approx 0$, corresponding to all of the light going into the lower and middle return fibers, respectively. At these settings, fluctuations in detected light are ascribed to vibration, with sensitivity to real magnetic fields reduced by at least a factor of 20. The final result was assembled by recording data at minimum sensitivity $\dd V/\dd B\approx 0$ at 4841 and \SI{4326}{\ampere}, weighting it by 0.5 and subtracting it from the maximum sensitivity $\dd V/\dd B\approx\SI{68}{\milli\volt/\tesla}$ data acquired at \SI{5124}{\ampere}. The result is shown in Fig.~\ref{fig:kicker_transient_plot}.

\begin{figure}[!hbt]
	\begin{subfigure}{0.48\textwidth}
		\centering
		\includegraphics[width=3.375in]{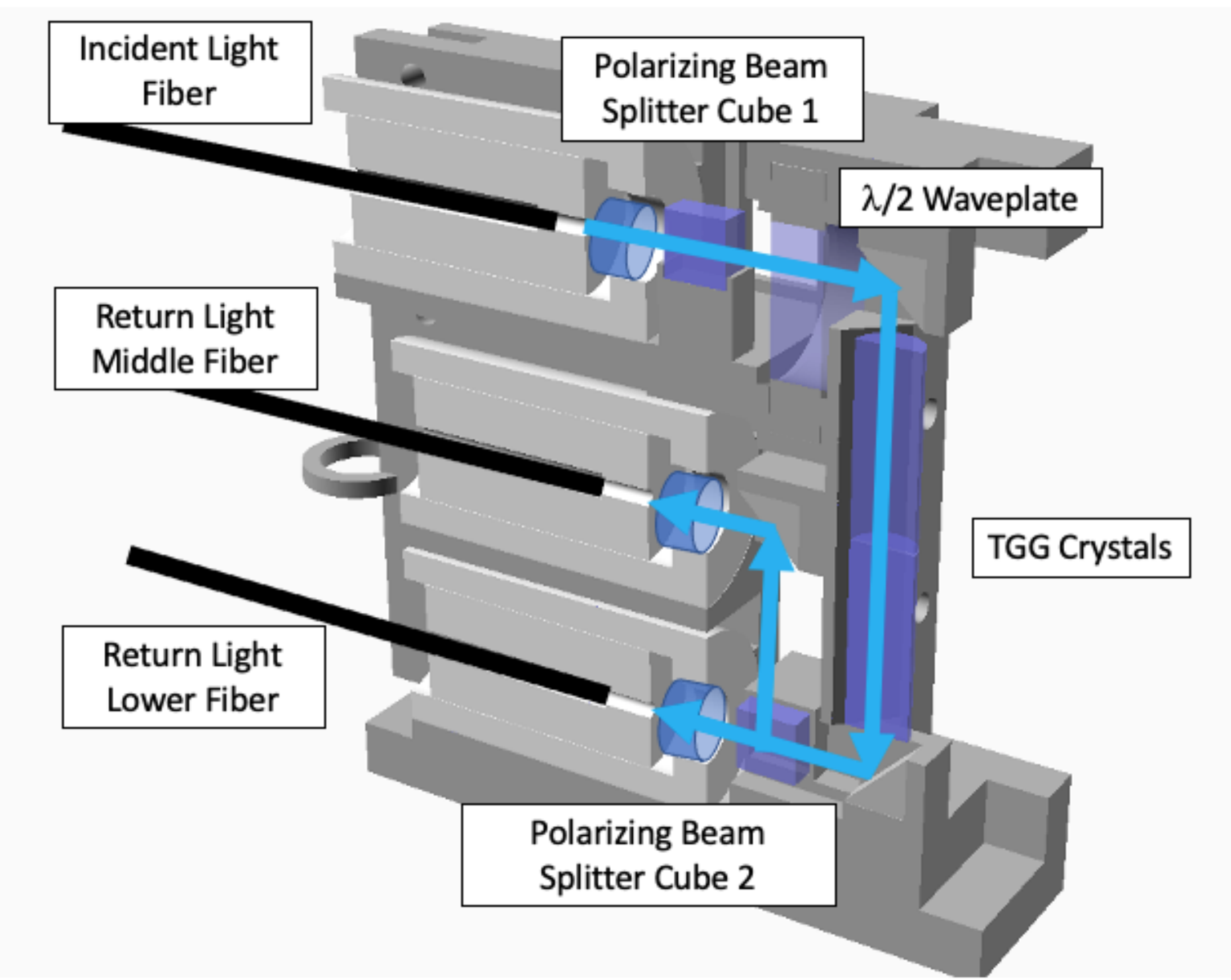}
		\caption{\label{fig:kicker_transient_magnetometer}}
	\end{subfigure}
	\begin{subfigure}{0.48\textwidth}
		\centering
		\includegraphics[width=3.375in]{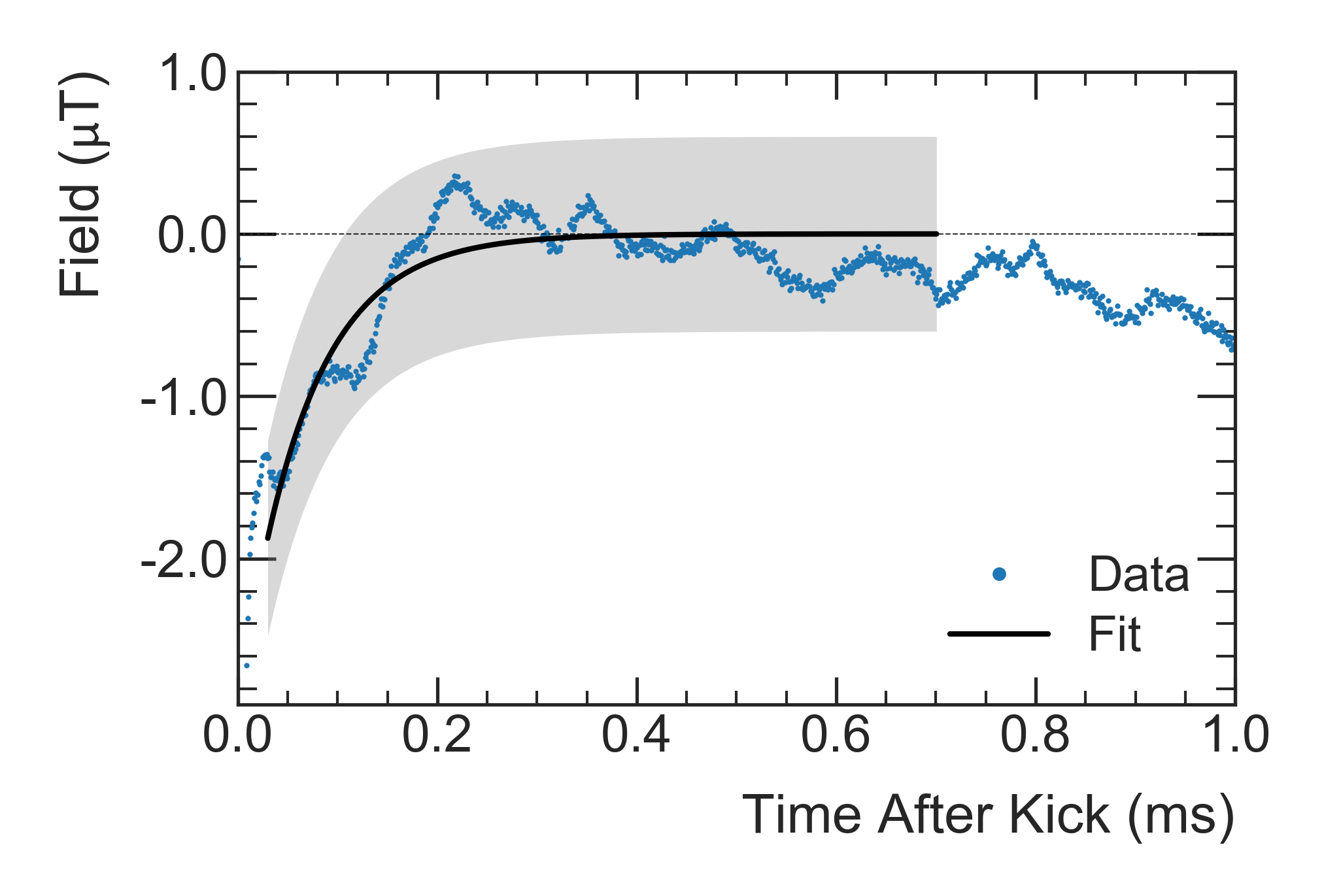}
		\caption{\label{fig:kicker_transient_plot}}
	\end{subfigure}
    \caption{(a) Schematic of the fiber magnetometer. The device is about \SI{6}{\centi\meter} tall. (b) The signal measured by the fiber magnetometer after subtracting the vibration background. The measurements and a fit to the transient are shown. The gray shaded band represents the associated uncertainty of $\pm$\SI{0.6}{\micro\tesla}. Muon data are fit from \SIrange{30}{700}{\micro\second} after the kick. }
	\label{fig:kicker_transient}
\end{figure}

\subsubsection{Analysis}

The data are fit to a decaying exponential from \SIrange{30}{700}{\micro\second} after the kick ($t=\SI{0}{\micro\second}$), corresponding to the nominal fit range of the \oa analysis. The total uncertainty includes those from calibration, fit uncertainty, and background subtraction. We estimate an uncertainty on the background subtraction of \SI{0.6}{\micro\tesla}.

The kickers subtend about \SI{8.5}{\percent} of the storage ring azimuth, so the results were scaled by $0.085$ to get the average kicker transient seen by the muons, assuming that the transients do not extend beyond the kicker plates. The magnetometer measured the field in the center of the storage volume. Simple models of the transient's spatial dependence suggest it drops off in the horizontal direction from the center but increases in the vertical when closer to the fields' sources. A weighted average of the muon distribution with this transient spatial dependence suggests the average muon sees a slightly smaller transient, reduced by a factor of 0.94.

For a field perturbation of the form $\Delta B(t)=\Delta B(t^{}_0)\exp(-(t-t^{}_0)/\tau_{k})$, the fractional effect on the muon anomalous precession frequency for a fit starting at $t=t^{}_{0}=\SI{30}{\micro\second}$ and ending at measurement time $t\gg(\gamma\tau_\mu)$ is
\begin{align}
	\frac{\Delta \omega_{a}}{\omega_{a}}&\approx\frac{\Delta B(t^{}_0)}{B(t^{}_0)}\left(\frac{\tau_{k}}{\tau_{k}+\gamma\tau_{\mu}}\right)^{2} \nonumber \\
	&\approx\frac{\SI{-1.87}{\micro\tesla}}{\SI{1.45}{\tesla}} \times \SI{8.5}{\percent} \times 0.94 \times \left(\frac{\SI{68}{\micro\second}}{\SI{68}{\micro\second} + \SI{64}{\micro\second}}\right)^{2} \nonumber \\
	&\approx \SI{-27(37)}{ppb}
\end{align}
The uncertainty on the correction is estimated from \SI{15}{\percent} on amplitude, 25\% on $\tau_{k}$ (\SI{17}{\micro\second}), \SI{25}{\percent} on azimuthal weighting factor, \SI{25}{\percent} on the transverse weighting factor, and $\pm$ \SI{0.6}{\micro\tesla} due to uncertainties on the vibrating background subtraction.

A second Faraday magnetometer gave consistent results. This magnetometer directed the light through open-air paths rather than optical fibers. It used \ac{TGG} crystals and a free-space laser propagation directed by mirrors inside the storage volume, while all other optical elements were on a breadboard outside the storage volume, allowing excellent control of systematic effects except for a weak sensitivity to vibration.

 \graphicspath{{./transients/figures/}}

\subsection{Electrostatic Quadrupole Transient Fields}\label{subsec:quadtransient}

During studies of the correlation between the fixed probe measurements and the muon injection time, a time-dependent, \SI{}{\micro\second}-scale transient magnetic field was discovered. Further studies revealed that the transient field is caused by mechanical vibrations of the charged plates induced by pulsing the \acp{ESQ}. The perturbation caused by this transient field is large enough to require precise measurements; however, the fixed probe system cannot directly measure the field to the required precision, primarily due to the skin depth effect of the aluminum vacuum chambers.

The \acp{ESQ} are arranged into four stations, each consisting of a short section, which subtends \SI{13}{\degree} in azimuth, and a long section approximately twice the length of a short section. The amplitude of the transient field generated by a short section is maximized near the section's azimuthal center. Observations showed that the long sections can be approximated as two short sections in series. In total, the \acp{ESQ} cover \ang{156} (\SI{43.3}{\percent} of the ring). Averaging the perturbation to the magnetic field over the whole ring reduces the total effect accordingly.

The dedicated transient measurements were performed at a lower \ac{HV} (\SI{18.2}{\kilo\volt}) than production runs (\SI{18.3}{\kilo\volt} and \SI{20.4}{\kilo\volt}). From first principles, the amplitude of the magnetic-field transient scales quadratically with the \ac{ESQ} voltage, which was confirmed with \emph{in situ} measurements in a range from \SIrange{0}{18.2}{\kilo\volt}. Therefore, we can correct the measurements to the HV setting used during any given production period.

\subsubsection{Measurement}\label{subsubsec:ESQmeasurements}

The dedicated measurements were made by a set of trolley NMR probes sealed inside \ac{PEEK} plastic tubes for vacuum compatibility and read out through the fixed probe NMR system.
The NMR system is synchronized with the \ac{ESQ} pulsing system; the \ac{ESQ} trigger usually precedes muon injection by \SI{23}{\micro\second}. The \ac{ESQ}s remain powered for the duration of the muon precession fit range, which ends \SI{650}{\micro\second} after beam injection, corresponding to \SI{673}{\micro\second} after the trigger. The \acp{ESQ} discharge \SI{700}{\micro\second} after the trigger. The beam is delivered in a series of eight such pulses spaced by \SI{10}{\milli\second}. The second series of eight pulses occurs \SI{266.7}{\milli\second} after the first series. The entire structure of 16 beam pulses repeats every $\approx$\SI{1.4}{\second}. Reading \ac{NMR} measurements from every fixed probe in the ring takes \SI{1.2}{\second}. The frequencies of the \acp{FID} are extracted in 0.4-ms-long fit windows. No additional frequency structures with fixed relations to the \ac{ESQ} pulsing time are observed within these windows.
\begin{figure}
	\begin{subfigure}{0.48\textwidth}
		\centering
\includegraphics[width=3.375in]{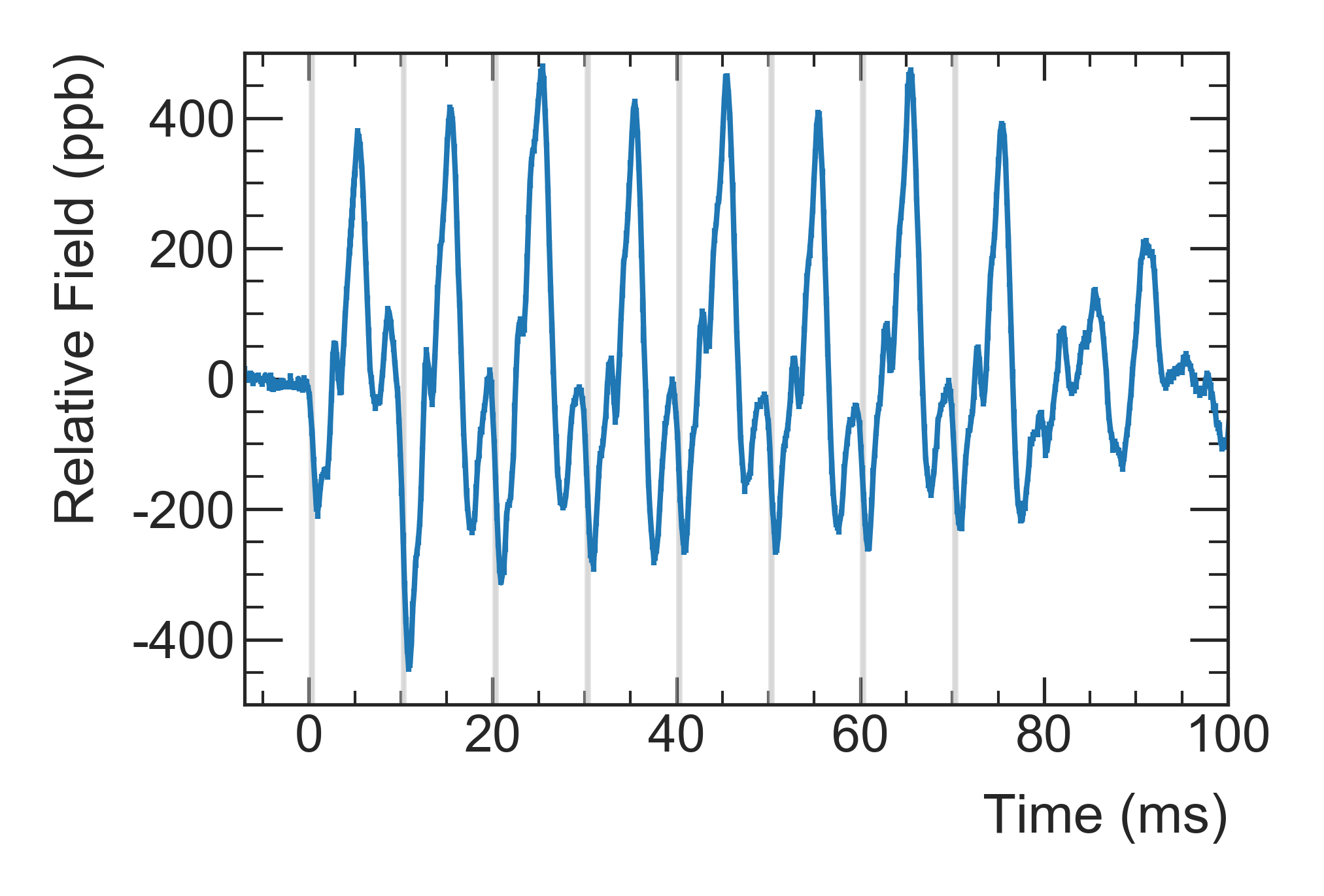}
		\caption{\label{fig:esq_time_structure_full}}
	\end{subfigure}
	\begin{subfigure}{0.48\textwidth}
		\centering
\includegraphics[width=3.375in]{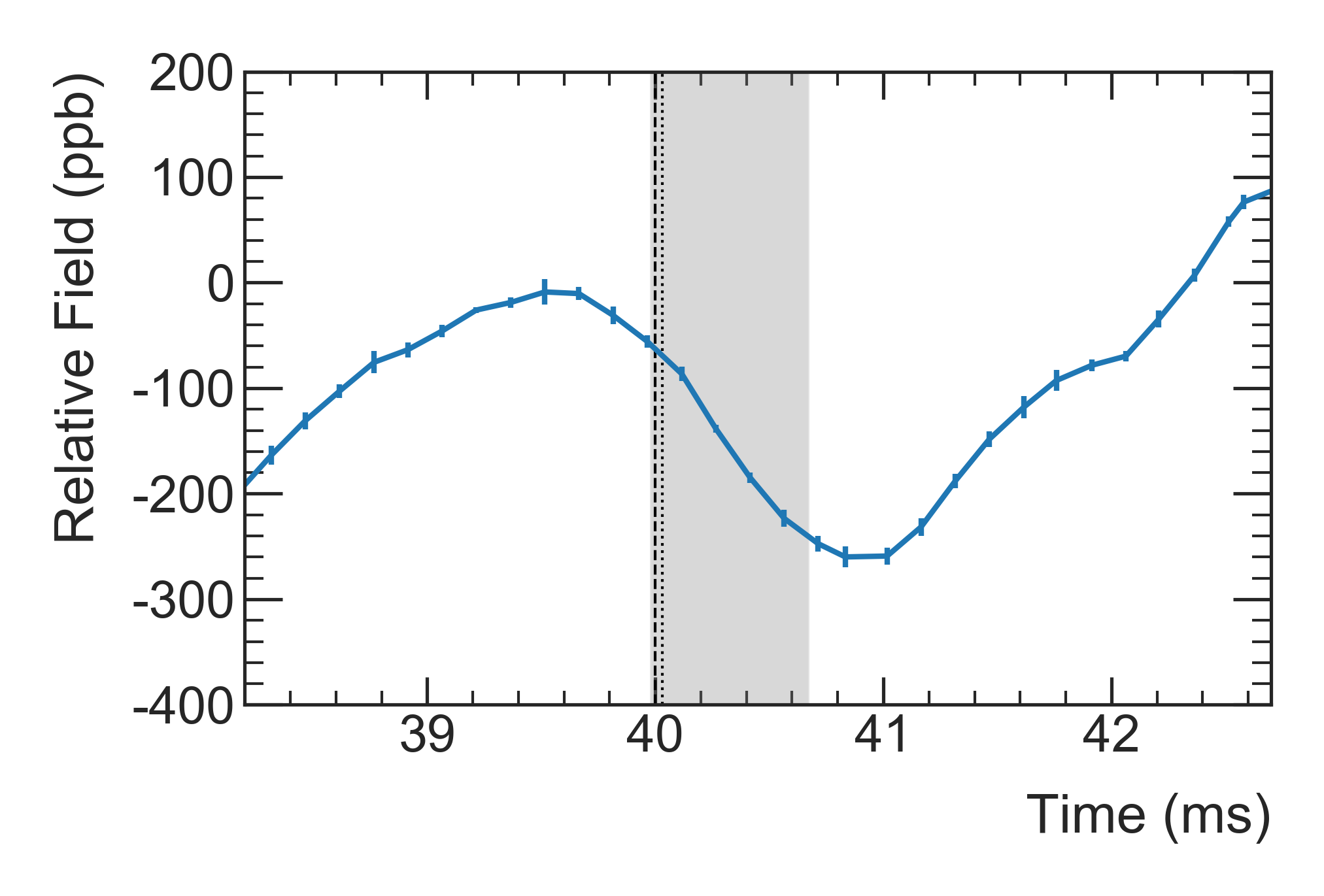}    
		\caption{\label{fig:esq_time_structure_zoom}}
	\end{subfigure}
    \caption{(a) The time structure of the \ac{ESQ} transient is determined by scanning the delay time between the pulse trigger and the \ac{NMR} measurement. The gray region corresponds to the time intervals in which the \ac{ESQ} are charged and muons can be used for the muon precession fits. (b) The same time structure zoomed in to a single beam pulse. The black dashed line indicates the time of the muon injection, the dotted line the earliest start of the precession fits.}
	\label{fig:esq_time_structure}
\end{figure}

The transient's time dependence was measured by varying the delay time between the \ac{ESQ} trigger and the \ac{NMR} measurement. Figure~\ref{fig:esq_time_structure_full} shows the time structure of the transient field, including a closeup in Fig.~\ref{fig:esq_time_structure_zoom} of the transient over a single pulse.

The measurements of the transient's dependence on the azimuthal position within an \ac{ESQ} were made in one half of a single long section, chosen for the accessibility in the vacuum chamber. The long sections are approximated by two identical short sections in series by pins \cite{SEMERTZIDIS2003458}. The transient was measured at seven positions along one \ac{ESQ} section as shown in Fig.~\ref{fig:esq_azi_structure}. Additional measurements were made with one probe in each of the eight sections.

\begin{figure}
	\centering
	\includegraphics[width=3.375in]{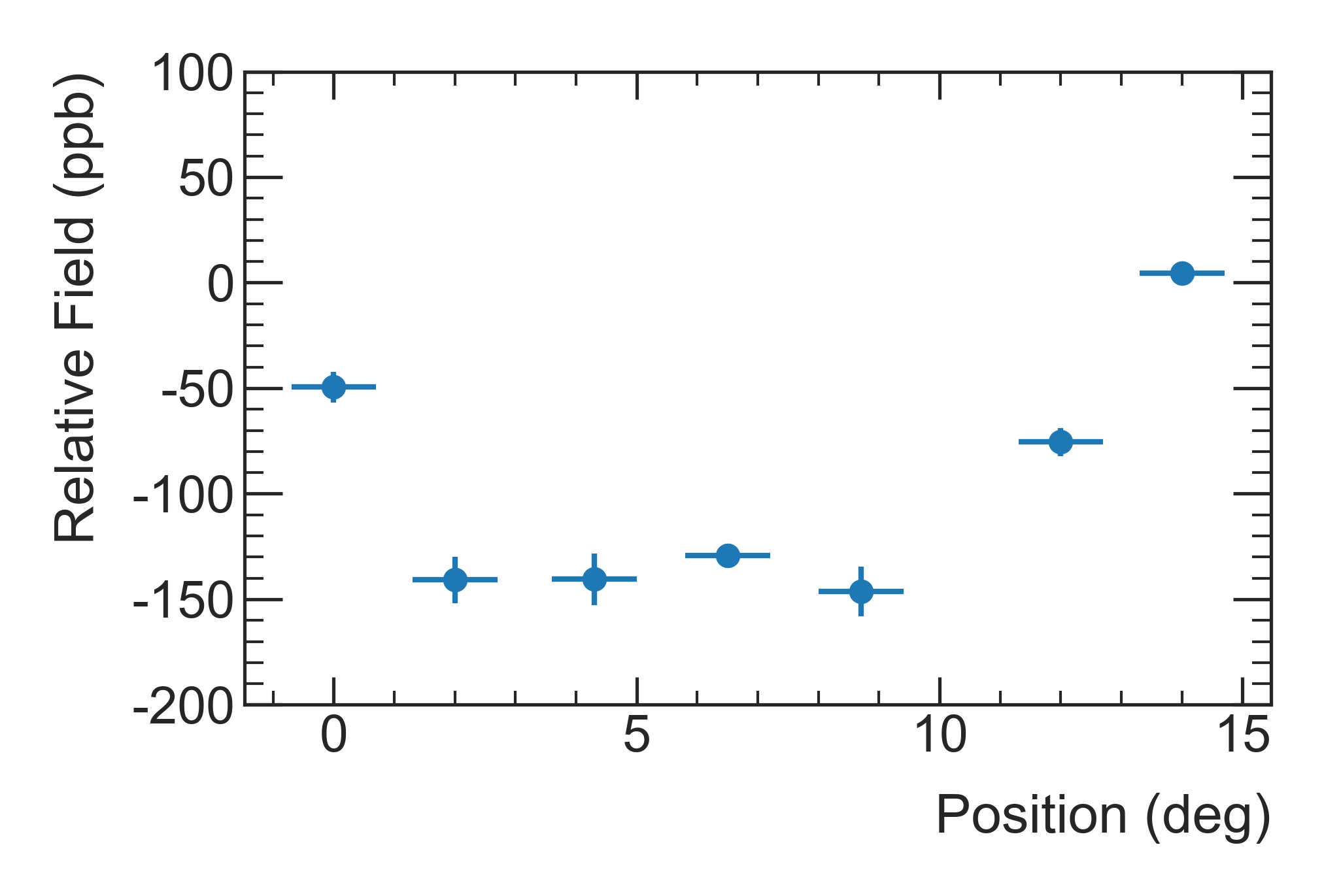}
    \caption{The relative field shift caused by the transient as a function of the azimuthal position with respect to the center ($\phi = 0\,$deg) of a long \ac{ESQ}. The transient is strongest in the center of the \ac{ESQ} section, falling off toward the edges.}
	\label{fig:esq_azi_structure}
\end{figure}

\subsubsection{Analysis}

Because of the skin depth effect at \SI{100}{\hertz} (the \acp{ESQ}' pulse rate), the fixed probes' sensitivities are reduced by \SI{70}{\percent}. Harmonics are attenuated even further, making the fixed probes mostly insensitive to the transient field's substructure. By comparison, the trolley probes used in the dedicated PEEK measurement system have a 0.5-mm-thick aluminum shell
and are attenuated by less than \SI{5}{\percent}. The fixed and PEEK probes also experience a phase delay due to the aluminum skin depth accounted for in the analysis.

Because each \ac{ESQ} pulse in the series causes a mechanical vibration in the plates, the transient's precise structure is affected by previous pulses. Figure~\ref{fig:esq_time_structure_zoom} shows the time structure for the fourth pulse in the series as an example. The transient field, which was extensively mapped in a single section, is then averaged over its azimuthal extent. The transient at the center of all sections was measured, but the azimuthal and transverse variations were only measured in half of one long \ac{ESQ}; the volume average of the effect over this section was scaled by the measurement at the center of each of other seven sections.

The transient is not constant over the time of a muon injection. Different weighting methods were developed to model how the muons sample the transient field over their lifetime. Each method is propagated through the analysis as a systematic check. The final determination produces an accurate correction to \opprimetilde and assigns a very conservative estimate of the uncertainty due to this effect.

\subsubsection{Systematic Effects}

\begin{table}
	\centering
	\begin{tabular}{lc}\hline\hline 
		\rule{0pt}{1em}Systematic Source & Uncertainty (\ac{ppb})\\
		\hline 
		\rule{0pt}{1em}Time and Azimuthal Structure & 77\\
		Second Pulse Train & 14\\
		Repeatability & 13\\
		Skin Depth & 13\\
		Field Drift & 10\\
		Frequency Extraction & 5\\
		Radial Dependence & 4\\
		Probe Positioning & 2\\
		\hline 
		\rule{0pt}{1em}Total ESQ-Transient Uncertainty & 82\\
		\hline
		\hline
	\end{tabular}
	\caption{The sources of uncertainty in the determination of the \ac{ESQ} transient measured at ESQ \ac{HV}$=$\SI{18.2}{\kilo\volt}. The total ESQ-Transient uncertainty is the dominant uncertainty in the determination of the \opprimetilde uncertainties for \RunOne. }
	\label{tab:esq_systematic}
\end{table}

\begin{figure}
	\centering
\includegraphics[width=3.375in]{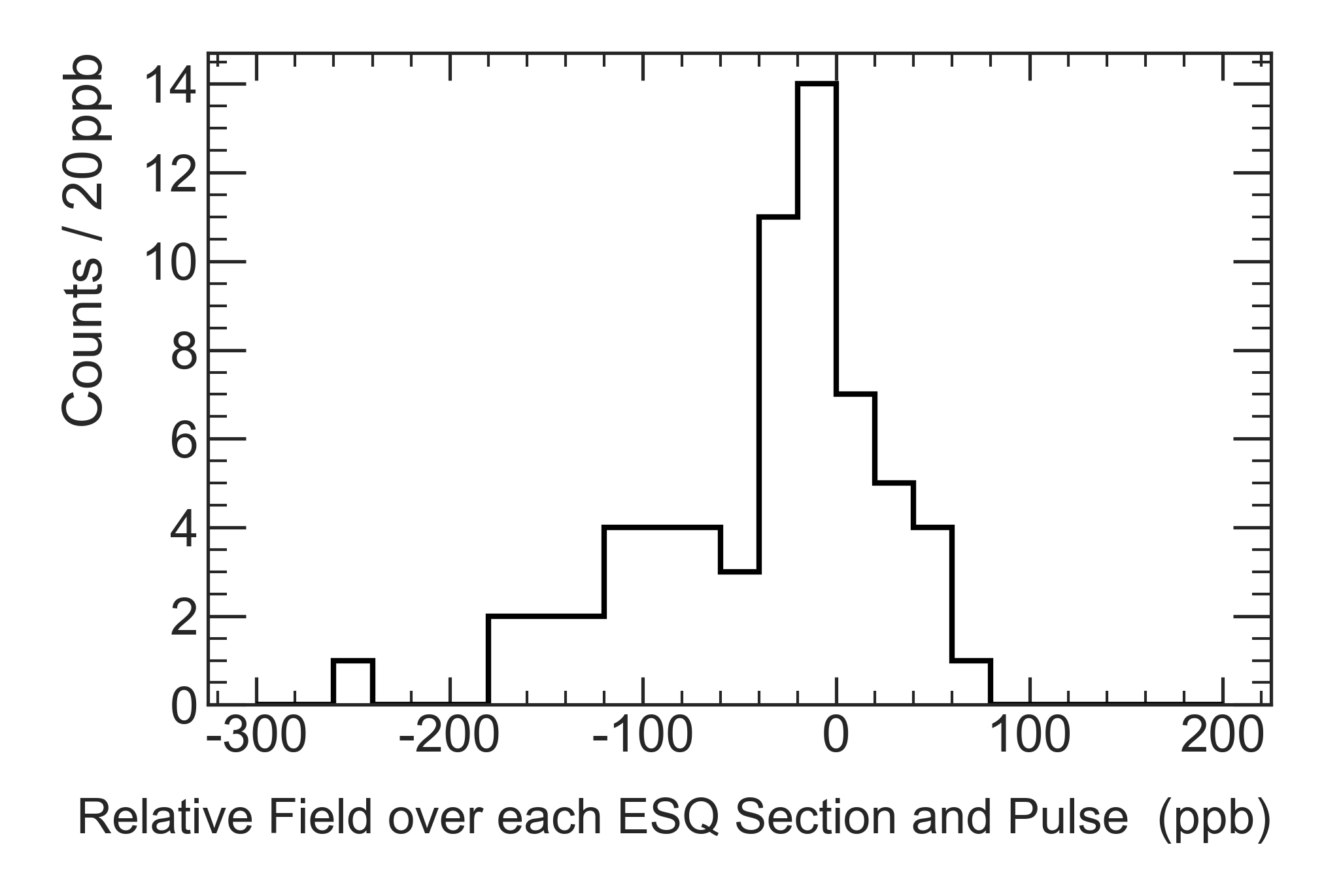}
	\caption{The distribution of the observed \ac{ESQ} transient effect over all stations and sections. The full width of the distribution is used as the uncertainty ($\pm$\SI{178}{ppb}) in the \ac{ESQ} region and scaled down by the geometrical coverage factor of the \ac{ESQ} in the storage ring (0.433).}
	\label{fig:esq_hist}
\end{figure}

The uncertainty sources for the \RunOneA data subset are summarized in Table~\ref{tab:esq_systematic}. The substructure in time and azimuth of the \ac{ESQ} transient is the dominant uncertainty. It arises because the azimuthal dependence of only one of the 12 \ac{ESQ} sections was measured, and the substructure of the \ac{ESQ} transient was not measured until Run-3. Because of the length of time between these measurements and \RunOne, we applied a very conservative estimate of the uncertainty to the \RunOne data. Figure~\ref{fig:esq_hist} shows the distribution of the observed ESQ transients for the four \ac{ESQ} stations.

The \ac{ESQ} transient studies were performed using the first train of eight beam pulses. It is expected that the second group of eight pulses behaves like the first group because the vibrations and the field transient completely die out before the next set of pulses begin. A simple study was conducted that confirmed this expectation. The average transient from the first train and a second train agreed to within 14 ppb, which is used to estimate the uncertainty.

Other systematic checks include measuring the transient beyond the azimuthal extent of the \ac{ESQ} sections and the radial dependence of the transient.
Both of these observed variations are added as uncertainties.
The measurements were checked for repeatability, which was found to be at the 13\,\ac{ppb} level. This number was conservatively assigned as an uncertainty. Linear drift in the dipole field during the measurement is removed by tracking the local fixed probe drift. Higher-order drift is small on the time scales of these measurements (3 hours). Estimates of this drift are made using \ac{PEEK} probe measurements outside of the \ac{ESQ} stations.

\begin{table}
	\centering
	\begin{tabular}{ccc}\hline\hline
		\rule{0pt}{1em}Data subset {\hspace{1pc}} & Correction (\ac{ppb}) {\hspace{1pc}} & Uncertainty (\ac{ppb})\\
		\hline
		\rule{0pt}{1em}\RunOneA & -15 & 83\\
		\RunOneB & -19 & 103\\
		\RunOneC & -19 & 103\\
		\RunOneD & -15 & 83\\
		\hline
		\hline
	\end{tabular}
	\caption{The total correction and uncertainty on the determination of \opprimetilde from the \ac{ESQ} transient. The data subsets have different values because they had different \ac{ESQ} HV values. The dedicated measurements shown in Table~\ref{tab:esq_systematic} are scaled for each data subset using the known quadratic relation between transient amplitude and HV setting.}
	\label{tab:esq_results}
\end{table}

The final correction to each data subset and the uncertainty due to the \ac{ESQ} transient are shown in Table~\ref{tab:esq_results}. Note that the uncertainty values differ from Table~\ref{tab:esq_systematic} because they are scaled to the HV settings from each data subset individually.

     \section{Final Results}
\label{sec:final_results}

This paper has covered the full analysis chain for the determination of the magnetic field, \opprimetildeatTexp, for the Muon \gm Experiment. \RunOne was broken down into four data subsets, \RunOneA through \RunOneD, defined by the settings for the kicker and \acp{ESQ}. The values of \opprimetildeatTexp are combined with the corresponding values of \oa into the ratio $\frac{\oa}{\opprimetildeatTexp}$ for each data subset. The ratios are then combined into the single \RunOne value that is input into Eq.~\eqref{eq:amu} to calculate \amu \cite{prl2021}.

The instrumentation and measurements in this paper represent a significant improvement over the \ac{BNL} experiment. They are part of a well-studied chain of calibrations and synchronizations where all of our measurements are referenced to the absolute calibration of the water calibration probe that was cross-checked with a novel \hethree~probe.

Several key field analyses (the trolley calibration, field tracking, and muon weighting) were performed by at least two mutually blinded independent teams that, in all cases, found agreement below our total uncertainty.

The \ac{ESQ} transient discovery and measurement represents a significant effort to characterize each system, as well as interactions between systems. A dedicated measurement campaign quickly quantified the systematic correction to the measured field and the corresponding uncertainties. Additional measurements taken after \RunOne will further constrain the systematic effect of the transient.

The final results of the field analysis are summarized in Table~\ref{tab:final_table}. Tables~\ref{tab:run_unc} and~\ref{tab:subrun_unc} summarize the systematic corrections and uncertainties covered in this paper. The uncertainty is dominated by the \ac{ESQ} transient. These uncertainties are not strictly independent, leading to correlations between the four data subsets. Most uncertainties are treated as fully correlated between the data subsets; only the tracking error discussed in Sec.~\ref{subsec:interpolation_systematics} is treated as uncorrelated.

The total systematic error on $\opprimetildeatTexp$ for \RunOne is \SI{114}{ppb}. The contributions from calibration, field tracking, and muon weighting total \SI{56}{ppb}. The contribution from the \ac{ESQ} and kicker transients are, respectively, 92 and \SI{37}{ppb}. Most of the uncertainties in Tables~\ref{tab:run_unc} and~\ref{tab:subrun_unc} already meet the design goals. Improvements to the determination of the ESQ transient are expected in future analyses and combined with the improved temperature stability of the magnet after \RunOne, we expect to reduce the total uncertainty below the \SI{70}{ppb} target for $\opprimetildeatTexp$ in the future.

\renewcommand{\arraystretch}{1.25}
\begin{table}
	\begin{tabular}{ccc}
		\hline\hline
		Dataset &	$\opprimetildeatTexp/2\pi$ (\si{\hertz}) &	Uncertainty (\si{ppb}) \\
		\hline
		\RunOneA &	$61,791,871.2$ &	$115$ \\
		\RunOneB &	$61,791,937.8$ &	$127$ \\
		\RunOneC &	$61,791,845.4$ &	$125$ \\
		\RunOneD &	$61,792,003.4$ &	$108$ \\
		\hline\hline
		\multicolumn{3}{c}{Average Over All Datasets}\\
		\hline
		\multicolumn{2}{c}{Field Measurements} & 56 \\
		\multicolumn{2}{c}{\ac{ESQ} Transient} & 92 \\
		\multicolumn{2}{c}{Kicker Transient} & 37 \\
		\hline
		\multicolumn{2}{c}{Total} &	$114$ \\
		\hline\hline
	\end{tabular}
	\caption{The final result for \opprimetildeatTexp for each of the four datasets in \RunOne. These numbers represent the Larmor precession frequency of protons in a spherical water sample in the same magnetic field experienced by the muons. The uncertainties are in \si{ppb} of the measured value of \opprime.}
	\label{tab:final_table}
\end{table}

\begin{table*} 
	\small\begin{tabular}{cccc}
		\hline\hline\rule{0pt}{1em}
		Systematic &	Correction (\si{ppb}) &	Uncertainty (\si{ppb}) &	Reference \\
		\hline\rule{0pt}{1em}
		Absolute Calibration &	0 &	15 &	Sec.~\ref{sec:pp_syst} \\
		Trolley Calibration &	0 &	28 &	Sec.~\ref{sec:calib-systematic} \\
		Configuration &	-1 &	23 &	Sec.~\ref{subsubsec:Trolley_Other_Systematic_Effects} \\
		Trolley Baseline $\mtr(0)$ &	-13 &	25 &	Sec.~\ref{subsec:Trolley_Systematic_Effects} \\
		Fixed Probe Baseline $\mfp(0)$ &	0 &	8 &	Sec.~\ref{subsec:interpolation_systematics} \\
		Fixed Probe Runs $\mfp(t)$ &	0 &	1 &	Sec.~\ref{subsec:interpolation_systematics} \\
		\hline\rule{0pt}{1em}
		Total &	-14 &	48 & \\
		\hline\hline
	\end{tabular}\normalsize
	\caption{The systematic corrections and uncertainties on \opprimetildeatTexp that do not vary by dataset.}
	\label{tab:run_unc}
\end{table*}

\begin{table*}
	\footnotesize\begin{tabular}{cccccccccc}
		\hline\hline\rule{0pt}{1em}
		&	\multicolumn{2}{c}{\RunOneA} &	\multicolumn{2}{c}{\RunOneB} &	\multicolumn{2}{c}{\RunOneC} &	\multicolumn{2}{c}{\RunOneD} & \\
		Systematic &	Corr. (\si{ppb}) &	Unc. (\si{ppb}) &	Corr. (\si{ppb}) &	Unc. (\si{ppb}) &	Corr. (\si{ppb}) &	Unc. (\si{ppb}) &	Corr. (\si{ppb}) &	Unc. (\si{ppb}) &	Reference \\
		\hline\rule{0pt}{1em}
		Trolley Temp &	0 &	28 &	0 &	25 &	0 &	21 &	0 &	15 &	Sec.~\ref{subsubsec:Trolley_Temperature_Correction} \\
		Tracking Error &	0 &	43 &	0 &	34 &	0 &	25 &	0 &	22 &	Sec.~\ref{subsec:interpolation_systematics} \\
		Muon Weighting &	0 &	11 &	-1 &	14 &	1 &	16 &	-3 &	20 &	Sec.~\ref{sec:weighting-systematics} \\
		Transients &	-43 &	91 &	-46 &	110 &	-46 &	110 &	-43 &	91 &	Sec.~\ref{sec:transients} \\
		\hline\rule{0pt}{1em}
		Total &	-43 &	105 &	-47 &	118 &	-46 &	116 &	-45 &	97 & \\
		\hline\hline
	\end{tabular}\normalsize
	\caption{The systematic corrections and uncertainties on \opprimetildeatTexp that vary by dataset.}
	\label{tab:subrun_unc}
\end{table*}

    \section{Acknowledgments} 
We thank the Fermilab management and staff for their strong support of this experiment, as well as the tremendous support from our university and national laboratory engineers, technicians, and workshops.

The Muon \gm Experiment was performed at the Fermi National
Accelerator Laboratory, a U.S. Department of Energy, Office of
Science, HEP User Facility. Fermilab is managed by Fermi Research
Alliance, LLC (FRA), acting under Contract No. DE-AC02-07CH11359.
Additional support for the experiment was provided by the Department
of Energy offices of HEP and NP (USA), the National Science Foundation
(USA), the Istituto Nazionale di Fisica Nucleare (Italy), the Science
and Technology Facilities Council (UK), the Royal Society (UK), the
European Union's Horizon 2020 research and innovation programme under
the Marie Sk\l{}odowska-Curie grant agreements No. 690835,
No. 734303, the National Natural Science Foundation of China
(Grant No. 11975153, 12075151), MSIP, NRF and IBS-R017-D1 (Republic of Korea),
the German Research Foundation (DFG) through the Cluster of
Excellence PRISMA+ (EXC 2118/1, Project ID 39083149).

    \newpage
\graphicspath{{./appendix/figures/}}
\appendix
\appendixpage
\addappheadtotoc
\section{Data Quality Control}
\subsection{Instrument Failures} \label{subsec:Data_Quality_Control_Instrument_Failures}
If an instrumentation failure occurs in a single measurement, the corresponding value is dropped.
In the fixed probe system, such failures are caused by the absence of a proper \ac{RF} $\pi/2$ pulse needed for the \ac{NMR} sequence to rotate the sample magnetization, by an out-of-time triggered pulse or an out-of-time waveform digitization.

The absence of the RF pulse leads to a noise-only waveform that is detected by the signal amplitude and power of the \ac{FID}.
The switches in the multiplexer that swap between RF pulse and signal path trigger on the amplitude of the RF pulse.
If the amplitude of the RF pulse falls below a threshold, it is not propagated through the system.
The electronics components used show a small temperature dependence that can cause slight variations in the $\pi/2$-pulse amplitudes.
If the RF-pulse amplitude is close to the threshold, this can lead to isolated measurements with missing  $\pi/2$ pulses.

\begin{figure}
       \centering
       \includegraphics[width=0.48\textwidth]{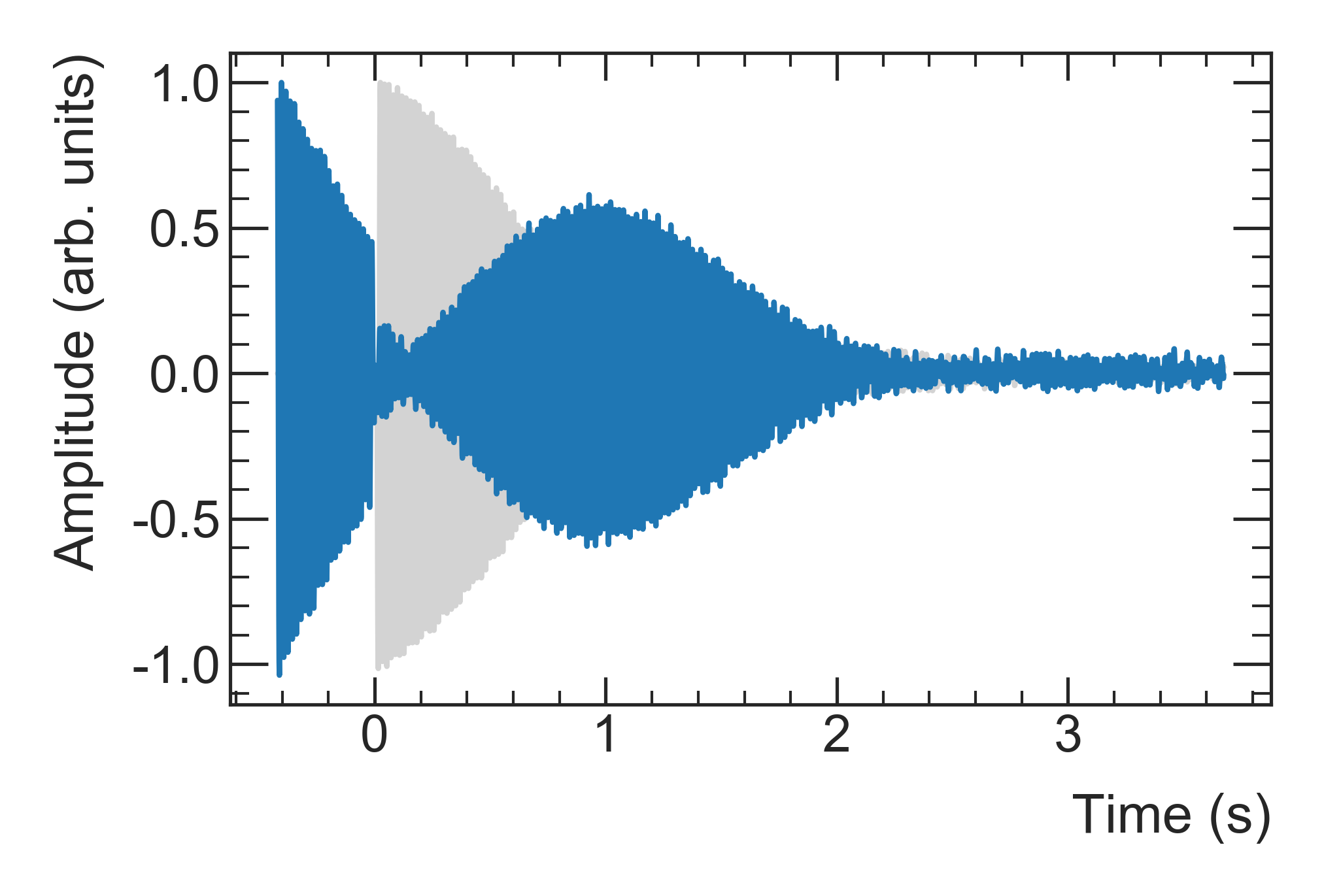}
       \caption{\ac{FID} with two RF pulses overlayed to a nominal FID (gray) with a single RF pulse for reference.}
       \label{fig:Data_Quality_Control_Instrument_Failures_Fid_Double}
    
   \label{fig:Data_Quality_Control_Instrument_Failures_Fid}

\end{figure}

A signal from the control board triggers the fixed probe RF pulse.
If interference from other pulsed systems in the experiment is picked up in these signal cables, an out-of-time $\pi/2$ pulse can be fired.
Depending on the relative timing, this can lead to the superposition of two RF pulses in the digitization window, or to reduced \ac{FID}-amplitude if the spins of the samples are not yet recovered fully.
Figure~\ref{fig:Data_Quality_Control_Instrument_Failures_Fid} shows a comparison between a waveform with a nominal \ac{FID} and a waveform with two $\pi/2$ pulses.
A second RF pulse, during the \ac{FID} of a previous pulse, has not yet decayed, and can lead to spin-echo-like behaviors of the system.
Such measurements are mainly detected by a spike in the power of the \ac{FID} of the corresponding waveform.
The power of a waveform is defined as the sum of the squared \ac{ADC}values. 
In Run-1, damaged resistors in the pulsed electrostatic quadrupole systems induced increased numbers of such false triggers.
The replacement of the resistors and improved shielding of the corresponding cables eliminated this issue.

Similar to false triggers of the $\pi/2$ pulses, the digitizer can also be affected by picked up interference signals.
This results in digitization outside of the time window of the \acs{FID} and with it in noise-only waveforms.

\subsection{Severe Field Instabilities}  \label{subsec:Data_Quality_Control_Severe_Field_Instabilities}

Periods around severe field instabilities are not used for the \amu determination.
Such instabilities are driven by magnetic field jumps, induced by the feedback systems, or failing hardware. 

In addition to the \ac{FID}-wise quality flags, sudden steps in the
magnetic fields are noted in the ``production'' indicating field
instabilities.  Such steps are identified by frequency changes larger than 7 times the resolution of a given probe,
over a time of up to \SI{8.5}{\second}, in at least \SI{40}{probes}.

\subsubsection{Field Steps}
Sudden field steps are either caused by external changes of the environment, for example, a magnetic connector moving into the proximity of the storage volume, or by internal changes of the magnet.
The latter are denoted field jumps.
It is believed that such jumps are caused by the physical motion of the magnet coils in the cryostat.
The coils are held in place by radial stops \cite{Danby:2001eh}. Field jumps are believed to be caused by the coils releasing tension by suddenly slipping radially.
Figure~\ref{fig:Data_Quality_Control_Severe_Field_Instabilities_Jumps} shows the effect of such field jumps as a function of azimuth. 
The positions of the jumps correlate with the radial stops. 
The azimuthal extent of the jumps is roughly \SI{130}{\degree}, and their integral over the whole ring typically cancels. 
\begin{figure}
    \centering
    \includegraphics[width=0.48\textwidth]{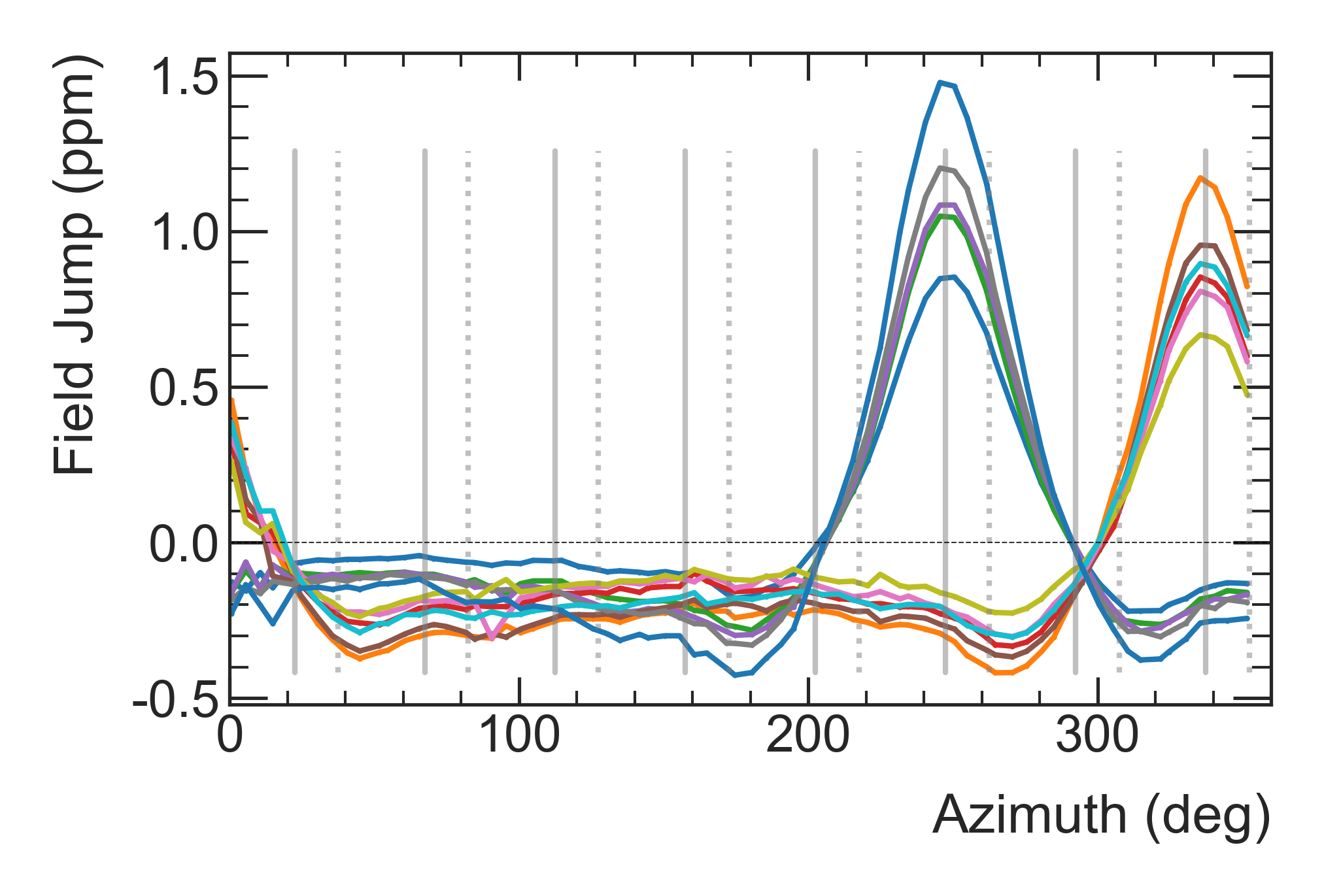}
    \caption{The field step size as a function of azimuth for all field jumps during the Run-1a dataset. 
    The vertical gray lines indicate the positions of the radial stops (solid: top, dashed: bottom). 
    }
    \label{fig:Data_Quality_Control_Severe_Field_Instabilities_Jumps}
\end{figure}

It has been shown that the fixed probes track the field equally well before and after a field jump. 
The period of \SI{120}{\second} before and after a jump is dropped from the \amu determination.

\subsubsection{Instabilities Caused by the Feedback System}
Instrumentation failures as described in Appendix~\ref{subsec:Data_Quality_Control_Instrument_Failures} can lead to non-physical frequency determinations in the online \ac{FID} analysis.
If this happens in a fixed probe that is part of the feedback system, the un-physical frequencies can impact the \ac{PID} loop. 
In such a case, the feedback reacts to the non-real change of the magnetic field, driving the mean magnetic field away from its set point.
The control loop takes some time to stabilize the field after such an excursion.

If the mean over 10 consecutive measurement cycles of the mean magnetic field as determined by the online \ac{FID}-analysis of selected probes is more than \SI{10}{\hertz} (\SI{162}{ppb}) away from the set point, the control loop switches to a more aggressive correction mode.
These time periods and \SI{24}{\second} before and \SI{240}{\second} afterward are dropped from the determination of \amu to guarantee stable conditions. 

In addition to instrumentation failures, field changes on a timescale faster than the reaction time of the feedback control loop can also cause the mean frequency to diverge from the set point.
During some periods, the magnetic field was affected by a roughly \SI{2}{\minute} oscillation of unknown origin.
In rare cases, the amplitude of these field changes crossed the above-mentioned threshold. 
The \acs{DQC} also vetoes these periods.
Adjustments to the time constants of the feedback loop mitigated these issues.

\section{The Jacobian Matrix}
\label{appendix:jacobian}

As covered in Sec.~\ref{subsubsec:Introduction_Trolley_Multipole_Expansion}, the trolley and fixed probe systems provide measurements of the $B_y$ field in different bases, respecting the different spatial symmetries of each set of probes. The two different sets of moments are equivalent if the moments can be calculated perfectly. However, because they are calculated as discrete approximations, the two sets are not identical; there is a change-of-basis matrix that takes moments from one basis to the other. This is the Jacobian matrix in Eq.~\eqref{eq:Taylor_form},
\begin{linenomath}\begin{equation}
	\BJst = \pdiff{\Bmtrst}{\Bmfpst},
	\label{eq:Jacobian}
\end{equation}\end{linenomath}
where it is important to note that the index $\st$ runs over the number of fixed probe stations (72). Because there are four different fixed probe configurations, there are several different Jacobian matrices.

Analytically, it is easier to calculate $\BJ^{-1}$, which represents how the measured Cartesian moments change as a function of the multipole moments. Because the field is linear in the moment strength parameter, the derivative with respect to the multipole strength is simply the measured Cartesian moments given a multipole moment strength of 1. The Cartesian moments are calculated from fixed probe measurements assuming a pure multipole input field. There are off-diagonal terms in the Jacobian matrix caused by asymmetries in the fixed probe positions.

\begin{figure}
	\centering
	\includegraphics[width=.95\linewidth]{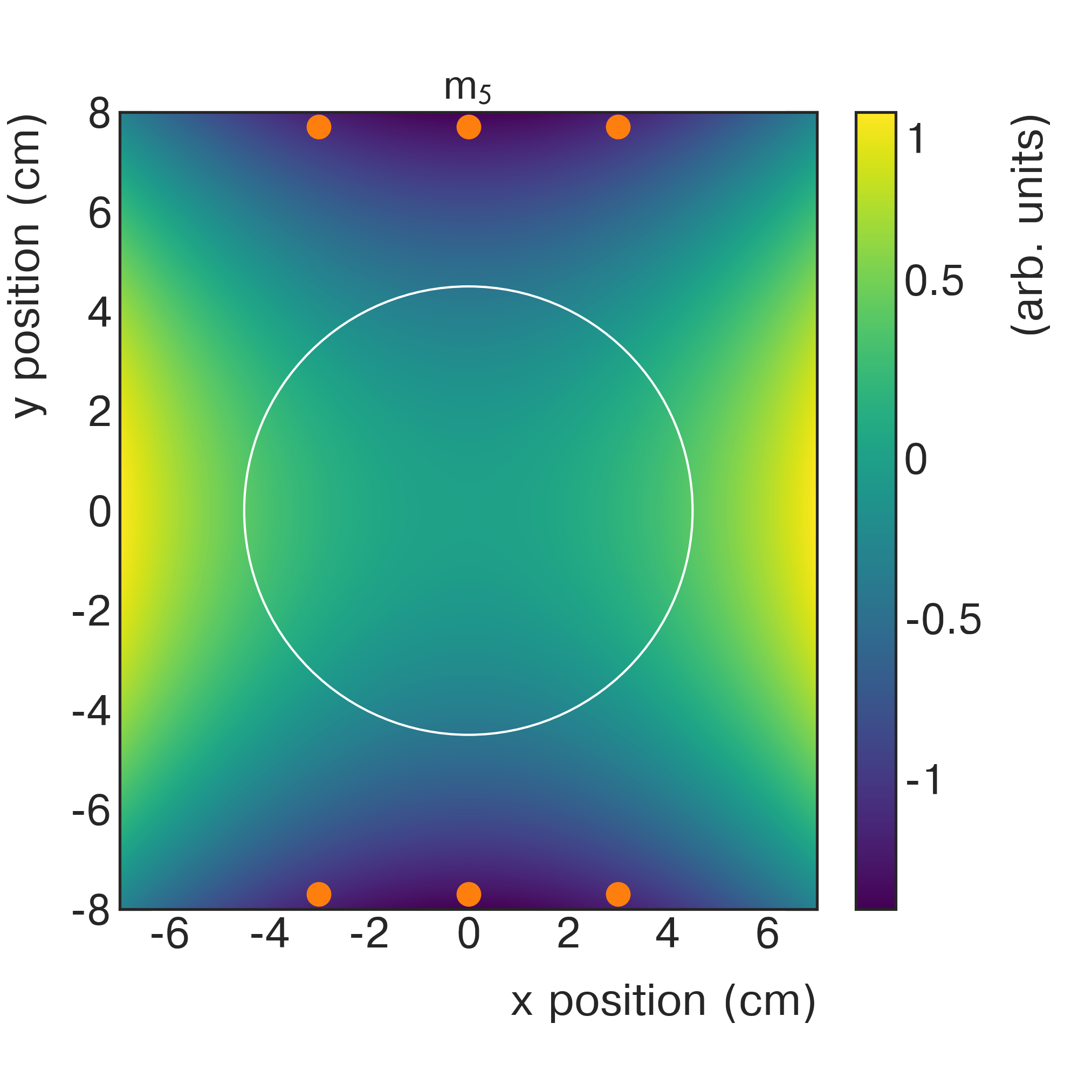}
	\caption{This shows a pure normal sextupole (\mfive) field. The fixed probes (light circles) are all located in a low region (dark). If we simply relate the dipole component in the muon storage region (depicted by the large circle) to the average of all six fixed probes, then the presence of a nonzero normal sextupole would bias the calculation.}
	\label{fig:sext_alias}
\end{figure}

There are two reasons that the two different bases are not identical. First, the \ac{NMR} probes' discrete nature can cause higher-order moments to alias into the extraction of lower-order moments. The fewer probes used to calculate a moment, the more this aliasing affects the measurement. For example, the normal sextupole \mfive causes a false dipole reading in the fixed probes. Because of their position above and below the muon storage region as shown in Fig. \ref{fig:sext_alias} the fixed probes are all located in regions where a shift due to a normal sextupole moment has the same sign. When the average of all six probes is taken, the contribution to the average field from a true normal sextupole moment will be nonzero, causing a biased magnetic dipole determination. In a six-probe station, the fixed probes can estimate the drift in the normal sextupole. The Jacobian is calculated to determine how much the sextupole aliases into the dipole measured by the fixed probes. Then the station's measurements of the normal sextupole moment are used to correct the dipole measurement. However, this correction procedure is impossible to repeat for higher-order moments that cannot be distinguished due to the fixed probes configurations.

The second reason that the two bases are not identical is that the fixed probes' position in a given station is not always symmetric. For example, in standard four-probe stations the position average is not at $(0,0)$ but at $(\SI{1.5}{\centi\meter}, 0)$. This radial shift means, for example, that a simple average of measurements from the four probes would be an approximation of the field at $(\SI{1.5}{\centi\meter}, 0)$, not at $(0,0)$. A correction would then need to be made that mixes the measured moments, using the horizontal gradient ($\mtwo$) to correct the field on center ($\mone$). There are other fixed probe stations with geometric configurations that are not already accounted for in the initial change of basis. For example, all the fixed probes in the beam injection vacuum chamber are translated radially inward by 1 cm with respect to the nominal configuration. This is shown in Fig.~\ref{fig:J_6_offset} together with the respective Jacobian that contains off-diagonal elements. Another example is the probe position in the four-probe stations of the vacuum chamber containing the trolley garage. These are not symmetric across the $x$ axis. Figure \ref{fig:all_jacobians} summarizes all relevant Jacobians for the various fixed probe station configurations present in the experiment.

\begin{figure}[htbp]
	\centering
	\begin{subfigure}[c]{0.95\linewidth}
		\centering
		\begin{tikzpicture}
			\node [draw, circle] (TI) at (-2.67, 2) {};
			\node [draw, circle] (TM) at (-0.67, 2) {};
			\node [draw, circle] (TO) at (1.33, 2) {};
			\node [draw, circle] (BI) at (-2.67, -2) {};
			\node [draw, circle] (BM) at (-0.67, -2) {};
			\node [draw, circle] (BO) at (1.33, -2) {};
			
			\path[<->]
				(TO) edge node[right, pos=0.33] {15.4 cm} (BO)
				(TI) edge node[above] {3 cm} (TM)
				(TM) edge node[above] {3 cm} (TO);
			
			\draw [dashed] (-2.5,0) -- (2.5,0);
			\draw [dashed] (0, -1.5) -- (0, 1.5);
			\node [pin={[outer sep=2pt]60:$\mathbf{O}$}] (O) at (0,0) {};
			
			\draw [thick, dotted] (-0.67, -1.5) -- (-0.67, 1.5);
			\draw [<->] (-0.67, 1) -- node[below] {\tiny$1\cm$} (0, 1);
			
			\draw [->] (-3.67,-3) -- node[below] {$\x$} (-1.5, -3);
			\draw [->] (-3.67,-3) -- node[left] {$\y$} (-3.67, -1.5);
		\end{tikzpicture}
	\end{subfigure}
	\hfill
	\begin{subfigure}[c]{0.95\linewidth}
		\centering
		\begin{equation*}
			\BJ = \left(
			\begin{matrix}
				1.0 & 0.222 & 0 & 0 & 2.681\\
				0 & 1.0 & 0 & 0 & 0.444\\
				0 & 0 & 1.0 & 0.444 & 0\\
				0 & 0 & 0 & 1.0 & 0\\
				0 & 0 & 0 & 0 & 1.0\\
			\end{matrix}
			\right )
		\end{equation*}
	\end{subfigure}
\caption{The geometry of an offset six-probe station and the corresponding Jacobian matrix. The change-of-basis matrix for offset stations is not corrected; instead, the correction for the offset is done with the Jacobian.}\label{fig:J_6_offset}\end{figure}
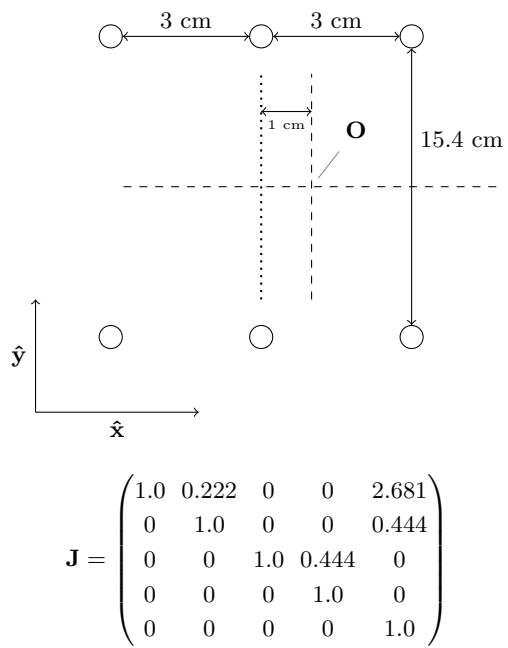
 
\begin{figure}
	\centering
	\begin{align*}
		\BJ_\text{6-probe} &= \left(
		\begin{matrix}
			1.0 & 0 & 0 & 0 & 2.632\\
			0 & 1.0 & 0 & 0 & 0\\
			0 & 0 & 1.0 & 0 & 0\\
			0 & 0 & 0 & 1.0 & 0\\
			0 & 0 & 0 & 0 & 1.0\\
		\end{matrix}
		\right )\\
		\BJ_\text{6-probe,offset} &= \left(
		\begin{matrix}
			1.0 & 0.222 & 0 & 0 & 2.681\\
			0 & 1.0 & 0 & 0 & 0.444\\
			0 & 0 & 1.0 & 0.444 & 0\\
			0 & 0 & 0 & 1.0 & 0\\
			0 & 0 & 0 & 0 & 1.0\\
		\end{matrix}
		\right )\\
		\BJ_\text{4-probe} &= \left(
		\begin{matrix}
			1.0 & 0 & 0 & 0 & 2.928\\
			0 & 1.0 & 0 & 0 & -0.667\\
			0 & 0 & 1.0 & -0.667 & 0\\
			0 & 0 & 0 & 1.0 & 0\\
			0 & 0 & 0 & 0 & 1.0\\
		\end{matrix}
		\right )\\
		\BJ_\text{4-probe,garage} &= \left(
		\begin{matrix}
			1.0 & 0 & 0 & 0 & 2.928\\
			0 & 1.0 & 0 & 0 & 0\\
			0 & -0.195 & 1.0 & 0 & 0\\
			0 & 0 & 0 & 1.0 & -0.195\\
			0 & 0 & 0 & 0 & 1.0\\
		\end{matrix}
		\right )
	\end{align*}
	\caption{The Jacobians for all four different fixed probe layouts present in the experiment. Recall that $\mfive$ at four-probe stations is estimated using the average of their nearest neighbors. These estimates are used to make corrections to the measured values, as seen in the $5\times5$ Jacobians for the four-probe stations.}
	\label{fig:all_jacobians}
\end{figure}

\section{Derivation of the Muon Distribution} \label{app:muon_dist}

This appendix details the derivation of the muon distribution used in Sec.~\ref{sec:MuonWeighting} from first principles.  As covered in Sec.~\ref{subsec:Introduction_Experiment}, the instantaneous anomalous spin-precession frequency of a muon in a magnetic field at position $(r, y, \phi)$ is
	\begin{linenomath}\begin{equation}
		\omega_a = -a_\mu \frac{q}{m} B(r, y, \phi).
	\end{equation}\end{linenomath}
The muon accumulates a phase as it travels around the ring until it decays at time $T$. As the decay times are short ($\SI{64}{\micro\second}$), the field drift is negligible over the time of the fill. The total phase accumulated by the muon from the beginning of the integration time until its decay is
	\begin{linenomath}\begin{equation}
		\Delta\varphi_k = a_\mu\frac{q}{m} \integral{B(\Br_{\mu_k}(t))}{t}{0}{T},
	\end{equation}\end{linenomath}
where $\Br_{\mu_k}(t)$ is the muon's position as a function of time. The subscript $k$ here indicates that this is a time average for the $k$th muon. The average frequency of the $k$th muon is
	\begin{linenomath}\begin{equation}
		\avg{\omega_a}_k = \frac{\Delta\varphi_k}{T}.
	\end{equation}\end{linenomath}

To convert this to an integral over azimuth instead of over time, the muon's path of a function of time is converted into the muon's $r-y$ position as a function of azimuth. The following substitutions are made, assuming that the path the muon follows is predominantly circular:
	\begin{linenomath}\begin{equation}
		\dd t = \frac{\dd l}{c} = \frac{r_{\mu_k}(\phi) \dd \phi}{c} ~ \implies ~ T = \frac{R}{c}\Phi.
	\end{equation}\end{linenomath}
In this equation, $\Phi$ is the accumulated azimuth; on average, it will approach values of thousands of radians. The term $r_{\mu_k}(\phi)$ is the radius of muon's path at a given azimuth, and $R = \avg{r_\mu}$ is the average radius. Making these substitutions,
	\begin{linenomath}\begin{equation}
		\avg{\omega_a}_k = a_\mu\frac{q}{m} \frac{1}{R\Phi}\integral{B(\Br_{\mu_k}(\phi)) r_{\mu_k}(\phi)}{\phi}{0}{\Phi}.
	\end{equation}\end{linenomath}

This integral can be extended to three dimensions by incorporating the muon path (in both the $r$ and $y$ directions) as delta functions and integrating $r$ and $y$ over the muon storage region:
	\begin{linenomath}\begin{align}
		\avg{\omega_a}_k &= a_\mu \frac{q}{m} \frac{1}{R \Phi} \int_0^\Phi\dd\phi \int_{r_1}^{r_2}\dd r \int_{-y_0}^{y_0}\dd y ~ \Bigl[ r B(r,y,\phi) \Bigr. \nonumber\\
		&\quad\times \Bigl. \delta(r - r_{\mu_k}(\phi)) \delta(y - y_{\mu_k}(\phi)) \Bigr].
	\end{align}\end{linenomath}
All the information about the muon's path is encoded in the delta functions so the field map and volume element $r$ can be integrated over 3D space. It is useful to split the integral over $\phi$ into a sum of integrals over single revolutions around the storage ring. These integrals are over $\phi\in[0,2\pi)$ and are parameterized by $n$, the number of cycles the muon makes. The muon makes $N + \Delta N$ total cycles. Going forward, the $\Delta N$ fractional cycle is neglected (it is, on average, less than 1\% of the total accumulated azimuth). Assuming that there are an integer number of cycles, the sum ranges from $n=0~\mathrm{to}~N-1$. Note that $\Phi \approx 2\pi N$. The only terms in the integral that depend on the parameter $n$ are the delta functions, so the sum can be included in the integrand, yielding
	\begin{linenomath}\begin{align}
			\avg{\omega_a}_k &= a_\mu \frac{q}{m}  \int_0^{2\pi}\dd\phi \int_{r_1}^{r_2}\dd r \int_{-y_0}^{y_0}\dd y ~ \Biggl[ r B(r,y,\phi) \Biggr. \nonumber\\
            &\quad \times \frac{1}{2\pi N R} \Biggl(\sum_{n=0}^{N-1}\delta[r - r_{\mu_k}(2\pi n + \phi)]\Biggr. \nonumber\\
            &\quad \times \Biggl.\Biggl. \delta[y - y_{\mu_k}(2\pi n + \phi)] \Biggr)\Biggr].
	\end{align}\end{linenomath}
	
The sum over delta functions is the distribution of the $k$th muon's position in the ring,
	\begin{linenomath}\begin{align}
        \rho_k(r,y,\phi) &= \frac{1}{2\pi N R}\Biggl( \sum_{n=0}^{N-1}\delta[r - r_{\mu_k}(2\pi n + \phi)] \Biggr. \nonumber\\
        &\quad \times \Biggl. \delta[y - y_{\mu_k}(2\pi n + \phi)] \Biggr).
	\label{eq:muon_dist}
	\end{align}\end{linenomath}
This is the normalized distribution with units of inverse volume, such that $\int\dd V~\rho_k = 1$. As a reminder, the subscript $k$ indicates that this is the average distribution for the $k$th muon. However, it is easy to see how this generalizes to the case of an average over many muons. The field map is constant for all muons in a fill, so the only averaging going from the case of a single muon to many will be averaging the distributions $\rho_k$ for each muon in the fill. Only muons that are included in the $\omega_a$ analysis are considered. Then, for a single fill, $\rho_k \rightarrow \rho^\mu$. As the total number of muon revolutions becomes very large, the distribution can be approximated as continuous because the muons average the field in the storage region. The final result is an average over all of the muons in a fill, $\avg{\omega_a}$, where
	\begin{linenomath}\small\begin{equation}
		\avg{\omega_a} = a_\mu \frac{q}{m} \int_0^{2\pi}\dd\phi \int_{r_1}^{r_2}\dd r \int_{-y_0}^{y_0}\dd y ~ r \rho^\mu(r,y,\phi) B(r,y,\phi).
	\label{eq:muon_fill_avg}
	\end{equation}\end{linenomath}\normalsize
        \,
 \newpage
    \bibliography{references}{}  
\end{document}